\documentclass[twocolumn]{aa}
\usepackage[varg]{txfonts}
\usepackage{graphicx}
\usepackage{amssymb}
\usepackage{amsfonts}
\usepackage{amsbsy}
\usepackage{amsmath}
\usepackage{subfig}

\begin{document}

\title{Unifying static analysis of gravitational structures with a scale-dependent scalar field gravity as an alternative to Dark Matter}
\titlerunning{Gravitation with a Scale-dependent Scalar Field as an Alternative to Dark Matter}
\authorrunning{Salzano et al.}

\author{V. Salzano\inst{\ref{inst1}}  \and D. F. Mota\inst{\ref{inst2}} \and S. Capozziello\inst{\ref{inst3}} \and N.R. Napolitano\inst{\ref{inst4}}}
\institute{Fisika Teorikoaren eta Zientziaren Historia Saila, Zientzia eta Teknologia Fakultatea, \\ Euskal
Herriko Unibertsitatea, 644 Posta Kutxatila, 48080 Bilbao, Spain \email{vincenzo.salzano@ehu.es}\label{inst1}
\and
Institute of Theoretical Astrophysics, University of Oslo, 0315 Oslo, Norway \email{D.F.Mota@astro.uio.no}\label{inst2}
\and
Dipartimento di Fisica, Universita' degli Studi di Napoli ``Federico II" and  INFN, Sezione di Napoli, Complesso
Universitario di Monte S. Angelo, Via Cinthia, Edificio N, 80126 Napoli, Italy \email{capozzie@na.infn.it}\label{inst3}
\and
INAF - Osservatorio Astronomico di Capodimonte, Salita Moiariello 16, I-80131 - Napoli, Italy \email{napolita@na.astro.it}\label{inst4}}

\date{}

\abstract {} {We investigated the gravitational effects of a scalar field within scalar-tensor gravity as an alternative to dark matter. Motivated by chameleon, symmetron and $f(R)$-gravity models, we studied a phenomenological scenario where the scalar field has both a mass (i.e. interaction length) and a coupling constant to the ordinary matter which scale with the local properties of the considered astrophysical system.} {We analysed the feasibility of this scenario using the modified gravitational potential obtained in its context and applied it to the galactic and hot gas/stellar dynamics in galaxy clusters and elliptical/spiral galaxies respectively. This is intended to be a first step in assessing the viability of this new approach in the context of ``alternative gravity'' models.} {The main results are: \textit{1.} the velocity dispersion of elliptical galaxies can be fitted remarkably well by the suggested scalar field, with model significance similar to a classical Navarro-Frenk-White dark halo profile; \textit{2.} the analysis of the stellar dynamics and the gas equilibrium in elliptical galaxies has shown that the scalar field can couple with ordinary matter with different strengths (different coupling constants) producing and/or depending on the different clustering state of matter components; \textit{3.} elliptical and spiral galaxies, combined with clusters of galaxies, show evident correlations among theory parameters which suggest the general validity of our results at all scales and a way toward a possible unification of the theory for all types of gravitational systems we considered. All these results demonstrate that the proposed scalar field scenario can work fairly well as an alternative to dark matter.}{}


\maketitle

\section{Introduction}
\label{sec:introduction}

Dark matter and dark energy are nowadays widely accepted as the main components of our Universe, although reliable clues about their origin, nature and properties are still missing.

There is a long list of possible dark matter candidates, ranging from standard to sterile neutrinos, from axions to super-symmetric candidates, from light to super-heavy scalar fields \citep[see e.g.][and references therein]{Bertone05,Krauss06}, while dark baryons seem to have a minor contribution. Concerning the nature of dark energy, the largest component in the mass-energy balance of the Universe with $\approx 69 \%$ from latest \textit{Planck} results \citep{planck}, a coherent cosmological model explaining all the open issues and the related observed phenomena has still to be found.

These two dark ingredients are the pillars of the current cosmological concordance model, the $\Lambda$-Cold Dark Matter ($\Lambda$CDM), where $\Lambda$ stands for the cosmological constant \citep{CarLam,Sahni} which is assumed to drive the Universe expansion. This model provides a suitable fit to most of the cosmological data \citep{planck,sanch05,Sel04,Teg03} but it is also well known that it is affected by many serious theoretical problems that motivate the search for alternatives.

On a phenomenological ground, alternative models to the dark matter have proposed a modification of the gravitational acceleration \citep{Milgrom83} within the context of MOND, which only later after his initial formulation was related to the context of the relativistic gravitation theory \citep{Bekenstein04,Bekenstein05,Sanders05}. On a more theoretical basis, more general candidates for the acceleration-driver counterpart have been proposed. Such models range from scalar fields rolling down self-interaction potentials to phantom fields, from phenomenological unified models of dark energy and dark matter to alternative theories of gravity \citep{Cap02,dfm3,dfm4,CapFra,Copeland06,dfm1,dfm2,Pad03,PB03,dfm5,mfrev}.

Among all these approaches, in \citet{MotaSalzano} (Paper I, hereafter) 
we have focused on a particular scenario, where a scalar field might be used to unify the cosmological scale acceleration of the Universe with the formation and dynamics of gravitational structures by mimicking dark matter on astrophysical scales.

Scalar fields play an important role in connecting cosmology and particle physics \citep{Binetruy,Linde}. In particular, theories like the chameleon fields \citep{Khoury04,shaw,Brax04}, $f(R)$ gravity \citep{fr1,fr2,fr3} and symmetron models \citep{sym1,sym2,sym3}, all share the fundamental ingredient of a scalar field which couples to matter via gravitational interaction, being short-ranged in highly dense regions, and long-ranged in low density regions. Such a scalar field would be relatively light on cosmological scales, thus describing cosmological evolution without differentiating much from the $\Lambda$CDM model, and, at the same time, it would also be able to satisfy local gravity constraints. Moreover, at different astrophysical scales, the effect of the scalar field would be suppressed or enhanced according to the local astrophysical density.

Starting from these broad properties, we propose a new parameterisation for a massive scalar field theory where we introduce a field mass (or an interaction length) and a coupling constant with ordinary matter which may change with scale. Our primary goal is to test whether this parametrization, based on the chameleon or symmetron theories, can account for different observational facts, regardless of the underlying physical mechanism producing such a scalar field. Some recent papers \citep{Oyaizu08,Brax12} analyse symmetron and f(R) gravity theories (of which chameleon models are an extension) in the context of structure formation, but they still consider the field as a background cosmology ingredient. Here we propose a scaling mechanism for which the field can both explain dark energy (on cosmological scales) and dark matter (on astrophysical scales).

In Paper I we studied the feasibility for this scale-dependent scalar field to work on different gravitational scales by using various cosmological indicators: type Ia Supernovae and their Hubble diagram, low surface brightness spiral galaxies and their rotation curves, clusters of galaxies and their mass profiles.

Here we want to go beyond Paper I, extend our analysis to elliptical galaxies and model their velocity dispersion profiles. As we will show in the following sections, this new test has provided us with new evidences that have motivated a modification of the original approach followed in Paper I and a new analysis of the gravitational objects studied in that same work (namely, spiral galaxies and clusters of galaxies). Finally we will unify all the results in a more general scheme.

The article is organised as follows: In \S\ref{sec:scalar_theory} we give a brief but exhaustive summary of all the main properties of the scalar field theory and describe all the main hypothesis underlying our work. In \S\ref{sec:obs_data} we accurately describe the used astrophysical data and the way we involved them in our analysis. In \S\ref{sec:results} we show results concerning elliptical galaxies and in \S\ref{sec:unified} we discussion the implications for a unified picture of all the gravitational structures we have considered. Conclusions are drawn in \S\ref{sec:conclusions}.

\section{The Scalar-Tensor field theory}
\label{sec:scalar_theory}

The most general action governing the dynamics of a scalar field can be written as \citep{Esposito01}:
\begin{eqnarray}\label{eq:scal-tens-action}
S &=& \frac{1}{16 \pi G_{\ast}} \int {\mathrm{d}}^{4}x \sqrt{-g} \left\{ F(\phi) {\mathcal{R}}
- Z(\phi) (\partial \phi)^{2} - V(\phi)\right\} \nonumber \\
&-& \int d^{4}x \, {\mathcal{L}}_{m}(\psi_{m}^{(i)}, g_{\mu \nu}^{i}) \; ,
\end{eqnarray}
where $g$ is the determinant of the metric $g_{\mu\nu}$, $\mathcal{R}$ is the Ricci scalar, $\psi_{m}^{(i)}$ are the various matter fields, ${\mathcal{L}}_{m}$ is the Lagrangian density of ordinary matter, $\phi$ is the scalar field, $F(\phi)$ and $Z(\phi)$ are some functions of the field which regulate its dynamics, and $V(\phi)$ is the scalar field potential. Depending on the expression of the given functions $F(\phi)$ and $Z(\phi)$ and of the potential $V(\phi)$ one can recover general scalar field theories, like the chameleon or the symmetron mechanisms. The main consequence of this action is easily seen when linear perturbations of matter are taken into account. A wide class of theories leads to a perturbation equation like this \citep{sym3,Brax04,Bertschinger08}:
\begin{equation}\label{eq:pert_eq}
\ddot{\delta_{m}} + 2 H \dot{\delta_{m}'} = \frac{3}{2} \Omega_{m} H^2 \frac{G_{eff}}{G_{N}} \delta_{m}\; ,
\end{equation}
where the dots mean time derivative, $k$ is the wave-number length, $a$ is the scale factor, $H$ is the Hubble function, $\delta_{m}$ is the matter density contrast and $\Omega_{m}$ is the matter density parameter. The quantity $G_{eff}$ can be interpreted as an \textit{effective} gravitational constant: in general relativity the gravitational coupling $G_{N}$ is a constant, while in many alternative theories of gravity the strength of gravity can vary with time and place. A time-varying gravitational coupling is a well known property of scalar-tensor theories and a generic feature of all modified gravity theories where the Newtonian potential and the spatial curvature potential are different \citep{Bertschinger08,Acquaviva05,Clifton05}. Within the context of massive scalar field models it has the general expression \citep{Gannouji09}:
\begin{equation}\label{eq:G_cham}
G_{eff}(a; \beta, m; k) = G_{N} \left( 1 + 2 \beta^{2}
\frac{\frac{k^{2}}{a^{2} m^{2}}}{1 + \frac{k^{2}}{a^{2} m^{2}}}
\right) \; ,
\end{equation}
where $\beta$ is the coupling constant of the scalar field with matter and, if the field is at the minimum, the scalar field mass is $m^2 = V_{,\phi\phi}$. In particular, the term proportional to $\beta^2$ results from the scalar field-mediated force, which is negligible if the physical length scale of the perturbation is much larger than the range of the scalar field-mediated force, namely, if $a/k \gg m^{-1}$. In this case 
matter fluctuations grow as in general relativity.

Taking the inverse Fourier transform of Eq.~(\ref{eq:G_cham}) it is straightforward to obtain the corresponding expression of the gravitational potential for a point mass distribution, $\psi(r)$. Remembering that a potential $\propto \frac{1}{r}$ in real space yields a $k^{-2}$ term in Fourier space, we can recognise in Eq.~(\ref{eq:G_cham}) the point-like gravitational potential per unit mass:
\begin{equation}\label{eq:grav_pot_point}
\psi (r) = -\frac{G}{r} \left( 1 + 2 \beta^2 e^{-m r}\right) = -\frac{G}{r} \left( 1 + 2 \beta^2 e^{-r/L}\right) \; ,
\end{equation}
where $m$ is the mass of the scalar field, $L \propto m^{-1}$ is the interaction range of the scalar field, and $\beta$ still being the coupling constant between matter and the scalar field. The gravitational potential given in Eq.~(\ref{eq:grav_pot_point}) has been calculated for a point-like source, and it has to be generalized to extended systems in a numerical way. Depending on the particular gravitational system we consider, we will adopt different geometrical hypothesis: for cluster of galaxies and elliptical galaxies we will adopt spherical symmetry, while spiral galaxies will be assumed as thin disks.

The point-like potential can be split in two terms. The \textit{Newtonian} component for a point-like mass $m$ is:
\begin{equation}
\psi_{N}(r) = -\frac{G m}{r} \; ,
\end{equation}
and its extended integral is the well-known expression:
\begin{equation}
\Psi_{N}(r) = -\frac{G M(<r)}{r} \; ,
\end{equation}
where $M(<r)$ is the mass enclosed in a sphere with radius $r$. The \textit{correction} term from the scalar field effect is:
\begin{equation}
\psi_{C}(r) = -\frac{G m}{r} \left(2 \beta^2 e^{-\frac{r}{L}}
\right) \; ,
\end{equation}
whose extended integral is given by:
\begin{equation}
\Psi_{C}(r) = \int_{0}^{\infty} r'^{2} dr' \int_{0}^{\pi} \sin \theta'
d\theta' \int_{0}^{2\pi} d\omega' \psi_{C}(r') \; ,
\end{equation}
where the angular part is analytically derivable, while the radial integral has to be numerically estimated once the mass density is given. A fundamental difference between the corrected and the Newtonian terms is that in the latter the matter outside the spherical shell of radius $r$ does not contribute to the potential, while in the former the external mass distribution enters into the potential integral, with a possible non-negligible contribution.

We also observe that a possible dependence of the coupling constant with scale, i.e. $\beta = \beta(r)$, should be considered when evaluating the integral or the physical observable quantities that we will define in following sections. As we do not know what is the possible analytical behaviour of $\beta(r)$, we will assume it as a constant or weakly depending on the scale, i.e. $\mathrm{d}\beta/\mathrm{d}r \approx 0$, as we will verify \textit{a posteriori}.

\subsection{Hypothesis}
\label{sec:hypothesis}

In this section we want to describe in more details what are the main properties and requirements of our parametrization and the motivations behind our approach 

In Paper I we assumed only one single scalar field working at each considered gravitational scale and/or object. In particular, we focussed on type Ia Supernovae, clusters of galaxies and spiral galaxies. The scalar field was characterized by: interaction length $L$ (or a mass), which should be related to the dimension of the gravitating structure under exam; and a coupling constant $\beta$, indicating the strength of the interaction between the field and the \textit{kind of matter} which constitutes the considered gravitational object. We worked under the hypothesis that the \textit{matter} pie was made only of the observed baryons (hot gas and galaxies in clusters of galaxies; gas and stars in spiral galaxies) with the scalar field generating a dynamical effect similar to the classical dark matter.
In practice, we replaced the eventual new and exotic dark matter component with an \textit{effective} mass 
induced by the modified gravitational interaction from the scalar field with ordinary matter.

First of all, we found that while type Ia Supernovae could be \textit{theoretically} used to detect an effective gravitational constant $G_{eff}$, because this can affect their light curves by changing both the thermonuclear energy release and the time scale of stellar explosion, \textit{actually} its effects are too weak to be clearly detected with present data.

More interestingly, in Paper I we showed that both the rotation curves of low surface brightness spiral galaxies and the matter profiles in clusters of galaxies, obtained using only visible galactic, stellar and gas mass components while substituting dark matter with the proposed scalar field, can be fairly fitted within our alternative scenario. The interaction length values of the scalar field are in turn consistent with the characteristic dimensions of the considered gravitational systems. On the other hand, the coupling between the field and ordinary baryonic matter is convincingly well constrained in ranges which scale quite well with the matter content of galaxies and/or clusters of galaxies.

All these results seemed to point towards the possibility of a unifying view of dark matter and dark energy via a scalar field with the properties we have assumed, at least at galactic and cluster scales. But it is important to stress that all these results from Paper I were obtained assuming the coupling constant $\beta$ to be \textit{unique} and fixed for all the intervening mass components.

As we will show in \S\ref{sec:results}, this turns out not to be the case for elliptical galaxies. For them we have to consider the possibility that the scalar field coupling constant shall have different values depending on the different mass components of a galaxy. This eventuality might have two implications: \textit{1.} we have separate values for the coupling constants, one for each baryonic mass component (i.e. stars and gas), or \textit{2.} we have only one coupling constant but its measurement might be affected by the matter \textit{clustering} state.

It is possible to state that these two options are not conflicting with each other. Indeed, in the first case, the scalar field theory predicts that the field can couple in different ways with different kinds of matter \citep[see e.g.]{Brax04}. This would mean that the scalar field can couple differently with \textit{ordinary} matter (baryons, neutrinos, quarks, and so on). In the second case, the different values could depend on the clustered states of the matter and be a consequence of a screening effect which can suppress the field effects and produce an apparent (measured) lower value of the coupling constant $\beta$. Such a screening effect is called \textit{thin-shell effect} in the chameleon theory and a similar effect is also present in the symmetron theory. It mainly affects gravitational systems where the inner value of the scalar field is different from the background and also reflects in a difference between the inner and the external matter density \citep{Brax04,tsuji,sym3}.

We \textit{suggest} carrying out a possible mixed scenario in order to make the two options above coexisting. Let us take the case of a cluster of galaxies and consider the galaxies and the gas inside it: if the scalar field scales with the density, we can argue that there are two scalar fields, one driving the formation and the dynamics of the cluster and another one driving the formation and the dynamics of the galaxies inside it. We can think that all the sub-structures inside the cluster experience the cluster-scale scalar field but, as long as the systems evolve, there will be a point where the (over--)densities representing the galaxies are large enough to turn on the screening effect, which from then on washes the cluster-scale scalar field effect out. This process would result in a suppression of the coupling constant of the cluster-scale scalar field with matter in galaxies, as long as these latter ones can be considered as clustered small structures within a larger structure.

In this picture, the diffuse hot gas in the cluster is spread through the cluster scale and can be considered as having only one typical scale, that of the cluster. This assumption is valid provided that the hydrostatic equilibrium is realized. In this case, the scalar field-gas coupling constant should contain information about the cluster--scale scalar field. The same assumption cannot be equally made for galaxy systems, like spiral and elliptical galaxies, where the gas can be strongly disturbed by local phenomena (as stellar winds, supernovae, radio jets by active nuclei etc.).

Within this framework, the total gravitational potential can be written in two equivalent ways.
We can define:
\begin{equation}\label{eq:total corrected potential1}
\Psi(r) = \Psi_{N}(r) + \Psi_{C}(r) \; ,
\end{equation}
if we want to highlight the separation between the correction that the scalar field provides to the gravitational potential and the classical Newtonian term. Alternatively, we can write:
\begin{equation}\label{eq:total corrected potential2}
\Psi(r) = \Psi_{star}(r) + \Psi_{gas}(r) \; ,
\end{equation}
where the suffix \textit{star} refers to the stellar component in galaxies, but it can be replaced by \textit{galaxy} when writing the cluster potentials. Each of the two terms is finally given by:
\begin{equation}\label{eq:total corrected potential3}
\Psi_{star}(r) = \Psi_{N,star}(r) + \Psi_{C,star}(r; \beta_{star}, L) \; ,
\end{equation}
and
\begin{equation}\label{eq:total corrected potential4}
\Psi_{gas}(r) = \Psi_{N,gas}(r) + \Psi_{C,gas}(r; \beta_{gas}, L) \; .
\end{equation}
We point out that the same scalar field, with mass $\propto L^{-1}$, can interact differently with ordinary matter (that is why we have two values for the coupling constant, $\beta_{star}$ and $\beta_{gas}$) depending on their clustering state.

Finally, we have to remark here that our approach is implicitly based on a \textit{static} assumption for the gravitational structures we are considering. Namely, we are ignoring that: \textit{1.} the values of the scalar field parameters could be subjected to temporal evolution, so that an analysis of how perturbations and over-density collapse work would be necessary; and \textit{2.} a dynamical analysis under the influence of a scalar field should be performed to verify the stability of such gravitational systems.

\section{Elliptical Galaxies: working model}
\label{sec:obs_data}

As pointed out in \S1, we want to extend the test of our scalar field hypothesis on galactic scales with elliptical galaxies.

\subsection{Preliminaries}

While spiral galaxies have easily interpretable flat rotation curves (that have been one of the first historical evidences for dark matter), elliptical galaxies are pressure supported systems dominated by hot random motions. The orbital distribution of stars is very difficult to model, and consequently the mass distribution is highly uncertain because of the well known mass--anisotropy degeneracy.

One way to have insight into the internal dynamics is to use the information stored in the line--of--sight velocity dispersion as a function of position by solving the Jeans equation. Under spherical symmetry, and assuming no-rotation, the only effective equation governing the galaxy equilibrium is the radial Jeans equation:
\begin{equation}\label{eq:jeans}
\frac{\mathrm{d} (\ell \, \sigma_{r}^2)}{\mathrm{d}r} + 2 \frac{\beta_{a}}{r} \, \ell \, \sigma_{r}^2 = - \ell \, \frac{\mathrm{d}\Psi(r)}{\mathrm{d}r} \; ,
\end{equation}
where $\ell(r)$ is the luminosity density of the galaxy, $\sigma_{r}(r)$ the radial velocity dispersion and $\Psi(r)$ the total gravitational potential. The anisotropy parameter $\beta_{a}$ is defined as:
\begin{equation}\label{eq:anisotropy}
\beta_{a} = 1 - \frac{\sigma_{t}^2}{\sigma_{r}^2}
\end{equation}
where $\sigma_t$ is the one-dimensional tangential velocity dispersion (defined as a combination of the two angular components of the velocity dispersion tensor, $\sigma_t^2=(\sigma_\theta^2+\sigma_\varphi^2)/2$) and $\sigma_{r}$ is the radial component.
When $\sigma_t=\sigma_r$, the system is called isotropic and $\beta_{a} = 0$; when $\beta_{a} = 1$ the system is fully radial anisotropic; while for $\beta_{a} \rightarrow -\infty$ it is fully tangential.

In the Eq.~(\ref{eq:jeans}), the unknown quantities are the anisotropy parameter and the mass which generates the potential, while $\ell (r)$ is given by the tracer distribution. Thus, different combinations of orbital anisotropy and radial distribution of the mass can produce the same observed dispersion profile. This mass--anisotropy degeneracy can be solved by using independent measurement for the mass. One possibility for that
is to use the information from X-ray emission from the hot gas \citep{Mathews03}, i.e. density and temperature, and to solve the hydrostatic balance within the galaxy potential, provided that the gas is at the hydrostatic equilibrium, but this is not always true in elliptical galaxies \citep{Diehl07,Humphrey06}.

Despite all these modelling complications, Jeans analysis has been extensively used in elliptical galaxies, taking great advantage of discrete kinematical tracers probing the gravitational potential out to many effective radii ($R_{\mathrm{eff}}$).
Globular clusters \citep{Puzia04,Bergond06,Romanowsky09,Shen10,Schuberth10,Woodley10}
or planetary nebulae (PNe, see e.g. \citealt{Napolitano01,Napolitano02,Mendez01,Peng04,Douglas07,Coccato09,Teo10,Napolitano11} )
have made it possible to extend kinematics up to $5$-$7$ $R_{\mathrm{eff}}$, from the $\approx R_{\mathrm{eff}}$ achievable with only stellar observations. Furthermore, the analysis of satellites orbiting around the galaxies could extend up to $50$-$500$ kpc ($\lesssim 10$ $R_{\mathrm{eff}}$) \citep{Klypin09}. However, globular clusters have classically been used as mass tracers for bright galaxies, but their samples are too small in ordinary elliptical galaxies. On the other side, PNe have been systematically used to map the mass profile of ellipticals (see e.g. \citealt{Romanoswky03, Napolitano09}, \citealt{Napolitano11}, N+11 hereafter, and references there in).

The selection of viable objects for our analysis has been made within the elliptical galaxy sample observed with the Planetary Nebula Spectrograph \citep{Douglas02} and presented in \citep[C+09 hereafter]{Coccato09}, where stellar kinematics of the central regions are combined with PNe kinematics of the galaxy regions outside the $R_{\mathrm{eff}}$. PNe data give strong hints about the mass profiles of elliptical galaxies, but also put forward much more questions. Nowadays many alternative scenarios are equally feasible. Two exemplary cases are \citealt{Romanoswky03} and \citealt{Dekel05}: in the former, using the PNe, the galaxies velocity dispersion profiles are found to decline with radius and dynamical modelling of the data indicates the presence of little if any dark matter in these galaxies halos; in the latter, starting from disc-galaxy merger simulations, the lower than expected velocities are in fact compatible with galaxy formation in dark matter halos, thus it depends on inner dynamics (elongated orbits) or on projection effects.

For our analysis, we are interested in galaxies that have both extended (stellar and/or PNe) kinematics and published X-rays observations (see e.g. \citealt{Fukazawa06}). We need both because we want to explore the coupling of the scalar field with \textit{all} the mass components of the gravitational systems under exam.

Among the 16 galaxies reported in C+09, the only galaxy which had a complete dataset for our purpose is NGC 4374, since other galaxies for which both long slit and PNe kinematics were available, like NGC 3377, NGC 3379 and NGC 4494, did not have a reliable de-projected X-ray emitting gas density profile\footnote{Fukazawa, private communication.}. All the required photometric properties of NGC 4374 are reported in Table~\ref{tab:tabdata}.

{\renewcommand{\tabcolsep}{2.mm}
{\renewcommand{\arraystretch}{1.5}
\begin{table*}
\begin{center}
 \caption{\textit{Elliptical galaxies.} Column 1: Galaxy name. Column 2: photometric band. Column 3: modulus distance from \cite{Tonry01} and shifted by $-0.16$ mag as explained in \cite{Coccato09}. Column 4: distance of galaxy derived from modulus distance. Columns 5: total B magnitude corrected for extinction and redshift. Columns 6: Sersic scale radius for a $n=4$ profile (classical De Vaucouleurs profile). Column 7: Sersic scale radius determined from the Sersic fit. Column 8: Sersic shape parameter. Column 9: Stellar Surface brightness at $a_{s}$. Column 10: maximum distance from the galaxy center of PNe derived kinematics (maximum distance from the galaxy center of PNe detections).\label{tab:tabdata}}
\begin{tabular}{cccccccccc}
  \hline
  name      &  band & $\mu_D$ & $D$     &  $B_{T}$ & $a^{(4)}_s$ & $a_s$    &  $m$   & $\mu_{s}$          & $R_{LAST}$            \\
            &       &         &(Mpc)    &          & (kpc)       & (kpc)    &        & (mag arcsec$^{-2}$)& (kpc)                 \\
  \hline
  \hline
  NGC4374   &  V    & $31.16 $& $17.06$ & $10.01$  &  $5.97$     & $9.34$  & $6.11$  & $23.1$ & $25.23 \; (34.07)$    \\
  \hline
  \hline
\end{tabular}
\end{center}
\end{table*}}}

\subsection{Line-of-sight velocity dispersion}

The general solution to the Jeans equation, Eq.~(\ref{eq:jeans}), is:
\begin{equation}\label{eq:sol_jeans}
\ell(r) \sigma_{r}^2 (r) = \frac{1}{f(r)} \int_{r}^{\infty} f(s) \, \ell(s) \frac{\mathrm{d}\Psi(s)}{\mathrm{d}s} \mathrm{d}s \; ,
\end{equation}
where the function $f$ is the solution to
\begin{equation}
\frac{\mathrm{d} \ln f}{\mathrm{d} \ln r} = 2 \beta_{a}(r) \; .
\end{equation}
By projecting the velocity ellipsoid along the line of sight one can obtain the line-of-sight velocity dispersion, which
is the kinematical quantity observed and reported in C+09:
\begin{equation}\label{eq:los_dispersion}
\sigma_{los}^2(R) = \frac{2}{I(R)} \left[ \int_{R}^{\infty} \frac{\ell \, \sigma_{r}^2 \, r}{\sqrt{r^2-R^{2}}} \mathrm{d}r
-  R^2 \int_{R}^{\infty} \frac{\beta_{a} \, \ell \, \sigma_{r}^2}{r \sqrt{r^2-R^2}}\mathrm{d}r \right] \; ,
\end{equation}
where $R$ is the projected distance from the center of the galaxy and $I(R)$ is the stellar surface brightness profile.
In order to calculate the line-of-sight velocity dispersion $\sigma_{los}$ one needs two ingredients: an analytical expression for the
anisotropy function and the total gravitational potential (which enters in $\sigma_{r}$).

Concerning the anisotropy function, the usual way of proceeding is to compare observations with profiles derived from cosmological N-body simulations. Many models can be used: the simplest isotropy ($\beta_{a} = 0$); a constant anisotropy profile; the Osipkov-Merritt model, however it provides a poor fit to the simulations. We decided to work with the anisotropy model given in \cite{MamonLokas1,MamonLokas2,MamonLokas3}:
\begin{equation}\label{eq:mamon_anisotropy}
\beta_{a}(r) = \frac{1}{2} \frac{r}{r + r_{a}} \; ,
\end{equation}
where $r_{a}$ is a typical anisotropy length, assumed to be $r_{a} \simeq 14 \, R_{\mathrm{eff}}$, this value\footnote{In order to be sure that our results are free of this particular choice, we have performed our analysis changing the length parameter $r_{a}$ from the chosen best value, $r_{a} \simeq 14 R_{eff}$, to $r_{a} \simeq 1.4 R_{\mathrm{eff}}$ (as discussed in \citep{MamonLokas2}), spanning a wide range of values. We can conclude that our results are completely unaffected by this choice.} thus providing a good fit to data from dissipation-less cosmological N-body simulations (for a more exhaustive discussion about reliable anisotropy models, see Fig.~2 and Section~3.2 of \cite{MamonLokas2}.

\subsection{Modelling galaxy components}

Finally, to calculate the gravitational potential, we need to model the galaxy components. In this case we have: stars, hot gas
and the central black hole.

The stellar luminosity density can be obtained by de-projecting the observed surface brightness profile $I(R)$; galaxies in our sample
are fitted with the mostly used Sersic profile:
\begin{equation}\label{eq:sersic}
I(R) = I_{0} \exp \left[ -\left( \frac{R}{a_{s}}\right)^{1/m}\right] \; ,
\end{equation}
where $I_{0}$ is the central surface brightness (in units of L$_{\odot}$ pc$^{-2}$), $a_{s}$ the Sersic scale parameter (in kpc) and $m$ the Sersic shape parameter. The luminosity density can be obtained by the approximation first proposed in \citep{Prugniel94}:
\begin{equation}\label{eq:luminosity_density}
\ell(r) \equiv \ell_{1} \widetilde{\ell}(r/a_{s}) \; ,
\end{equation}
with:
\begin{equation}
\widetilde{\ell}(x) \simeq x^{-p} \, \exp(-x^{1/m}) \; ,
\end{equation}
\begin{equation}
\ell_{1} = \frac{L_{tot}}{4 \pi \, m \, \Gamma[(3-p)m] a_{s}^{3}} \; ,
\end{equation}
where the function $p$ is defined in \citep{LimaNeto99} as:
\begin{equation}
p \simeq 1.0 - 0.6097/m + 0.05463/m^2 \; .
\end{equation}
The total galaxy luminosity (in solar units) in the V-band, where observations for NGC4374 were performed, is:
\begin{equation}
L_{tot}=10^{-0.4(B_{T}-\mu_{D}-C_{BV}-M_{B,\odot})} \; ,
\end{equation}
where $B_{T}$ is the galaxy B-band apparent magnitude; $M_{B,\odot}$ is the Sun absolute magnitude in the B-band; $C_{BV}$\footnote{It is obtained from the extragalactic database \textit{Hyperleda}\footnote{http://leda.univ-lyon1.fr/})} is the galaxy color, needed to convert all luminosity parameters from the band B to the band V; and $\mu_{D}$ is the galaxy distance modulus (see Table~\ref{tab:tabdata}). It is worth stressing that we also need:
\begin{equation}
I_{0} = \frac{L_{tot}}{2 \pi \, m \, \Gamma[2m] a_{s}^2} \, .
\end{equation}
which appears in Eq.~(\ref{eq:los_dispersion}) through the surface brightness expression $I(R)$, and the stellar mass-to-light ratio $Y_{\ast}$ in order to convert the luminosity density into the mass density which enters in the gravitational potential.

The central black hole is assumed to have a constant density inside its typical size, the Schwarzschild radius $r_{BH}$, which is:
\begin{equation}
r_{BH} = \frac{2 \, G_{N} \, M_{BH}}{c^2} \; ,
\end{equation}
where of course $G_{N}$ is the Newton gravitational constant and $c$ the light velocity. The black hole mass $M_{BH}$ is obtained using the relation which exists between a super massive black hole and the host galaxy luminosity \citep{Gultekin09}:
\begin{equation}
M_{BH} = 10^{8.95+1.11 \cdot \log[10, L_{tot}/10^{11}]} \; .
\end{equation}
Its density is assumed to be null outside $r_{BH}$.

The gas profiles are assumed to follow the traditionally used $\beta$-model\footnote{The $\beta$ that appears here does not have any relation
with the scalar field coupling constant.} \citep{Cavaliere78}:
\begin{equation}
\rho_{gas}(r) = \rho_{0}^{gas}\left(1 + \left(\frac{r}{a_{g}}\right)^2\right)^{3\beta_{g}/2} \; ,
\end{equation}
where the central gas density, $\rho_{gas,0}$, the gas core length, $a_{g}$, and the constant $\beta_{g}$ are provided by \citep{Fukazawa06}, where are  derived fitting mass profiles with hot gas X-ray emissions.

We remember that in the classical Newtonian approach the total dynamical mass is made of two counterparts, dark matter and baryons (stars, gas and
black hole), so that:
\begin{eqnarray}\label{eq:mass_classical}
M_{tot}(r) &=& \frac{r^2}{G} \frac{\mathrm{d} \Psi_{N}}{\mathrm{d}r} =\nonumber \\
&=& \frac{r^2}{G} \left( \frac{\mathrm{d} \Psi_{N,bar}}{\mathrm{d}r} + \frac{\mathrm{d} \Psi_{N,DM}}{\mathrm{d}r} \right) = \nonumber \\
&=& M_{bar}(r) + M_{DM}(r) \; ,
\end{eqnarray}
where $\Psi_{N}$ is the Newtonian potential and:
\begin{eqnarray}
\Psi_{N,DM} &\leftrightarrow& \rho_{DM} \\
\Psi_{N,bar} &\leftrightarrow& \rho_{bar} \sim \rho_{star} + \rho_{gas} + \rho_{BH}\; . \nonumber
\end{eqnarray}
As we pointed out in the \S~(\ref{sec:scalar_theory}), in our  approach the total gravitational potential is made of a Newtonian term and a corrective one:
\begin{equation}
\Psi = \Psi_{N} + \Psi_{C} \; ,
\end{equation}
so that the dynamical mass is:
\begin{eqnarray}\label{eq:mass_chameleon}
M_{tot}(r) &=& \frac{r^2}{G} \frac{\mathrm{d} \Psi}{\mathrm{d}r} =\nonumber \\
&=& \frac{r^2}{G} \left( \frac{\mathrm{d} \Psi_{N}}{\mathrm{d}r} + \frac{\mathrm{d} \Psi_{C}}{\mathrm{d}r} \right) = \nonumber \\
&=& M_{bar}(r) + M_{eff}(r)  \; ,
\end{eqnarray}
where
\begin{eqnarray}
\Psi_{N} &\leftrightarrow& \rho_{bar} \sim \rho_{star} + \rho_{gas} + \rho_{BH} \\
\Psi_{C} &\leftrightarrow& \rho_{star} + \rho_{gas} + \rho_{BH} + \mathrm{field \; correction} \; . \nonumber
\end{eqnarray}
So, the \textit{effective} mass $M_{eff}$ is due to the modification of gravity produced by the scalar field, instead of
requiring a new kind of matter as the dark one. Of course, the term $\Psi_{C}$ has contributions only from visible baryonic
mass, i.e. stars, gas and black hole plus correction induced by the scalar field.

By comparing Eq.~(\ref{eq:mass_classical}) and Eq.~(\ref{eq:mass_chameleon}) it is straightforward to observe that if we want
that the scalar field fits data as well as dark matter we need:
\begin{equation}
M_{eff}(r) \sim M_{DM}(r) \; .
\end{equation}
For this reason we also realize a fit of our data in the classical context of dark matter in order to perform a comparison between the two approaches.
We use the classical Navarro-Frenk-White (NFW) model density given by the relation \citep{NFW96} :
\begin{equation}\label{eq:NFW}
\rho_{DM}(r) = \rho_{0}^{DM} \left(\frac{r}{a_{d}}\right)^{-1} \left[1+\frac{r}{a_{d}} \right]^{-2} \; .
\end{equation}
In \citep{MamonLokas2} more dark matter models are considered: the generalised NFW model introduced by \citep{JingSuto}, with inner slope $-3/2$ instead of $-1$ as it is in the classical NFW profile; the convergent model of \citep{Navarro04}, with an inner slope that is a power-law function of radius. We have verified that using these two models does not give any substantial change in the general mass profiles and in the fitting of velocity dispersion curves; thus, the classical NFW model is sufficient for our requirements. Generally, one is used to convert the quantities appearing in Eq.~(\ref{eq:NFW}), i.e. $(\rho_{0}^{DM}, \, a_{d})$ in more useful quantities as the virial radius, $r_{v}$, namely the radius enclosing a mass whose mean density is $\approx 100$ times the critical density of the Universe, and the concentration parameter, $c_{vir} \equiv \frac{r_{v}}{a_{d}}$.

Finally, with all these ingredients, we are able to derive the observed quantity, i.e. the line-of-sight velocity dispersion, $\sigma_{los}$. We underline here that the data we are considering are derived from the sum in quadrature of two terms: the line-of-sight velocity dispersion and of the rotation velocity. The final quantity, $\sqrt{\sigma_{los}^2 + v^2}$ (see N+11), is a more efficient indicator of the total kinetic energy and is essentially $\approx \sigma_{los}$, because the rotation velocity in NGC4374 is not dynamically significant compared to random motion, being $\sim 50$ km$/$s$^{-1}$ against a velocity dispersion of $\sim 200-250$ km$/$s$^{-1}$.

Depending on the approach we consider, $\sigma_{los}$ will be a function of different sets of parameters. On one hand, in the classical approach with a NFW density profile for the dark matter component it will be $\sigma_{los} = \sigma_{los}(R; \rho_{0}^{DM}, a_{d}, Y_{\ast})$. On the other hand, when the modified gravity approach with a scalar field is considered, it will be $\sigma_{los} = \sigma_{los}(R; \beta, L, Y_{\ast})$. The statistical analysis to search for the parameters values which best fit our working model will be based on the minimisation of the chi-square function, defined as:
\begin{equation}
\chi^{2} = \sum_{j=1}^{\mathcal{N}}\frac{(\sigma_{los,th}(R_{i}; \{\boldsymbol{\theta}_{i}\})-\sigma_{los,obs}(R_{i}))^2}{\sigma_{i}^{2}}
\end{equation}
where $\mathcal{N}$ is the number of data points, $\sigma_{i}^{2}$ are the observationally-derived measurement variances and $\{\boldsymbol{\theta}_{i}\}$ is the parameters theory vector i.e., respectively, $\{\boldsymbol{\theta}_{i}\}= \{\rho_{0}^{DM}, a_{d}, Y_{\ast}\}$ in the dark matter approach, and $\{\boldsymbol{\theta}_{i}\}= \{\beta, L, Y_{\ast}\}$ in the scalar field approach.

To minimize the $\chi^2$ we use the Markov Chains Monte Carlo Method (MCMC) and test their convergence with the method described by \citep{Dunkley05}. The MCMC method makes it possible to fix some priors on the fitting parameters. As a conservative choice, we decided to leave them as free as possible: we have $\beta>0$ (given that in all the expressions above we always have $\beta^2$, we do not really have the possibility to distinguish between a positive or negative value; moreover, the scalar field theory predicts it to be positive), and $L$, $\rho_{0}^{DM}$, $a_{d}$ and $Y_{\ast}>0$ are all positive definite quantities.

\section{Elliptical Galaxies: analysis and results}
\label{sec:results}

\subsection{Mock galaxy test}

In order to check the validity of our analysis and how much the observed velocity dispersion is fitted to derive clues about the scalar field properties, we start by performing an ideal-case study, using a mock galaxy with all its intrinsic quantities fixed following the same prescriptions of \citep{MamonLokas2}.

\begin{figure*}
\centering
\hspace{-0.8cm}
  \includegraphics[width=90mm]{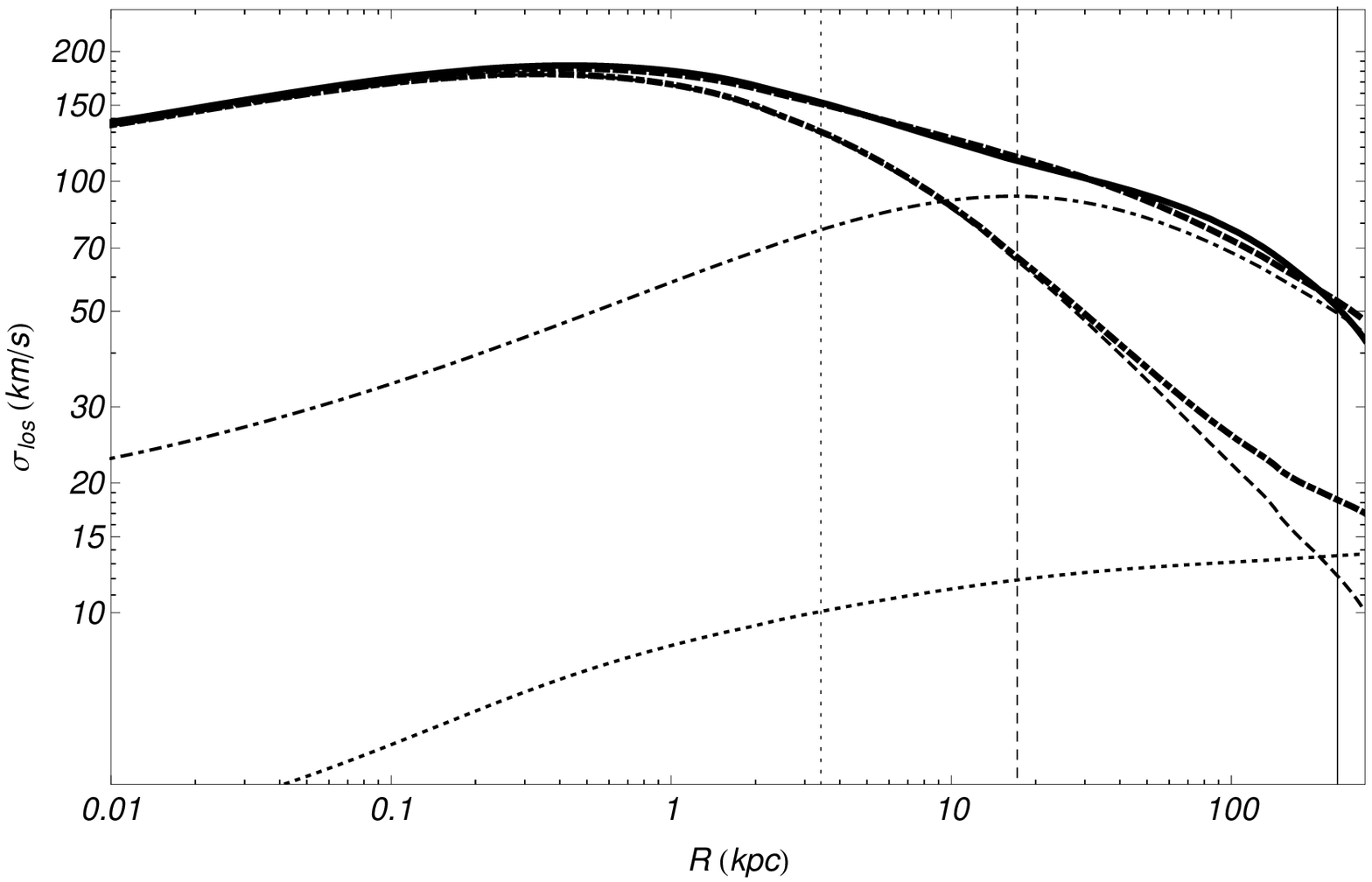}
  \includegraphics[width=90mm]{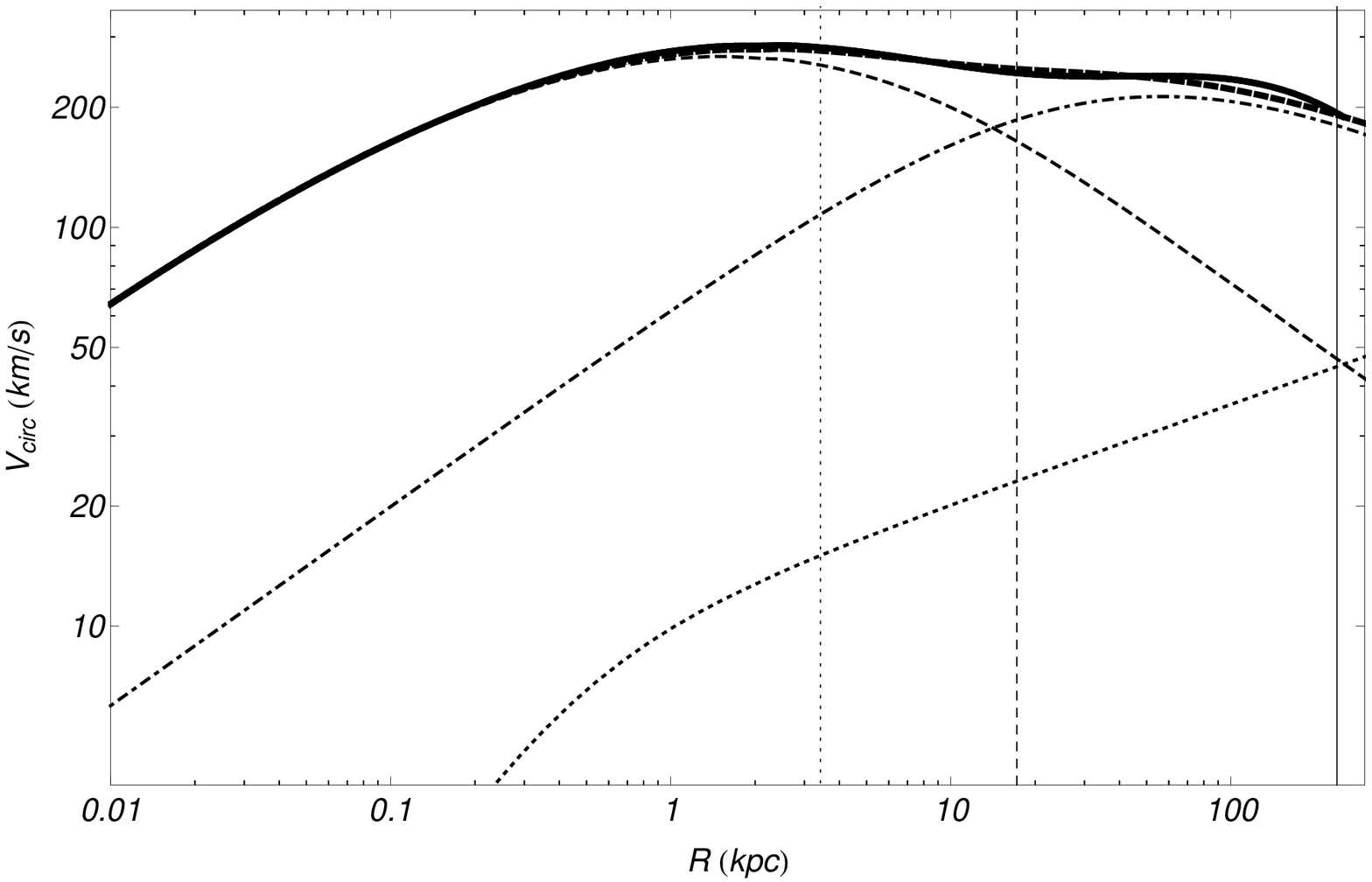}
  \caption{Mock elliptical galaxy analysis. \textit{Left panel}: velocity dispersion profile; \textit{Right panel}: circular velocity profile. Dark lines: solid - scalar field with two coupling constants; dot-dashed - scalar field with one coupling constant; dashed - classical NFW dark matter profile. Light lines: dashed - stellar velocity dispersion (circular velocity) in the classical dark matter approach; dotted - gas velocity dispersion (circular velocity) in the classical dark matter approach; dot-dashed - dark matter velocity dispersion (circular velocity). Vertical lines: dotted - effective radius R$_{eff}$, limit achievable with only stellar photometry; dashed - $5-6$ R$_{eff}$, limit achievable with PNe reconstructed kinematics; solid: virial radius $r_{vir}$. \label{fig:NFW-Scalar}}
\end{figure*}

According to their prescriptions we have considered an elliptical galaxy with a baryonic and a dark matter component characterised by the following parameters:
\begin{itemize}
  \item B-band luminosity $L_{\ast,B}= 1.88 \times 10^{10}$ $h_{70}^{-2}$ $L_{\odot}$
       (where $h_{70} = 1$ if $H_{0} \equiv 100 \cdot h = 70$ km s$^{-1}$ Mpc$^{-1}$), from which, using all the relations given in Section~2 of \citep{MamonLokas2}, we obtained a Sersic shape parameter $m = 3.12$ and a Sersic length $a_{s}= 11.6$ $h_{70}^{-2}$ kpc;
  \item a typical stellar mass-to-light ratio $Y_{\ast}=6.5$;
  \item a total mass-to-light ratio $Y=100$ corresponding to a virial radius, $r_{vir} = 79~R_{\mathrm{eff}}$ and to a concentration parameter $c=9.70$;
  \item a black hole to stellar mass ratio $M_{BH}/M_{\ast} = 0.0015$;
  \item a $\beta$-model for the gas component with index $\beta_{g} = -1.5$ and core radius $r_{c} = R_{\mathrm{eff}}/10$.
\end{itemize}

Using Eq.~(\ref{eq:los_dispersion}), and assuming the dark matter halo described by quantities at point $3$ above, we obtain the total velocity dispersion of this mock galaxy, shown in the left panel of Fig.~\ref{fig:NFW-Scalar} as a dark dashed line. This has a slightly decreasing trend with the radius which eventually changes its slope outside 50--60 kpc, well beyond typical radial coverage by PN kinematics (vertical dashed line).

Then we tried to recover (only in a qualitatively way) the total velocity dispersion profile with a scalar field. As shown in the left panel of Fig.~\ref{fig:NFW-Scalar} in dark dot-dashed line, the scalar field prediction with a single coupling constant for all mass components, and theory parameters $\beta = 0.05$ and $L \approx 1000$ kpc, does not match the velocity dispersion profile of the dark matter case. The values for these two parameters is severely limited because the velocity dispersion in the inner region is completely dominated by the stellar component; it works like a sort of normalization factor and strongly constrains the value that $\beta$ can have. If we give a value for $\beta$ which is too much different and higher than $0.05$, we will have a completely wrong velocity dispersion reconstruction. In particular, in the scalar field approach, where no dark matter is considered, the total velocity dispersion is almost equivalent to the only-stellar velocity dispersion well beyond the radial extent of PNe measurements (vertical dashed line), eventually rising only after this limit. Thus, we can conclude that a scalar field with only one coupling constant cannot simulate a dark matter profile (light dot-dashed line) in a consistent and sufficient way.

Things change drastically if one considers the possibility for the scalar field to have different coupling constants for each intervening mass component. In this case, we have stars and gas, and if we assume $\beta_{star} = 0.05, \; \beta_{gas} = 5.6$ and the common length $L \approx 90$ kpc, it is possible to decouple the effects from any of them. In fact, the model which includes two coupling constants (dark solid line) nicely reproduces the  velocity dispersion profiles of the classical case, i.e. the scalar field can mimic the profile of a NFW dark halo.

In the right panel of Fig.~\ref{fig:NFW-Scalar} we instead show the circular velocity curve, calculated from the relation $v^{2}_{c} = r \; d \Psi /d r$. We can see how it is rather flat out to 100 kpc, clearly showing how dominant is the dark matter (light dot-dashed line) in the total galaxy potential. This latter property is what we expect to be able to reproduce with the scalar field. We finally stress that if we had adopted the circular velocity as observational quantity (as it is done, for example, in spiral galaxies), we would not have any chance to discriminate between the two approaches, i.e., one or two coupling constants for the scalar field with the baryonic matter. This is clearly shown in the right panel of Fig.~\ref{fig:NFW-Scalar} where we can see how the two different cases for the scalar field give two equivalent reproductions of the NFW profile at least in the depicted distance range: the two representing lines, the dark solid and dot-dashed one, respectively, are indistinguishable as they are perfectly overlapping. This is an important issue to be considered when evaluating results for spiral galaxies in next sections.

\subsection{Real data: NGC4374}

Starting from these preliminary considerations, we can move to the analysis of a real system: NGC 4374~(see Table~\ref{tab:tabdata}), which is our test case for elliptical galaxy dynamics. This system has been shown to possess a standard NFW halo profile (N+11), thus it will be important to see whether its dynamics can be interpreted equally well with the scalar field potential.

We will make use of the PNe sample discussed in N+11 which we refer the reader to for more details of the PNe sample properties and the derivation of the kinematical profiles we will use in our analysis; the PNe dispersion profile extends out to $\approx 5 \, R_{\mathrm{eff}}$, which is a distance large enough to explore any deviation of the galaxy dynamics from a pure Newtonian no-dark matter behaviour.

This galaxy has been also analysed in the context of $f(R)$ theories \citep{Napolitano12}, where a Yukawa-like modification of the gravitational potential is adopted as alternative to dark matter. This approach is different from the one we are adopting here, because the possibility to break the contribution of all mass components (stars and gas) in the gravitational budget is a peculiar feature of our theoretical scenario that cannot be included in their physical model.

Before we go on with the dynamical model, we pay some attention to the modelling of the stellar component because in both cases ($\Lambda$CDM framework or scalar field) it is the one that dominates the velocity dispersion profile in the inner region, and with lower uncertainties with respect to PNe data at larger distances from the center, and this implies a stronger weight on the global fit to the velocity dispersion. For NGC 4374 there are different literature models of its stellar photometry (mainly depending on the extension of the adopted datasets): a typical De Vaucouleurs profile (a Sersic profile with index $m=4$), with $R_{\mathrm{eff}}= 5.97$ kpc \citep{Cappellari06}; a Sersic profile with $R_{\mathrm{eff}}=11.69$ kpc and $m=7.98$ \citep{Kormendy09}; and a Sersic profile with $R_{\mathrm{eff}} = 9.34$ kpc and $m=6.11$ (N+11). The first model has some problem in fitting data at low distances from the center and can also affect large radii data, as the De Vaucouleurs profile is not always able to describe all the intrinsic features of an elliptical galaxy. The second model allows a better fit to the stellar profile but fails to recover the behaviour of surface brightness at small radii ($R\lessapprox0.5$ kpc). Finally, the third model gives a very good fit of the stellar profile and in a wider range than the previous one, namely, $0.09\lesssim R \lesssim38$ kpc, out to the distances covered by PNe observations. For this reason we decided to adopt the model profile from N+11 in the following dynamical analysis. Furthermore, we adopted two different approaches in order to optimize at best the goodness of the surface brightness reconstruction: we have considered, one after the other, all the available data points and only data points with $R>0.09$ kpc.

{\renewcommand{\tabcolsep}{2.mm}
{\renewcommand{\arraystretch}{1.5}
\begin{table*}
\begin{center}
\caption{\textit{Elliptical galaxies: NFW dark matter.} Column 1: $\chi^2$ type. Column 2: central NFW density.
  Column 3: NFW radius. Columns 4: stellar mass-to-light ratio in the observation photometric band. Column 5: NFW concentration parameter.
  Columns 6: virial radius.  Column 7: virial mass. Column 8: stellar mass-to-light ratio in the B band. \label{tab:elldata1}}
\begin{tabular}{ccccc|ccc}
  \hline
  & $\chi^2/d.o.f.$ & $\rho_{0}^{DM}$                       & $a_{d}$ & $Y_{\ast}$    & $c_{vir}$ & $\log M_{vir}$ & $r_{vir}$ \\
  &                & $(10^6 \, M_{\odot}/\mathrm{kpc}^3)$  & (kpc)   & $(Y_{\odot})$ &           & $(M_{\odot})$  &   (kpc)   \\
  \hline
  \hline
  $\chi_{all}^2$ & $39.68/37$ & $0.361^{+0.296}_{-0.240}$ & $316.37^{+693.01}_{-116.73}$ & $6.67^{+0.051}_{-0.056}$ & $4.01^{+1.22}_{-1.61}$ & $13.34^{+2.58}_{-0.54}$ & $1269.65^{+4016.29}_{-790.49}$ \\
  $\chi_{>0.09}^2$ & $29.63/33$ & $0.402^{+0.393}_{-0.203}$ & $290.00^{+299.24}_{-119.81}$ & $6.66^{+0.060}_{-0.062}$ & $4.21^{+1.48}_{-1.16}$ & $14.02^{+1.31}_{-1.12}$ & $1221.66^{+2130.79}_{-702.81}$ \\
  $\chi_{>1}^2$ & $9.65/17$ & $0.841^{+1.552}_{-0.506}$ & $179.61^{+200.67}_{-93.74}$ & $6.28^{+0.263}_{-0.282}$ & $5.83^{+3.23}_{-1.95}$ & $13.81^{+1.55}_{-1.49}$ & $1047.03^{+2397.70}_{-713.89}$ \\
  \hline
  \hline
\end{tabular}
\end{center}
\end{table*}}}

{\renewcommand{\tabcolsep}{2.mm}
{\renewcommand{\arraystretch}{1.5}
\begin{table*}
\begin{center}
\caption{\textit{Elliptical galaxies: Scalar field.} Column 1: $\chi^2$ type. Column 2: Coupling constant of scalar field and star component.
  Column 3: Coupling constant of scalar field and gas component. Column 4: Coupling constant of scalar field and black hole component. Columns 5: Scalar field interaction length. Column 6: stellar mass-to-light ratio in the related band. \label{tab:elldata2}}
\begin{tabular}{ccccccc}
  \hline
  & $\chi^2/d.o.f.$ & $\beta_{star}$ & $\beta_{gas}$ & $\beta_{BH}$ & $L$  & $Y_{star}$ \\
  &                 &                &               &              &(kpc) & $(Y_{\odot})$   \\
  \hline
  \hline
  $\chi_{all}^2$ & $38.16/34$ & $0.100^{+0.229}_{-0.069}$ & $10.002^{+0.465}_{-0.528}$ & $0.644^{+0.079}_{-0.090}$ & $658.26^{+1328.51}_{-410.70}$  & $6.099^{+0.166}_{-0.992}$ \\
  $\chi_{>0.09}^2$ & $19.16/31$ & $0.086^{+0.154}_{-0.065}$ & $11.089^{+0.648}_{-0.561}$ & $1.167^{+0.116}_{-0.125}$ & $183.18^{+2127.44}_{-96.40}$  & $5.633^{+0.220}_{-0.499}$ \\
  \hline
  \hline
\end{tabular}
\end{center}
\end{table*}}}
\begin{figure*}
\centering
  \includegraphics[width=88mm]{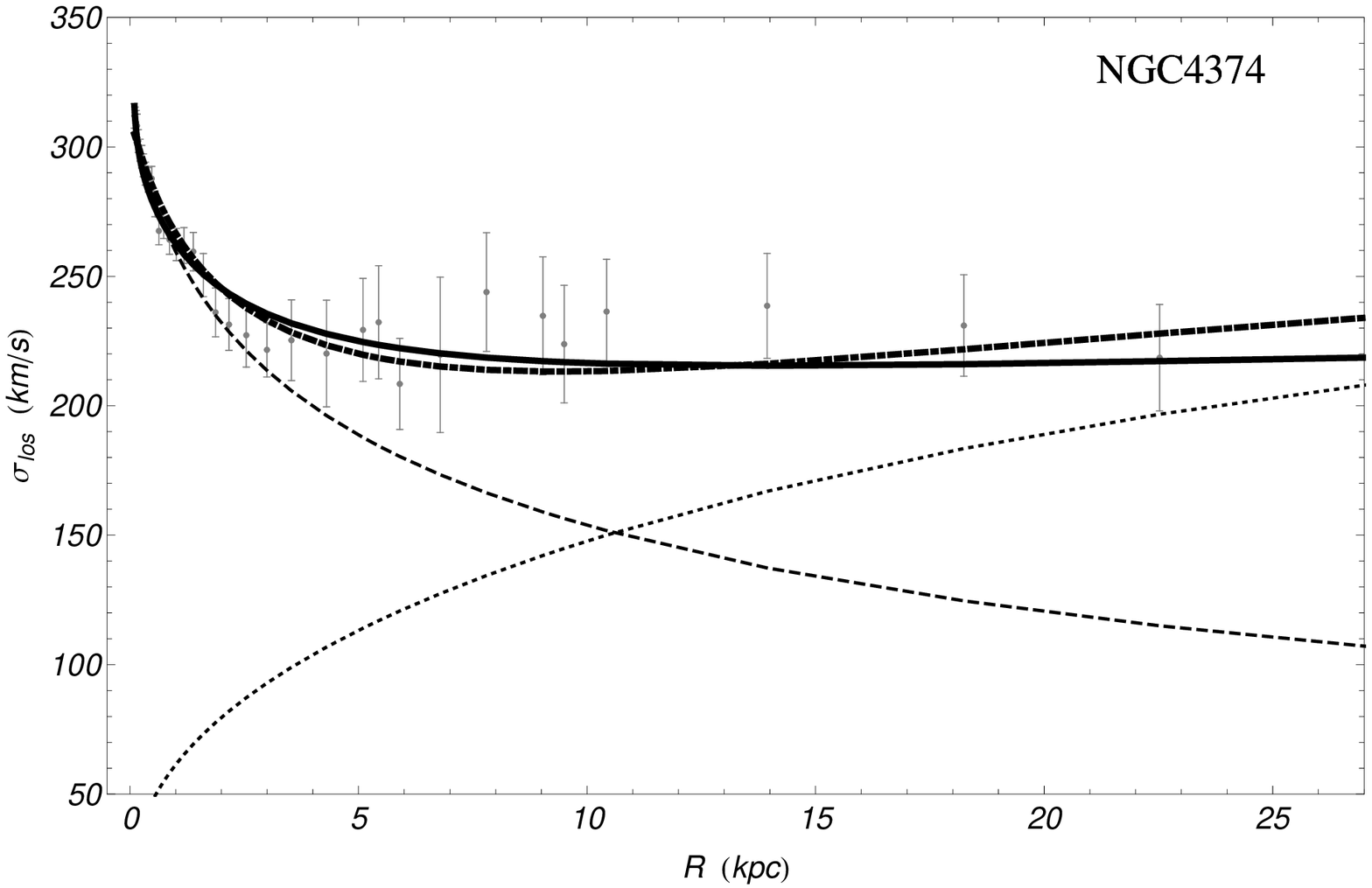}~~~
  \includegraphics[width=88mm]{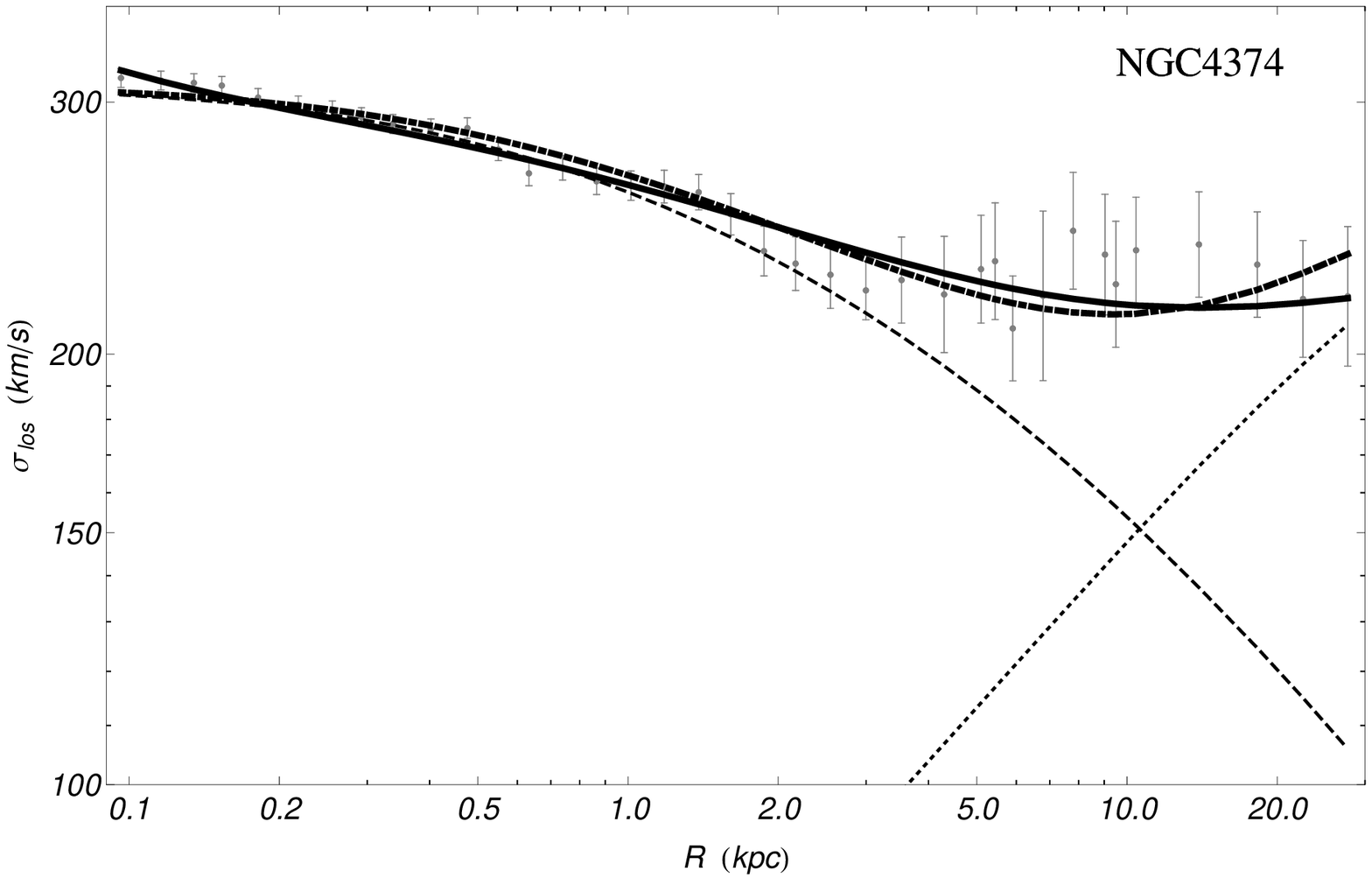}
  \caption{\textit{Left panel.} Velocity dispersion curve of NGC4374. Dark lines: solid line is the total velocity dispersion in the scalar field approach, i.e. scalar field + baryons; dot-dashed line is the total velocity dispersion in the classical approach, i.e. dark matter + baryons. Light lines: dashed line is the stellar velocity dispersion in the NFW approach; dotted line is the dark matter velocity dispersion. \textit{Right panel.} The same as before but in logarithmic scale. \label{fig:NGC4374}}
\end{figure*}

\subsection{Dark Matter}

We have started by using our model machinery assuming the standard Newton dynamics and a NFW dark halo and cross-checking our results with the ones presented in N+11. Our results are in Table~\ref{tab:elldata1} and somehow differ with the results discussed in their \S~3.5.1. Before we go into the details of the discrepancy, we need to remark that there are some critical differences between the two approaches. Here we use some parametrized stellar surface density, while in N+11 they use an interpolated function. Furthermore, we are assuming here an anisotropy profile which is rather different from the one constrained by N+11, where they have also used the kurtosis information. These two main differences can produce some substantial divergence in the modelling of the very central data points thus affecting the parameters which are more sensitive to the small radii fit. For this reason we have repeated the models including all the kinematics data point, and excluding the data at $R>0.09$ kpc and $R>1$ kpc, finally finding substantial differences, as shown in Table~\ref{tab:elldata1}).

In particular, we have obtained a lower value of the central density $\rho_{0}^{DM}$ and a higher value for the NFW radius $a_{d}$ with no statistically significant difference between the total and the $R>0.09$ cut sample; but if we consider the expected relation between these two parameters derivable from a collisionless $\Lambda$CDM universe with WMAP5 parameters (Eq.~(13) and blue contours of Fig.~7 in N+11), we have correspondence at $1\sigma$ level. The concentration parameter, $c \approx 4$, matches with the lower limit of most of the cases shown in the Table~2 of N+11, and in particular with their assumed best reference model (i.e., an adiabatically contracted NFW profile with an anisotropy distribution different from the one adopted here, which should result in a more concentrated dark matter density profile).

The virial radius is notably larger, while the virial mass is perfectly consistent with their results even if showing a wider confidence level extending preferentially to higher values. If we compare these results with the Fig.~11 in N+11, we see that our value for the parameters $(c_{vir},\log M_{vir})$ fall in the region limited by the results inferred from late-type galaxies dynamics and from weak lensing of all type of galaxies and groups \citep{Napolitano09}. In particular the values coming from the $r>0.09$ kpc sub-sample perfectly match with this last curve.

Concerning the luminous stellar counterpart, we found values slightly higher for the stellar mass-to-light ratio, $Y_{\ast} \approx 6.6$, that is however perfectly compatible with a Salpeter Initial mass function (Fig.~5 in N+11).

Looking to the left panel of Fig.~\ref{fig:NGC4374} we see how the stellar component dominates the velocity dispersion profile only in the very central regions ($R \lesssim 1.0$ kpc), with the black hole contribution being important only for very small scales, and the NFW dark matter becoming dominant in the line-of-sight velocity dispersion profile at $\approx 10$ kpc, which corresponds to $\approx 1-2$ $R_{\mathrm{eff}}$.

When using only data with $R > 1.0$ kpc we have been able to recover results more similar to N+11 for what concerns the NFW profile: an higher central density parameter for the dark matter profile (even if it is again lower than the one measured in N+11); a smaller value for the NFW length (but still higher than N+11); the couple $(c_{vir},\log M_{vir})$ now compatible with both weak lensing inferred trend and with the relation derived from WMAP5-based simulations; finally, the virial radius is now only a $30 \%$ higher than their value.

\subsection{Scalar Field}

If we now move to the scalar field alternative approach, it is clear (right panel of Fig.~\ref{fig:NGC4374}) that this is as successful as the classical dark matter approach in modelling the dispersion profile. The largest differences with respect to the NFW profile are found: at the very small radii, where the scalar field model shows a steeper slope while the NFW one seems to reach a plateau; and at the very outer region, where the scalar field model stays flatter than the NFW profile, although more extended data would allow us to adjust the two models better at the largest distance and possibly to recover a better agreement also at shorter scales. However, as seen by the $\chi^2$ results in Table~\ref{tab:elldata2}, the best fit is also in this case very good: while the NFW and the scalar field approaches are almost equivalent when all the data points are used, if we consider the value of the reduced $\chi^2$ (with the NFW $\chi^2$ being slightly smaller than the scalar field one) for the best--fit to the data points with $R>0.09$, the scalar field turns out to provide a far better significance of the fit with respect to the NFW model. In this case, the stellar mass-to-light ratio is smaller than the NFW-based one and it is now more compatible with the Kroupa IMF \citep{Kroupa01} values found in N+11.

Even more importantly, the scalar field parameters turned out to be consistent with what we argued for the double coupling constant hypothesis in \S\ref{sec:hypothesis}. First, the stellar mass component shows the lowest coupling constant among all the mass components, even lower than the one related to the black hole. As anticipated, this can be the consequence of the average effect of the screening action made by the scalar field on the stellar component. The same does not happen to the black hole, since this represents a singularity and it is difficult to detail the change in the field from inside to outside and the comparison with its \textit{classical} Newtonian force. Second, the coupling constant which refers to the gas seems to be mostly correlated with the galaxy gravitational potential and the scalar field mass.

To conclude, we stress that both the classical dark matter and the scalar field approaches seem to be unable to describe the small shoulder in the dispersion profile that is present at $R \approx 10$kpc, precisely where the PNe data overlap the only stellar kinematics. One may think that this can be a consequence of some unaccounted orbital anisotropy. In fact, this mainly depends on the choice we made to use a parametric form of the star density $\ell(r)$ instead of a smoothed light profile like in N+11 where it was possible to recover all the details of the dispersion profile (with a small degree of radial anisotropy). This means that the observed kinematics is somehow strongly sensitive to the tracer space density distribution.

\section{Unifying the scenario}
\label{sec:unified}

The results obtained with the elliptical galaxy NGC 4374 are not meant to have the sufficient generality to drive any conclusion on the newly proposed multi--coupling constant scenario. However, it is not surprising that the proper handling of physically more complex, multicomponent systems like ellipticals has opened new perspectives in the scalar field approach. If our scenario is correct, this shall be in fact made more evident on gravitational systems where the evolutionary status is different (it is the case of spiral galaxies, as we will show in next sections) or where matter shows very different phases. But a more detailed analysis in this sense is out of the purpose of our work and will be developed in forthcoming papers.

Since we do not expect to extend the elliptical sample soon, due to the difficulty in finding galaxies having both extended stellar (PNe) kinematics and not faint gas profiles, in this section we try to re-interpret the results of Paper I in terms of the new evidences found for NGC 4374, i.e. the possibility of disentangling the various mass components with respect to the coupling constant with the scalar field. In this respect we aim at drawing a common scenario for all the gravitating systems (galaxies: spirals and ellipticals, and clusters) in a consistent way.

\subsection{Clusters of Galaxies}\label{sec:gal_clus}
\begin{figure*}
\centering
  \includegraphics[width=80mm]{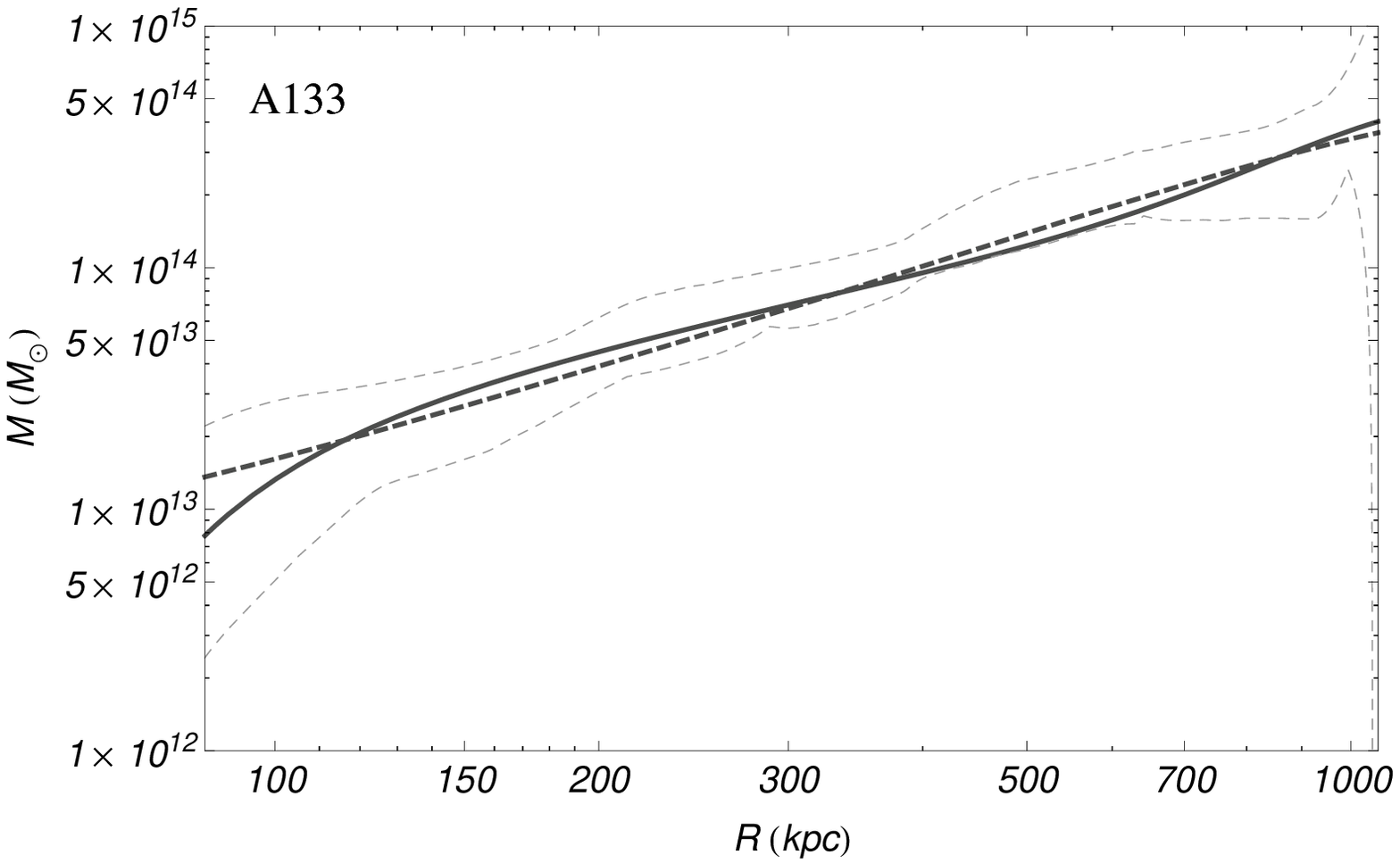}~~~
  \includegraphics[width=80mm]{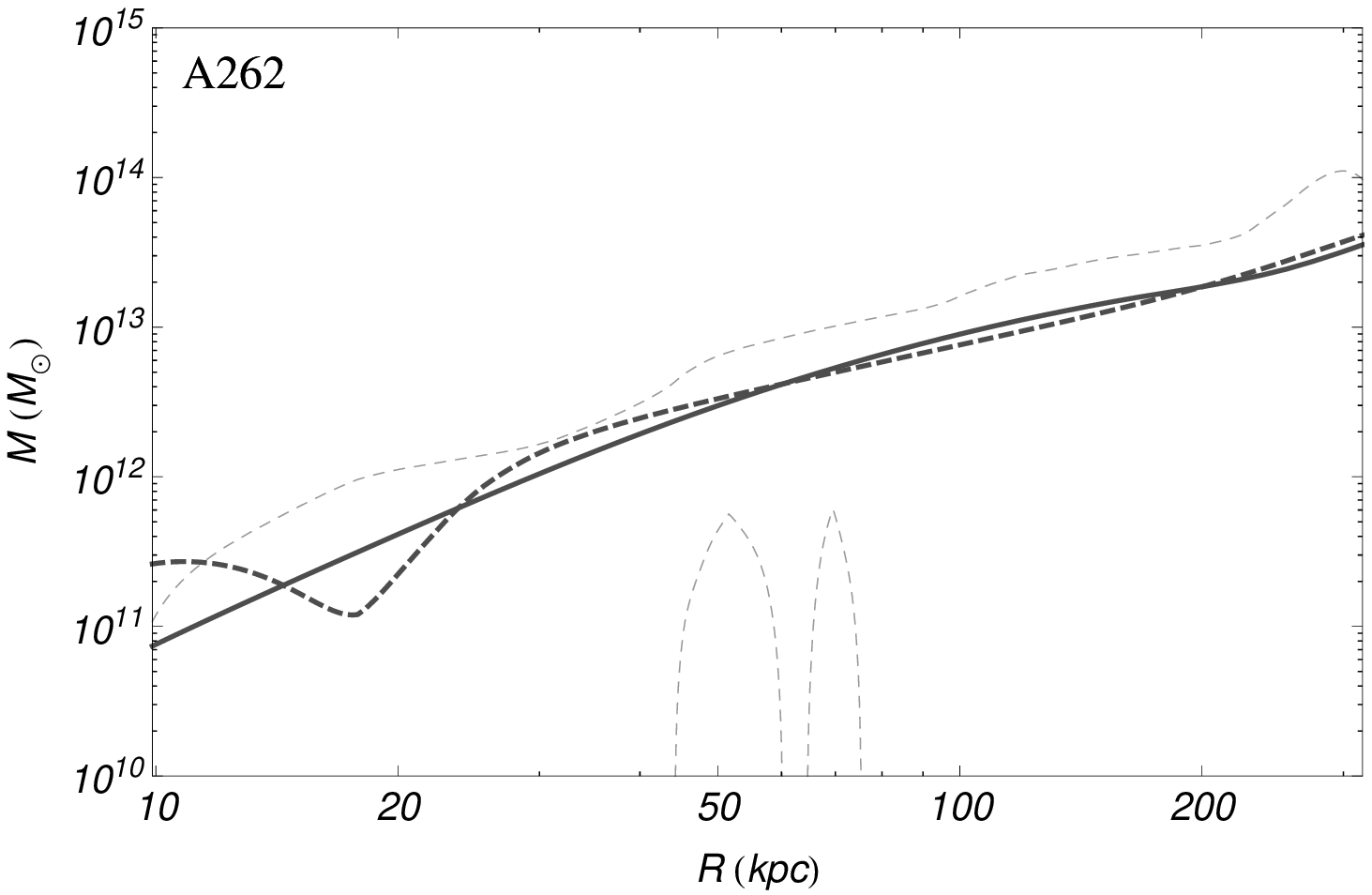}
  \includegraphics[width=80mm]{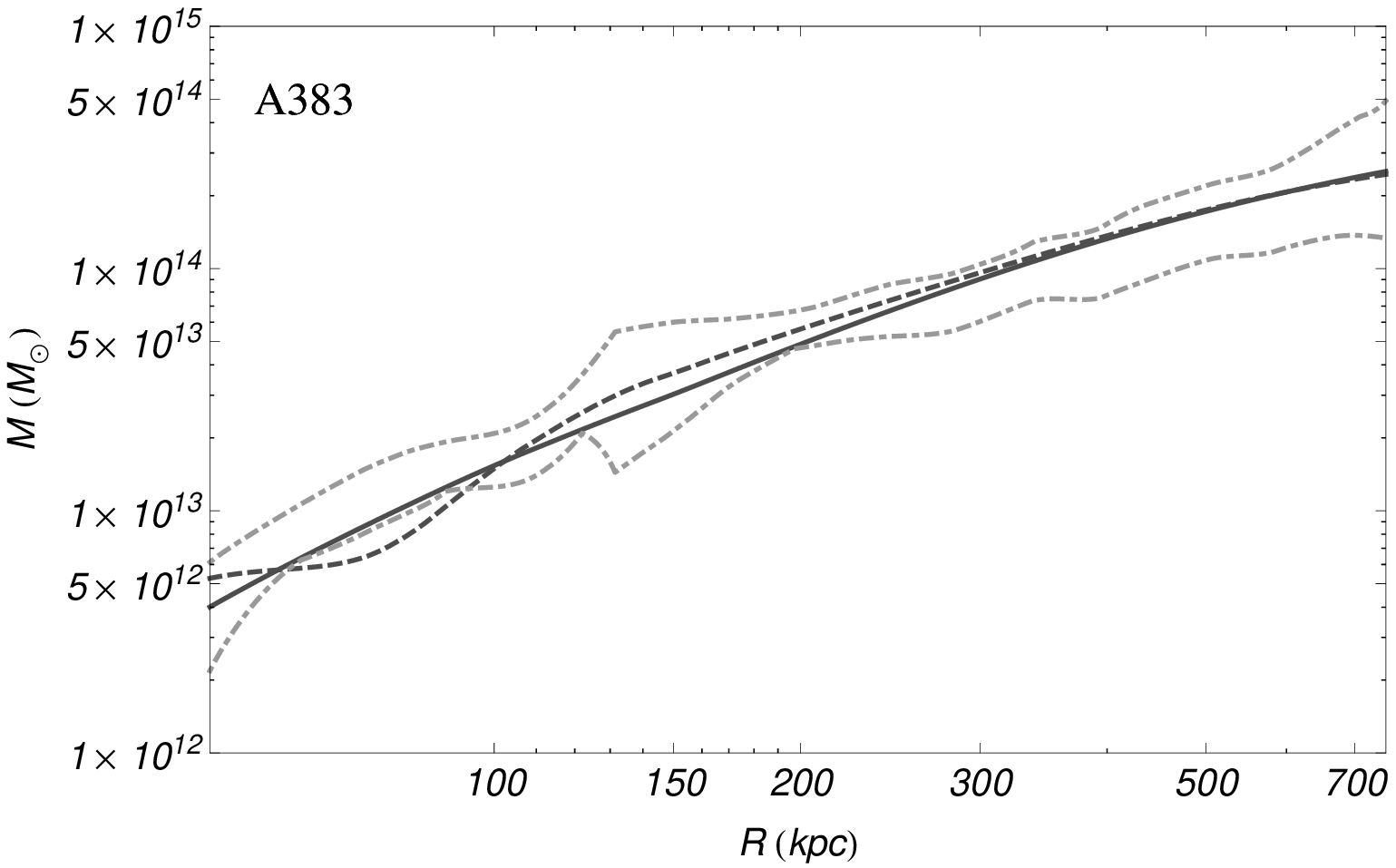}~~~
  \includegraphics[width=80mm]{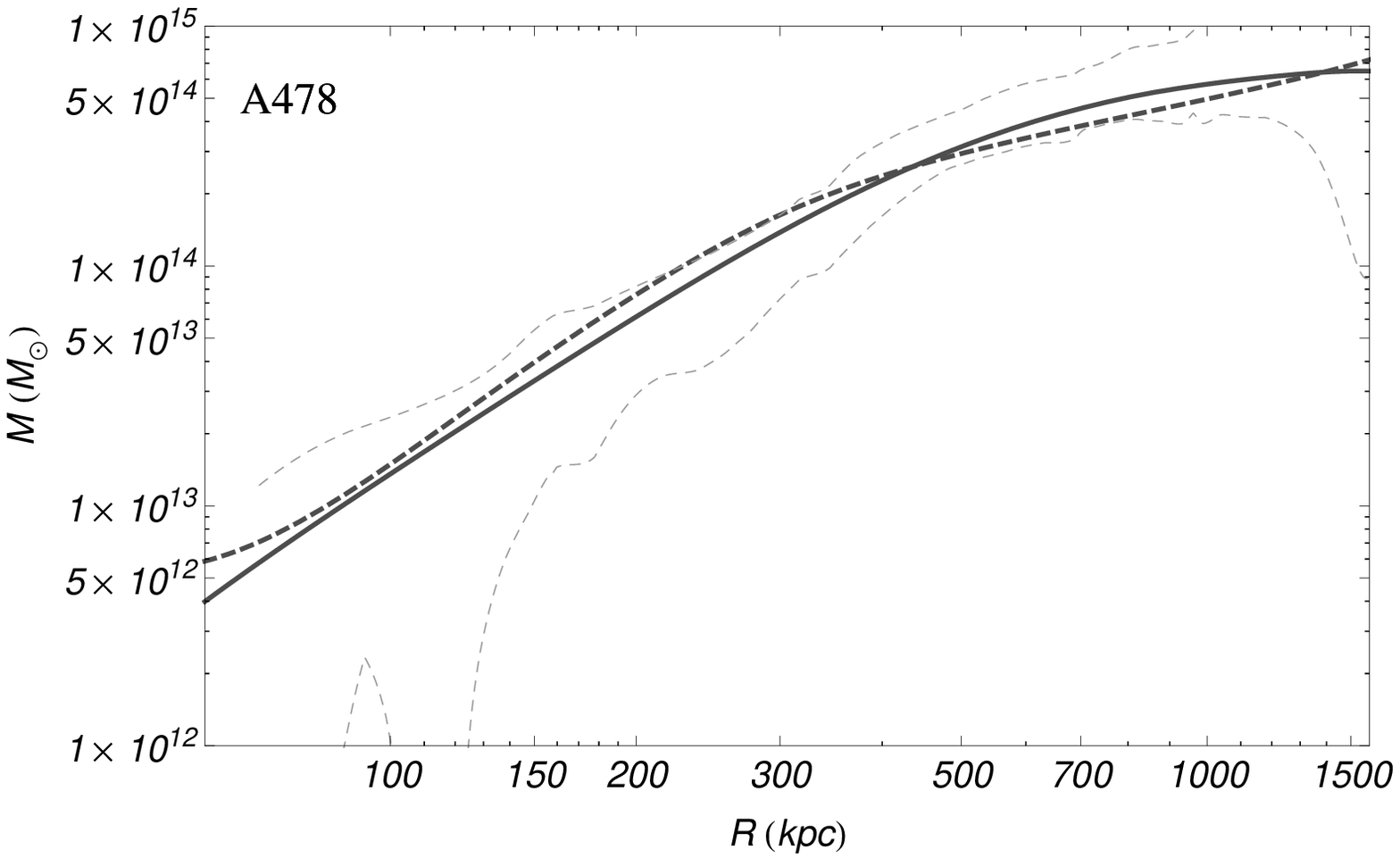}
  \includegraphics[width=80mm]{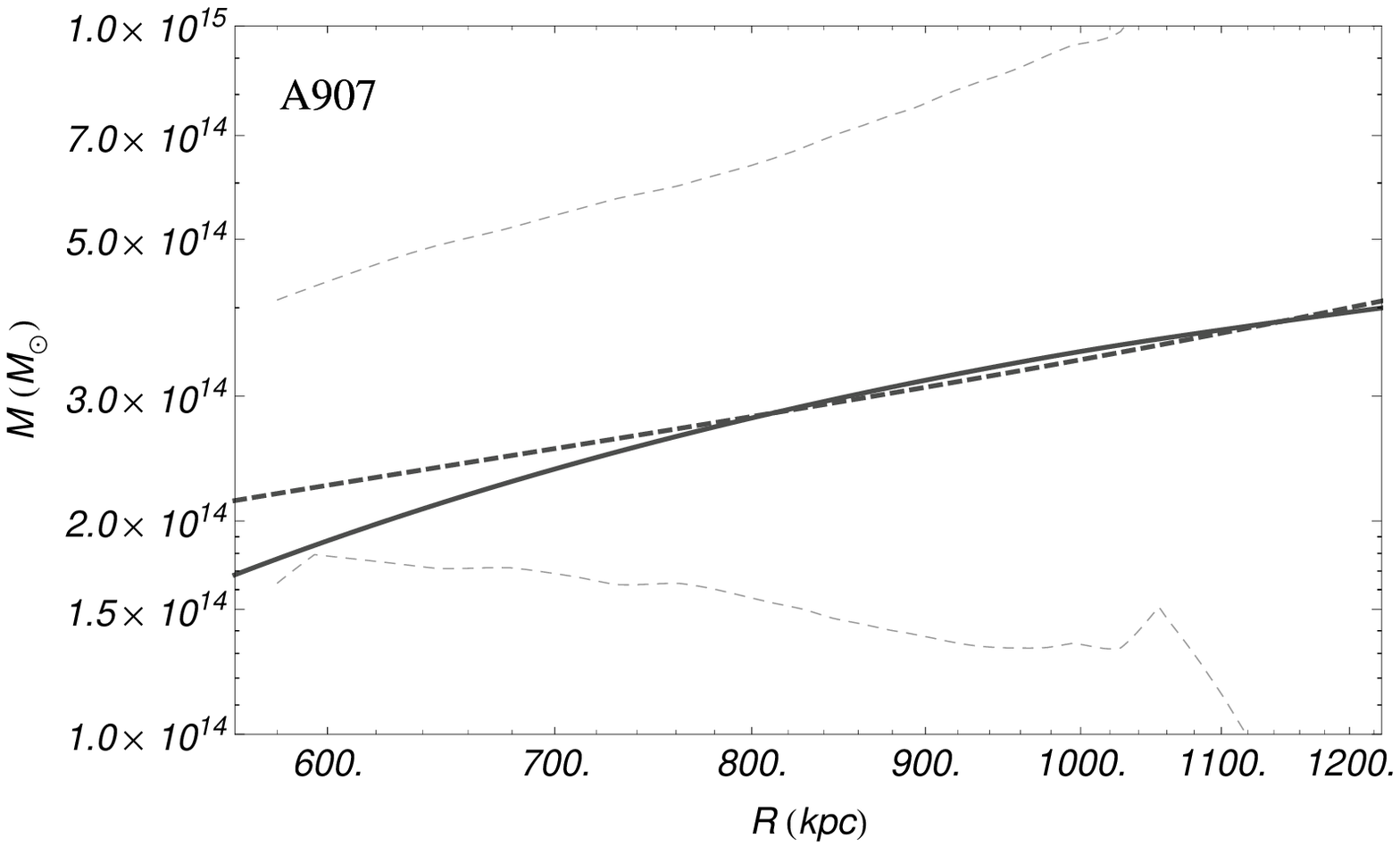}~~~
  \includegraphics[width=80mm]{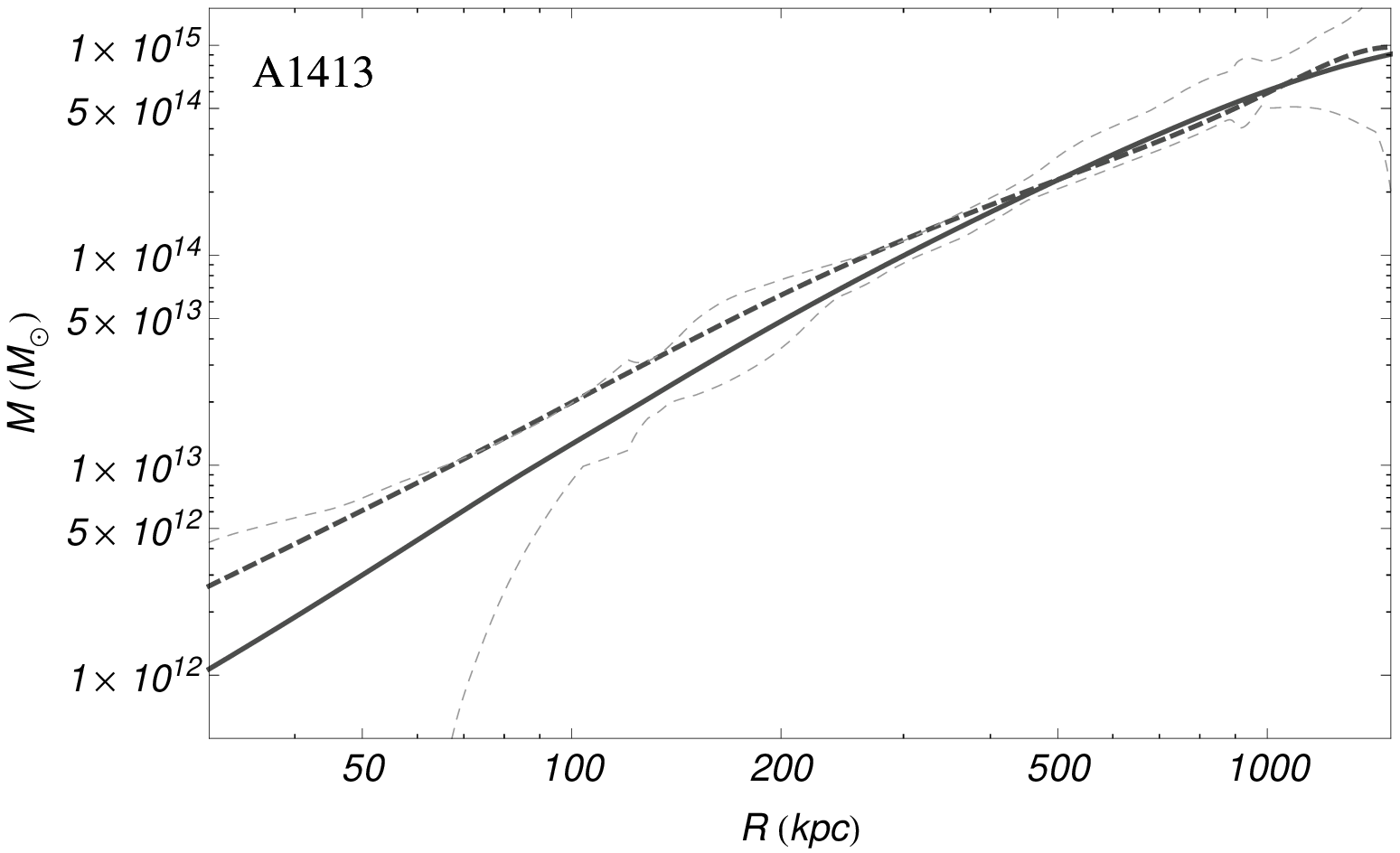}
  \caption{Dark matter profile vs radii for clusters of galaxies. Dashed line is the observationally derived estimation of dark matter; solid line is the theoretical estimation for the effective dark matter component; dot-dashed lines are the 1-$\sigma$ confidence levels given by errors on fitting parameters plus statistical errors on mass profiles.\label{fig:cham_cl1}}
\end{figure*}
\begin{figure*}
\ContinuedFloat
\centering
  \includegraphics[width=83mm]{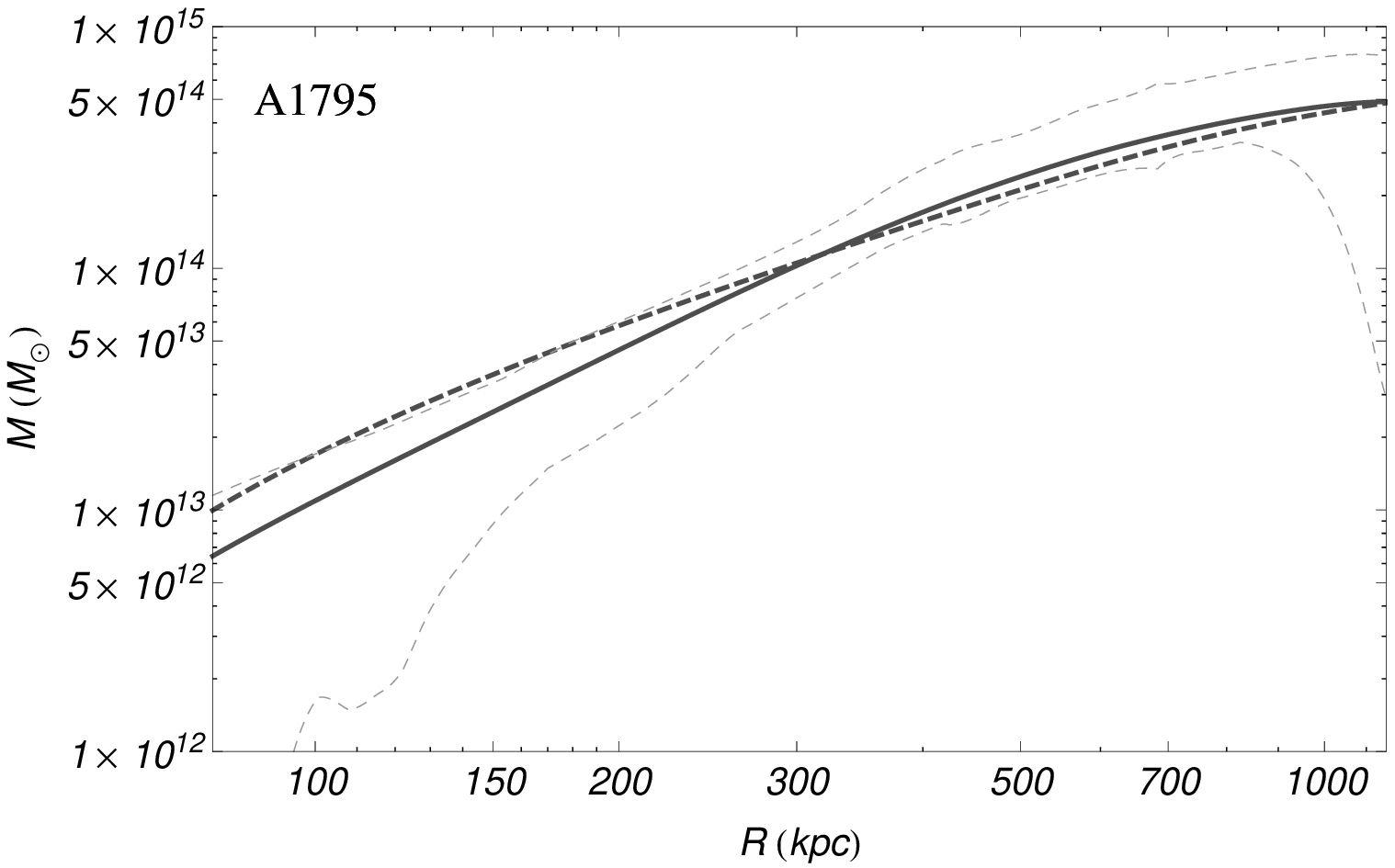}~~~
  \includegraphics[width=80mm]{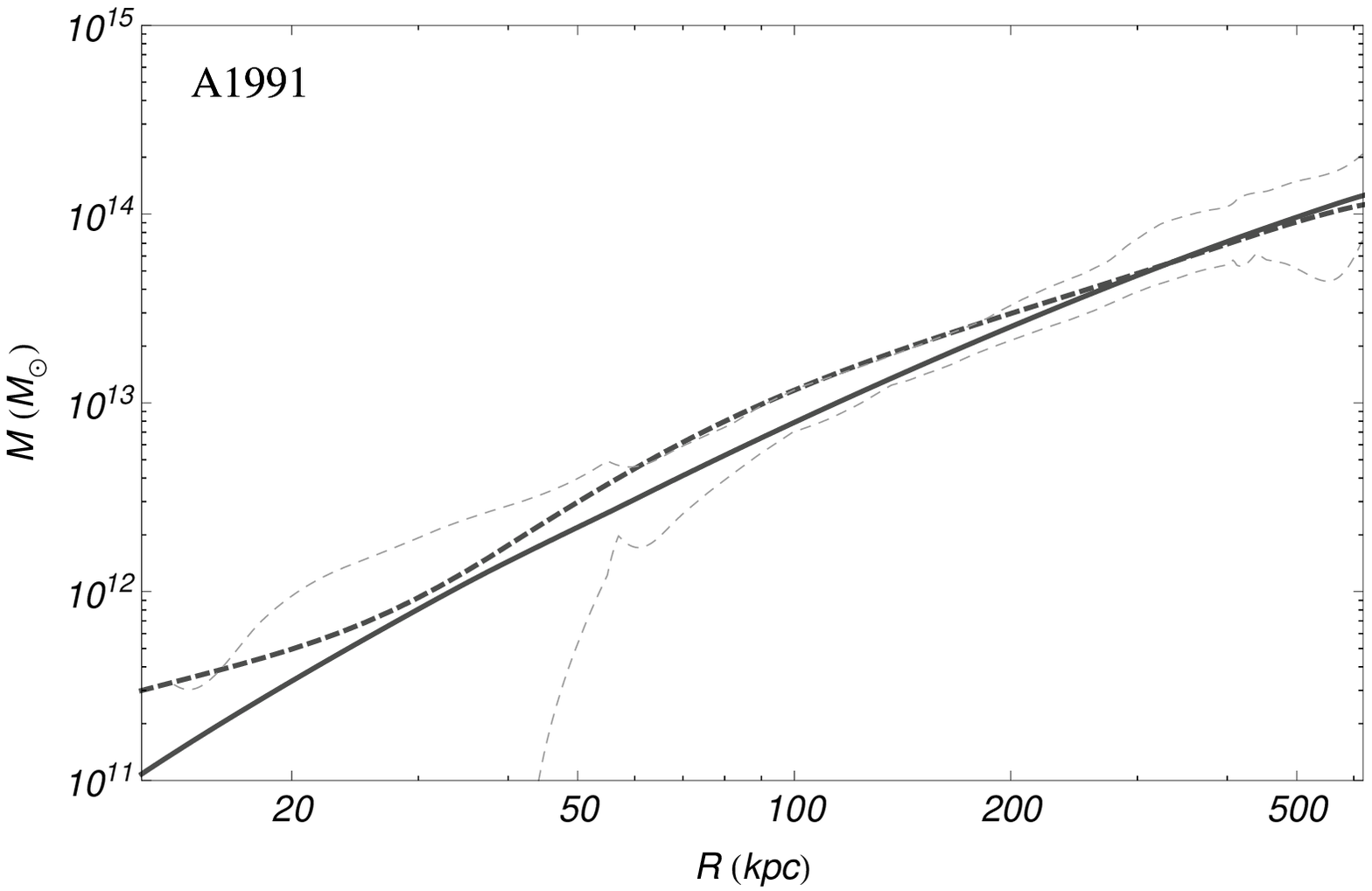}
  \includegraphics[width=80mm]{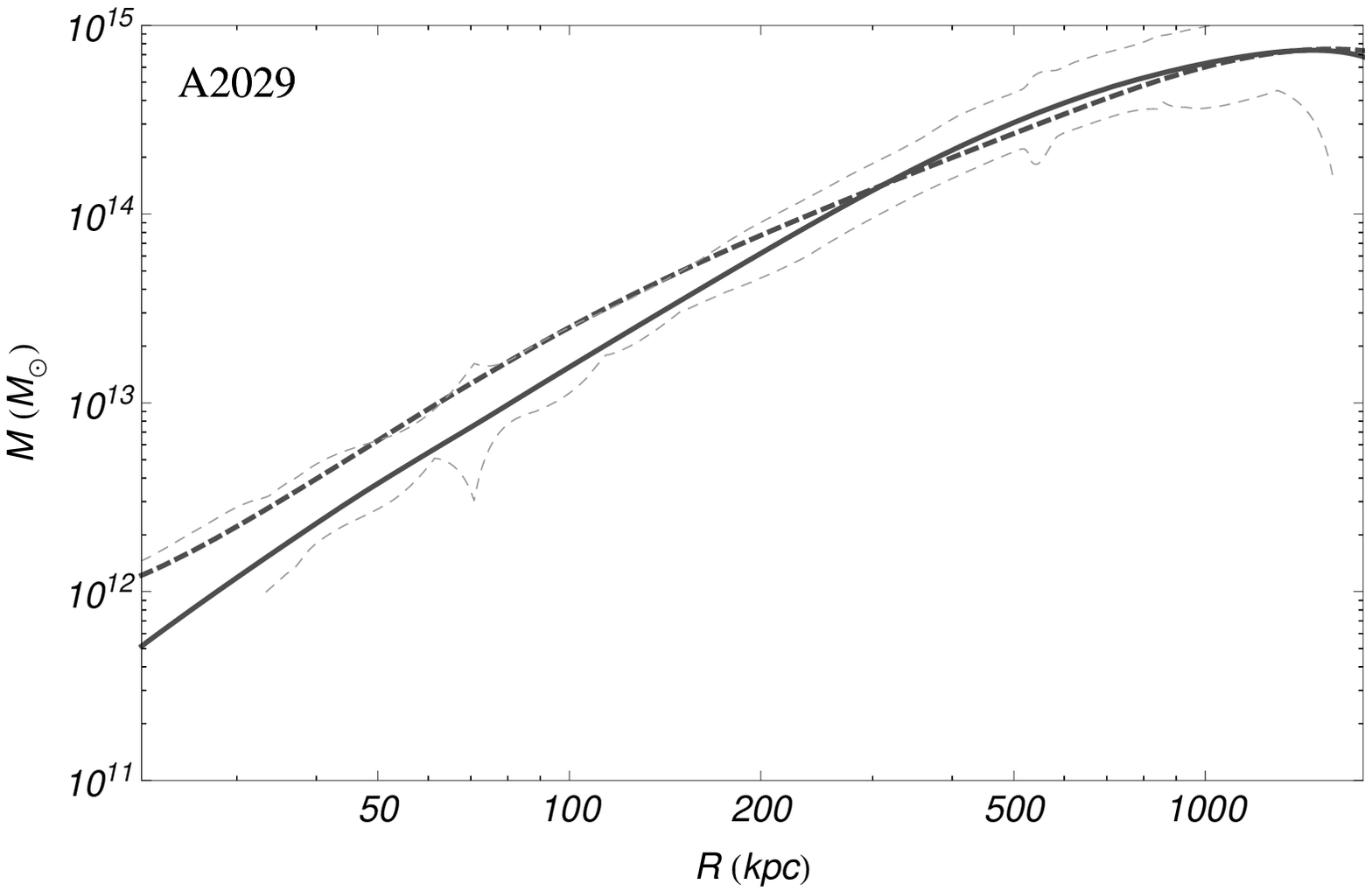}~~~
  \includegraphics[width=83mm]{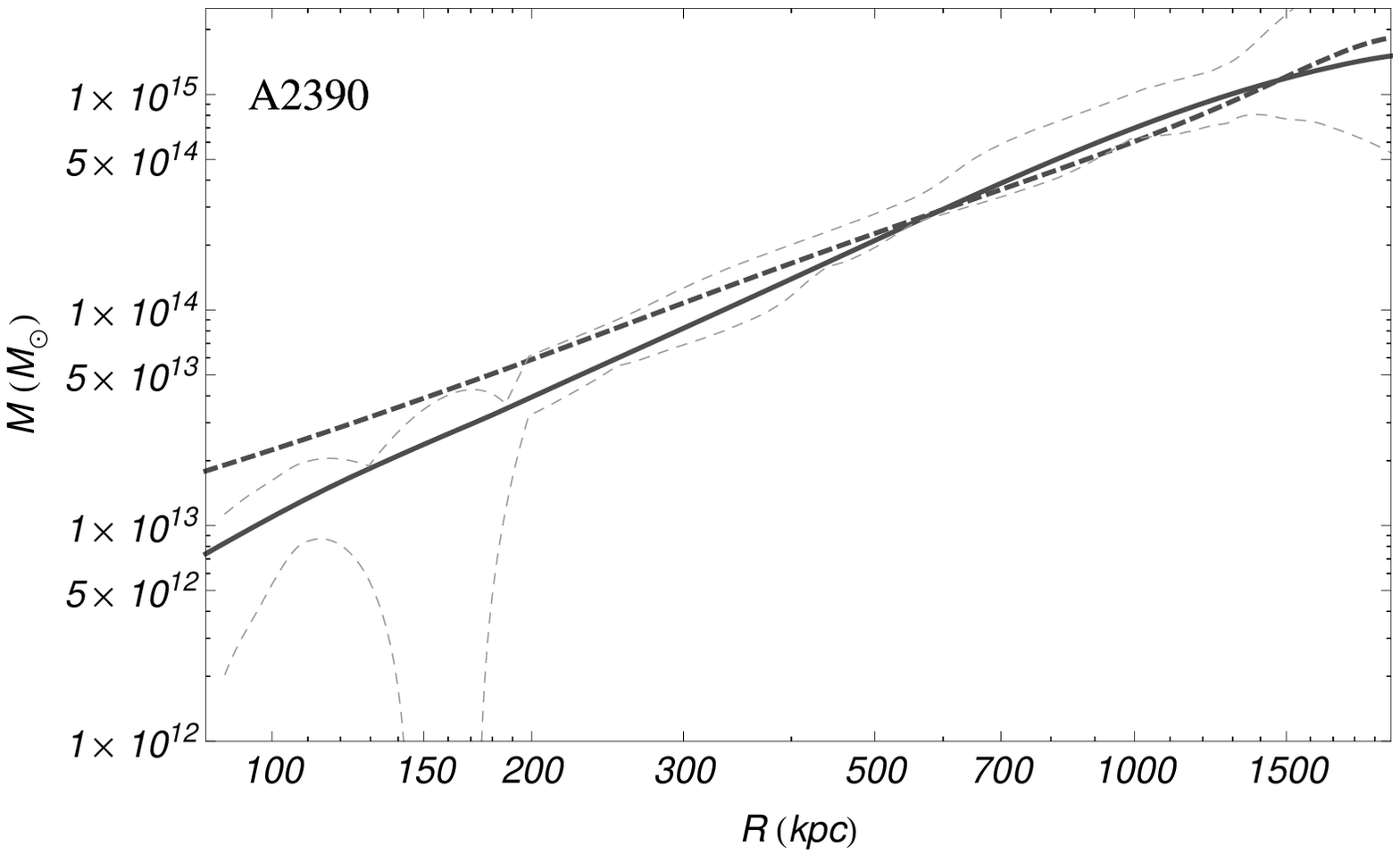}
  \includegraphics[width=80mm]{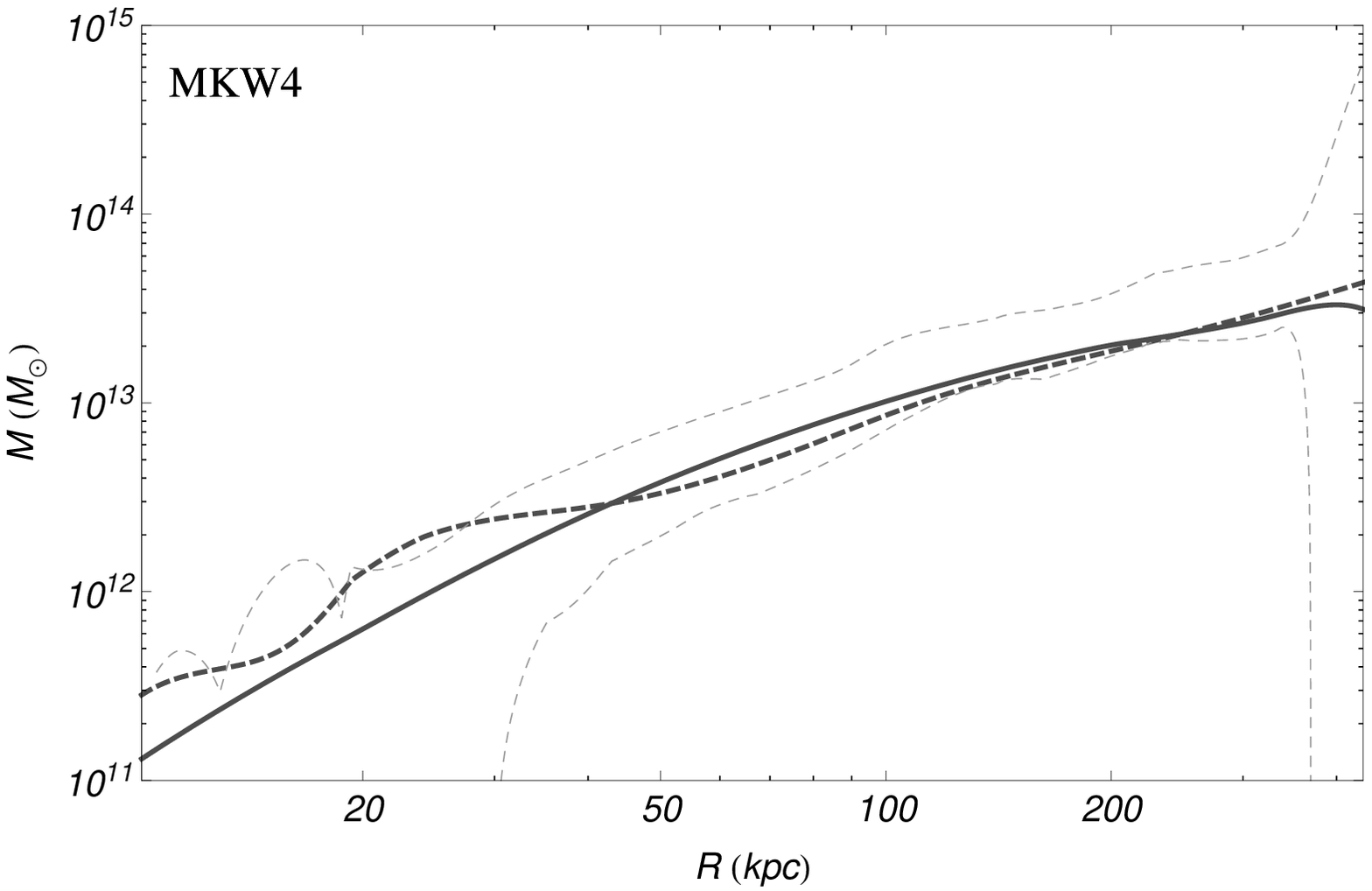}~~~
  \includegraphics[width=80mm]{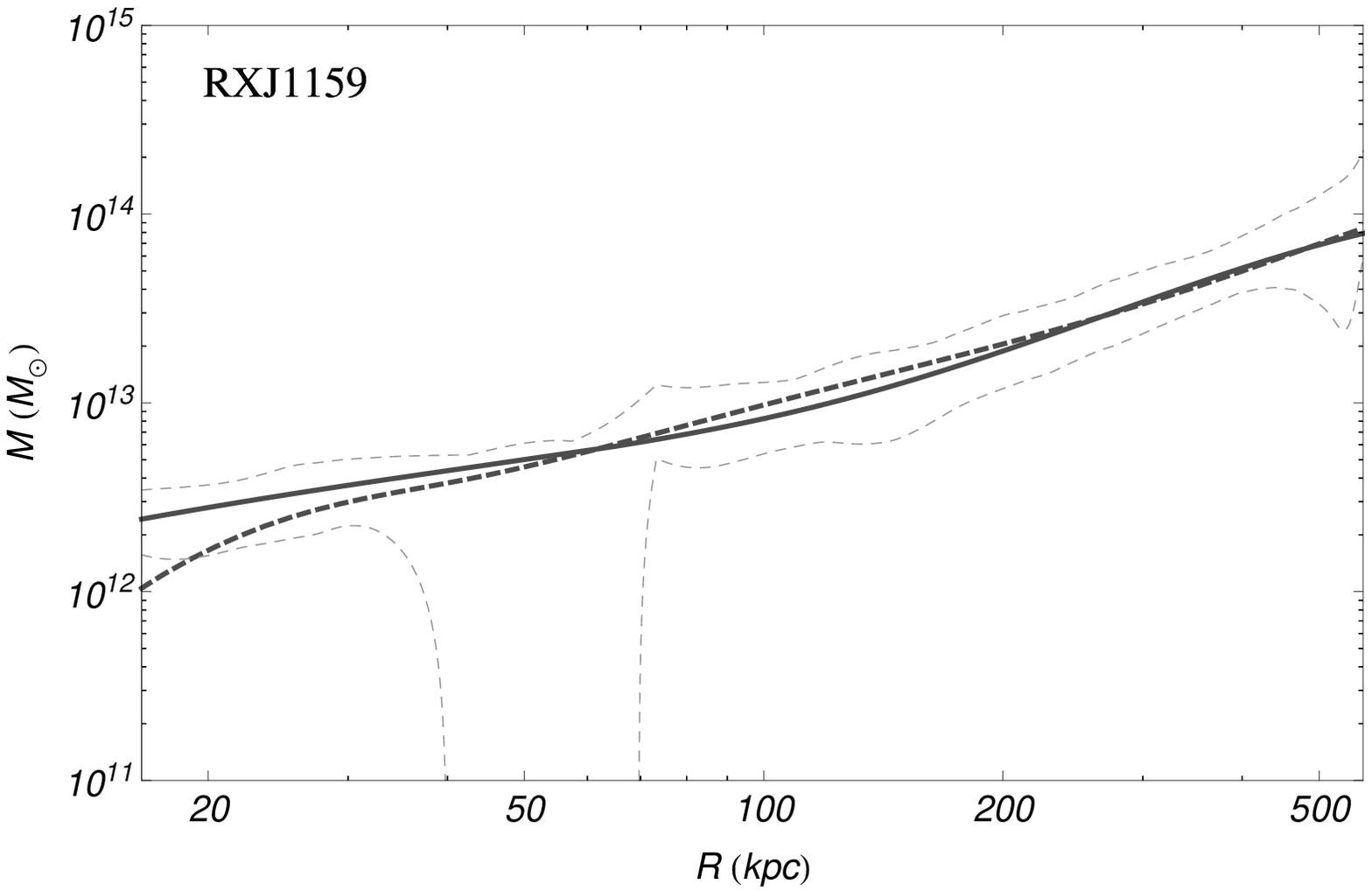}
  \caption{Continued.}
\end{figure*}

By comparing Fig.~\ref{fig:cham_cl1} with the corresponding figures in Paper I (Figs.~3~-~4 in that paper), we see that the the new approach of using the two mass components has allowed us to improve the model fit in $9$ cases, especially at small radii. In one case (RXJ1159) the new approach has produced a remarkable fit where the model in Paper I failed completely. Only for $3$ clusters the mass profile reconstruction has not got any benefit by the two component approach, although the deviation of the best fit from the data is never worse than at $1\sigma$ confidence level.

Looking at Table~\ref{tab:clusterdata} we can quantify better the differences with Paper I: the interaction length slightly diminishes in most of the objects, still staying compatible with typical cluster scales ($\approx 1$ Mpc; see also Eq.~(36)), while the coupling constant with gas is generally larger than previous estimates. Two objects in particular have very large coupling constant values, i.e. clusters A262 and MKW4.

On the other side, the coupling constant related to galaxies is always much lower than the one related to the gas (as it is expected from the discussion in \S\ \ref{sec:hypothesis}, except for two cases, i.e. clusters A133 and RXJ1159. It is interesting to note that among all these exceptions, two of them (MKW4 and RXJ1159) are considered more similar to groups than to real clusters of galaxies or even as extended elliptical galaxies \citep{Vikhlinin05}, thus for all these systems the argument adopted for normal clusters of galaxies might not apply straightforwardly.

For the other two objects no particular features were found in literature to explain the anomalies.
We can only remark that A262 seems to be as small as MKW4 (i.e. they share the same data extension) and differently from the latter it is believed to be a normal cluster of galaxies in literature.
A possible relation between the cluster scale and the gas coupling constant is possible: if we think about this last one as a concentration parameter, we can expect that smaller structures exhibit a higher value for it. For A133 nothing peculiar was found in literature so to justify the high value for the galaxy coupling constant we found; we can only verify that the fit with data is really good.

\begin{figure*}
\centering
  \includegraphics[width=85mm]{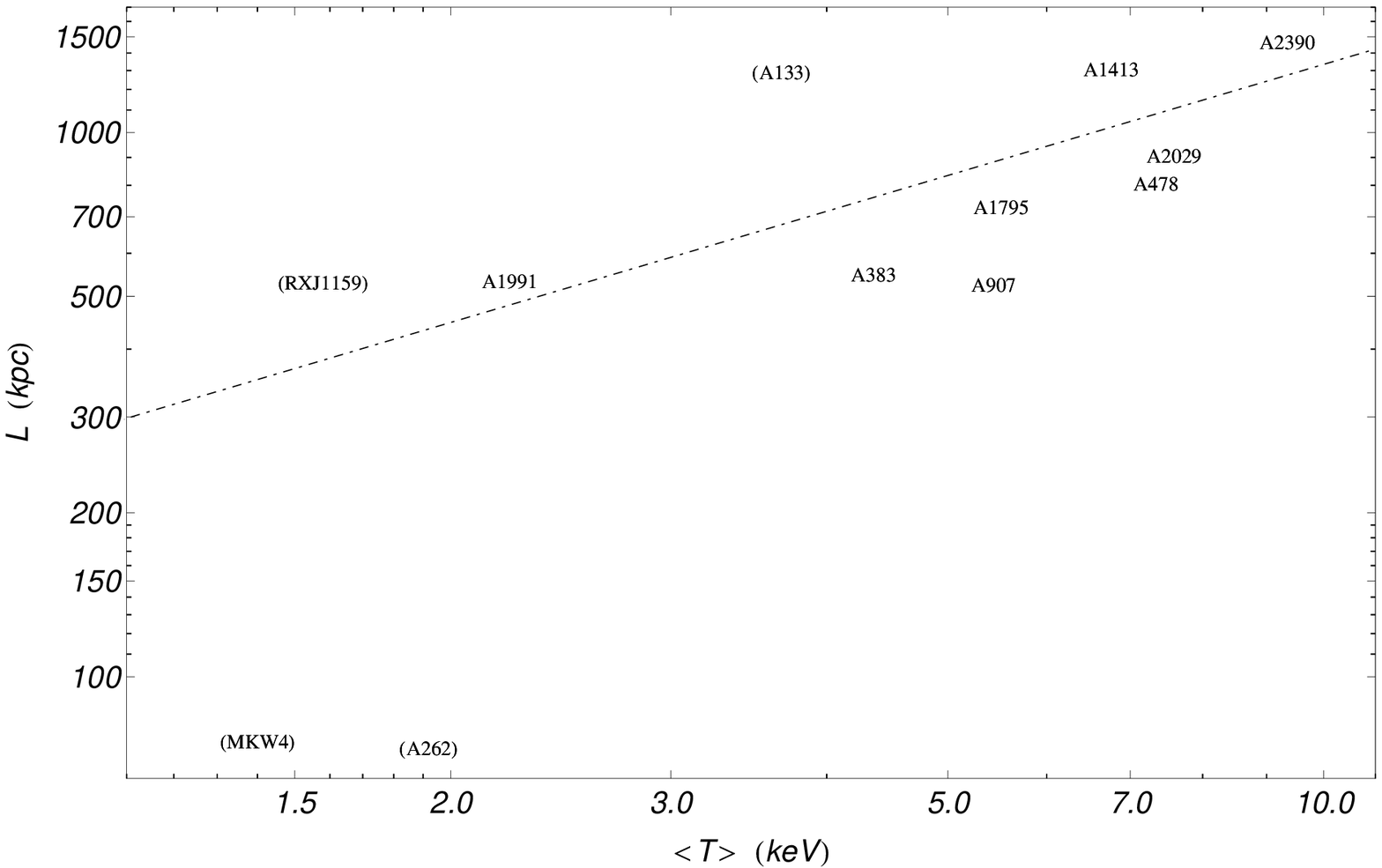}~~~
  \includegraphics[width=85mm]{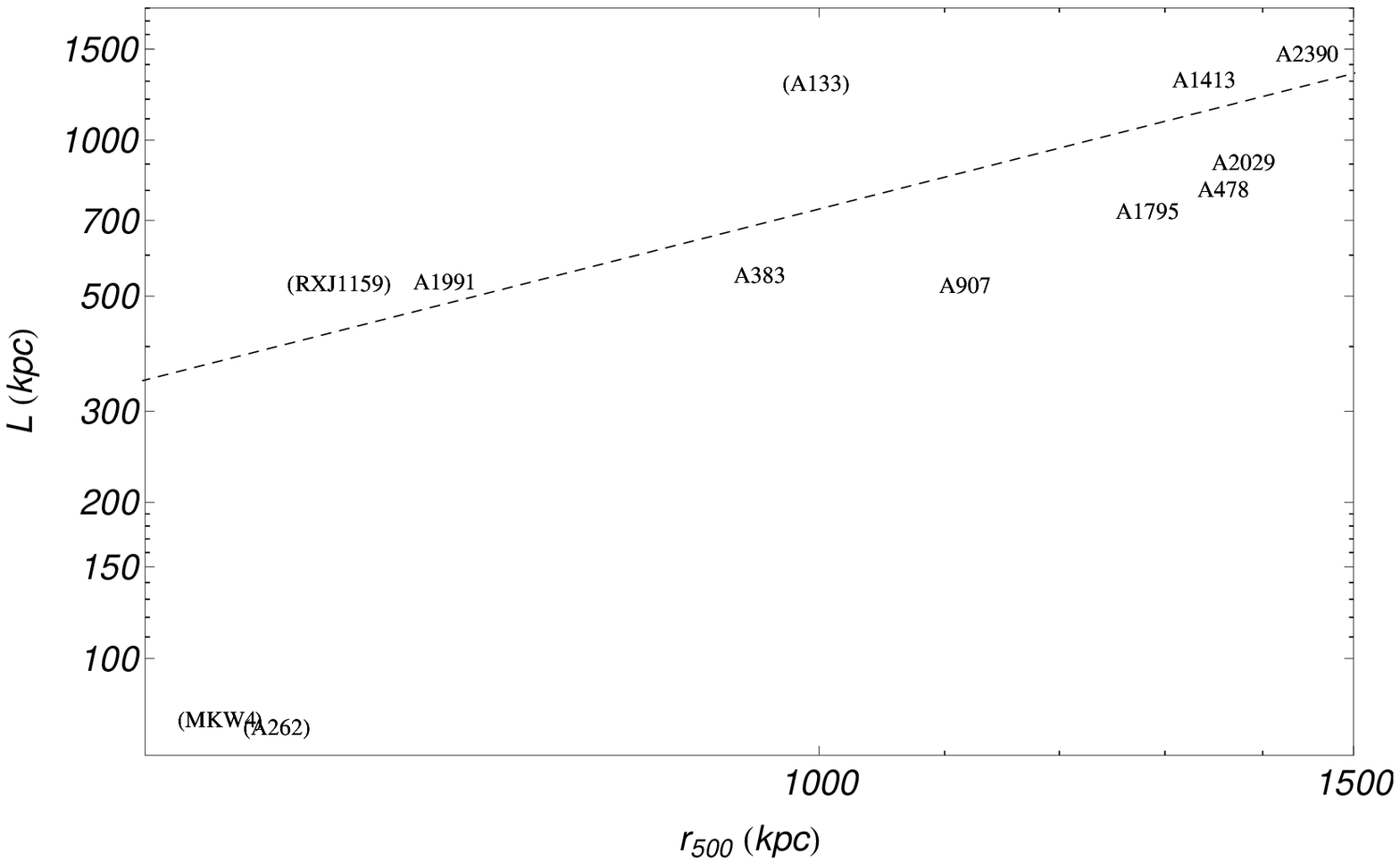}
  \caption{Scalar field length plotted versus mean (gas-density weighted) cluster temperature and the radius $r_{500}$ defined in the text.\label{fig:cluster_L}}
\end{figure*}

As we did in Paper I we want to find some relations among the scalar field parameters and the physical properties of the considered gravitational systems to establish if this alternative scenario can be a valid alternative to general relativity and dark matter. First of all, in the right panel of Fig.~\ref{fig:cluster_L}, we have a relation between the interaction length, $L$, and the radius $r_{500}$, the distance from the center corresponding to an overdensity $\approx500$ times relative to the critical density at the cluster redshift. We prefer this quantity with respect to the virial radius $r_{vir} $ used in Paper I because while the latter is derived using a relation coming from cosmological simulations \citep{Bryan98,Evrard96} and thus depending on the gas-weighted average temperature of the cluster, the former is derived in \citep{Vikhlinin05} directly from observational data by using the hydrostatic equilibrium equation. We have also verified that the two distances are proportional, having $r_{500} \propto r_{vir}^{1.039}$.

The relation between the scalar field length $L$ and the radius $r_{500}$ is derived from an error weighted fit excluding the previously described four peculiar clusters (right panel of Fig.~\ref{fig:cluster_L}):
\begin{equation}
\log L = (-1.59\pm0.82) + (1.49\pm0.27) \cdot \log r_{500} \; .
\end{equation}
We have also found a relation between $L$ and the average gas-weighted temperature $<T>$, left panel of Fig.~\ref{fig:cluster_L}:
\begin{equation}
\log L = (2.45\pm0.11) + (0.68\pm0.17) \cdot \log <T> \; .
\end{equation}

\begin{figure*}
\centering
  \includegraphics[width=85mm]{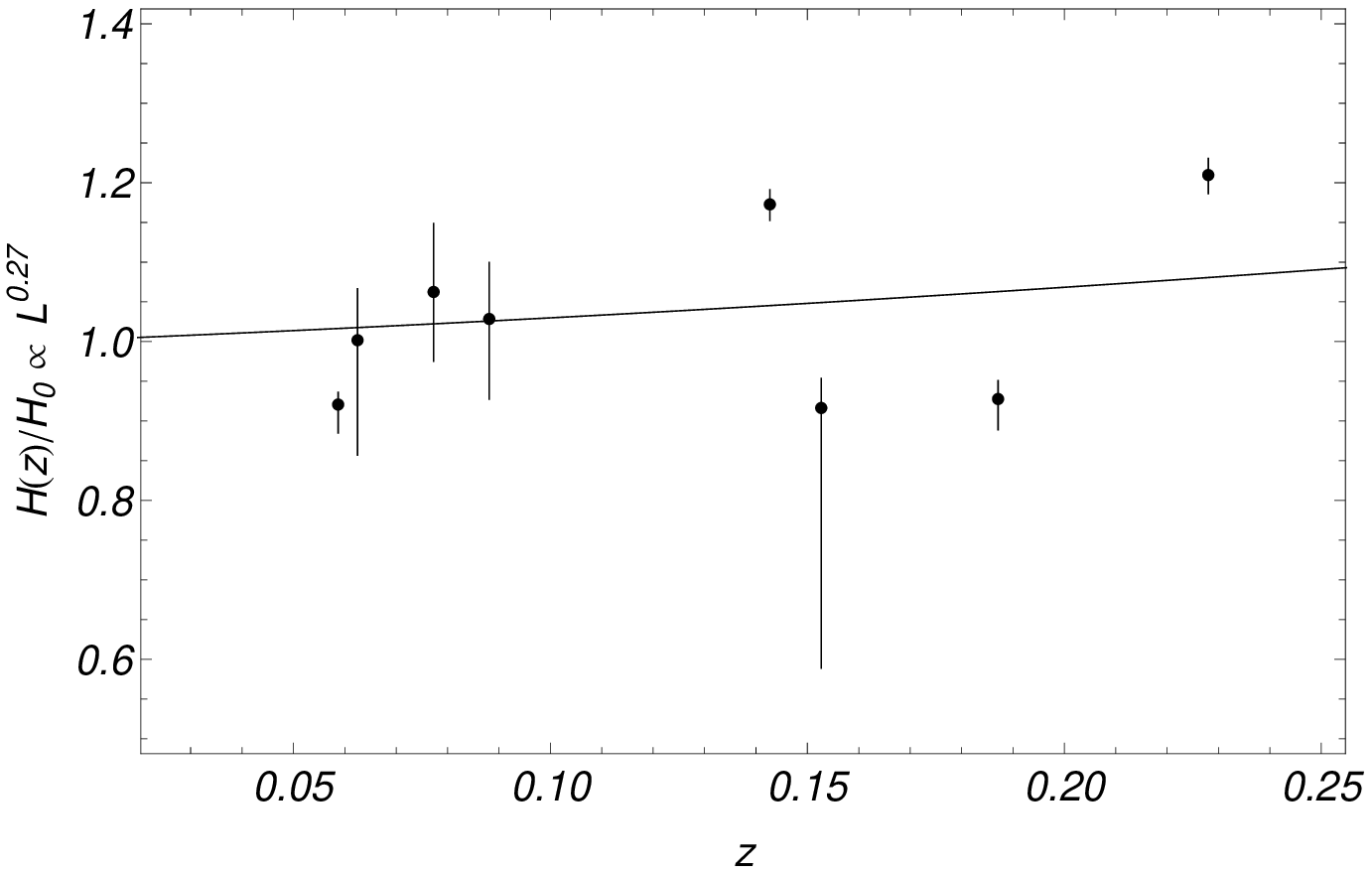}
  \caption{Comparison between the Hubble function $H(z)$ calculated from the fiducial model described in the text and the Hubble function obtained by the empirical relation $H(z) \propto L^{0.27}$. Error bars are calculated from the errors on the interaction length $L$. \label{fig:cluster_H}}
\end{figure*}

As we pointed out in Paper I, there exists a relation between the cluster mass and average temperature, i.e.:
\begin{equation}
M_{\Delta}/T^{3/2} \propto H_{0}/H(z) \; ,
\end{equation}
where $\Delta$ is the overdensity level relative to the critical density at the cluster redshift, so that $M_{180} = M_{vir}$. With the previous phenomenological expressions holding, this mass-temperature relation can be easily converted in:
\begin{equation}
L^{0.27} \propto H(z)/H_{0} \, .
\end{equation}
In Fig.~\ref{fig:cluster_H} we can verify that assuming a fiducial WMAP quintessence model\footnote{http://lambda.gsfc.nasa.gov/product/map/current/\\parameters.cfm}, with $\Omega_m = 0.259$ and $w=-1.12$, and assuming $H_{0} = 72.4$ km s$^{-1}$ Mpc$^{-1}$, the previous relation is able to match the chosen fiducial model following the relation $H(z)/H_{0} = (0.169 \pm 0.006) L^{0.27}$.

\begin{figure*}
\centering
  \includegraphics[width=85mm]{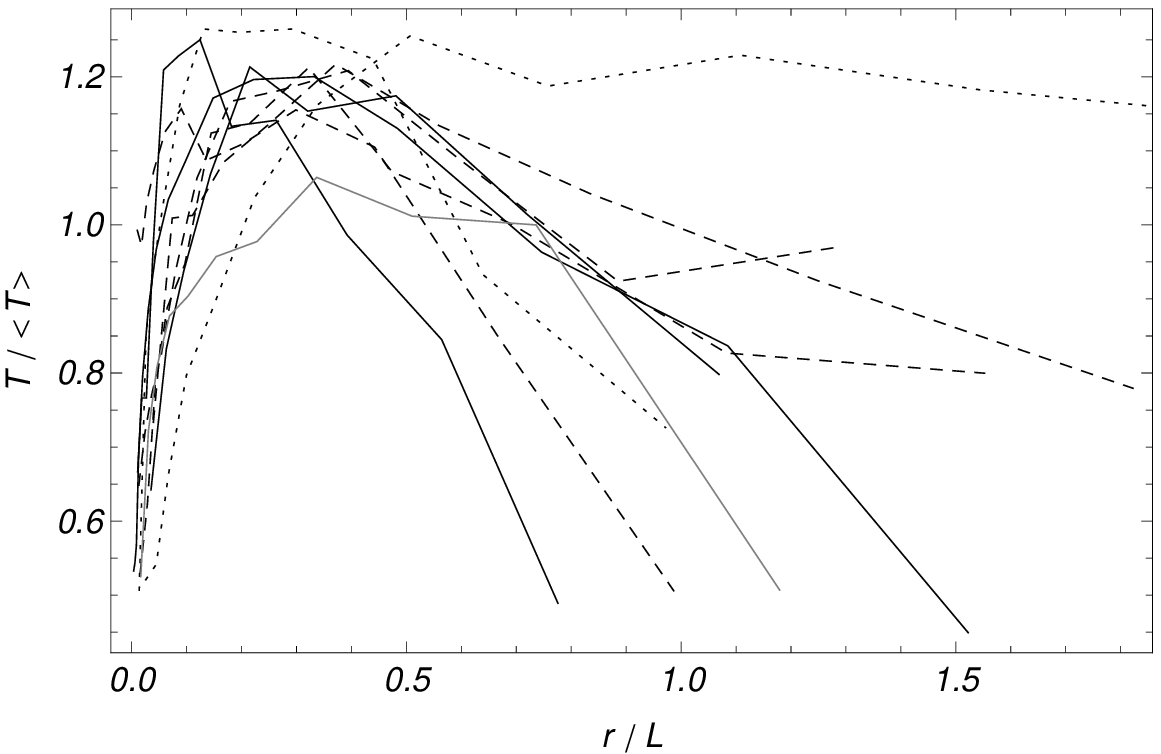}~~~
  \includegraphics[width=85mm]{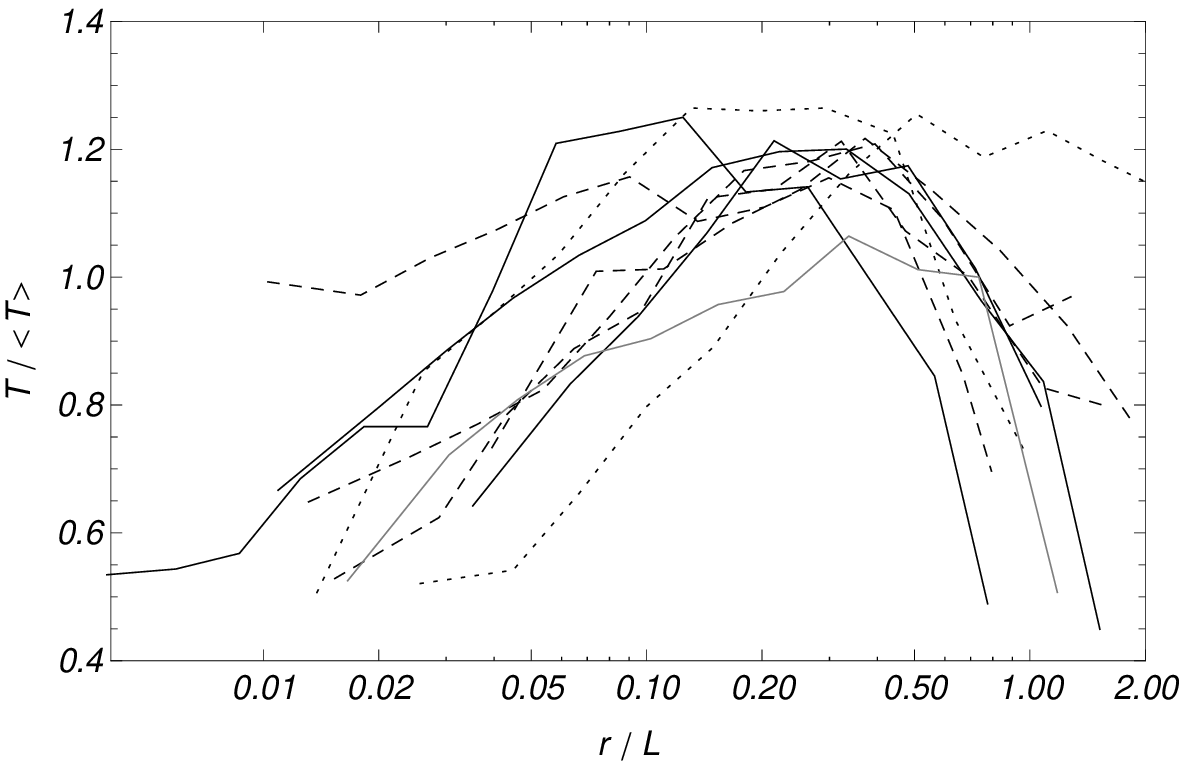}
  \caption{Temperature profiles for all clusters plotted as a function of distance from the center and in units of the scalar field length $L$. The temperatures are scaled to the mean (gas-density weighted) cluster temperature. Dashed line is for high temperature clusters ($<T> \; > 5$ keV); solid line is for intermediate temperature clusters ($2.5< \; <T> \; <5$ keV); dotted line is for low temperature clusters ($<T> \; < 2.5$ keV).\label{fig:cluser_T}}
\end{figure*}

In Figs.~\ref{fig:cluser_T} we also represent the scaled temperature profiles versus the distance from the center of any cluster scaled with respect to the scalar field length obtained by fit, showing the same good reproduction of the self-similarity that characterizes the classical dark matter approach. Moreover, as in Paper I, we detected the absence of the subgroups which the clusters are divided into depending on the mean temperature values \citep{Vikhlinin05}. The only exception with respect to Paper I is A133, which now appears to be to more stretched down, even if no peculiar elements have been found in literature and no problems in performing its analysis have been encountered.

\subsection{LSB spiral galaxies}

We now consider the case of spiral galaxies. The modelling of these systems involves the same number of parameters of elliptical galaxies, i.e. the parameters associated to the scalar field and the stellar mass-to-light ratio, $Y_{\ast}$. As discussed for the elliptical case, the latter parameter cannot vary arbitrarily but has to be consistent with stellar population models. As for spiral systems sample we have adopted in this work one can expect $Y_{\ast}$ to vary in the range $\approx 0.5$ and $\approx 2$, we have decided to be conservative and leave the model parameters to vary in the interval $[0;5]$ (for any further and detailed description see Paper I).

Before we proceed to illustrate the results, we need to remark that the two component approach, for the spiral galaxies, can suffer from some additional noise source with respect to the analysis performed on Paper I and can somehow affect the results. As we have exhaustively discussed in Paper I, the gas data are very noisy and in some cases they can also show particular features which are unavoidable due to intrinsic dynamical properties (e.g. negative velocities due to counter rotating discs, large scatter from non-circular motion, etc.). While these effects have been mitigated in Paper I, where the use of stars and gas together allowed the more circular velocity of stars to dilute the noisy features of the gas, here the coupling of the gas component to the scalar field might result in highly uncertain and irregular profiles. In order to improve the accuracy of the fit for the gas component, we have tried in this work to obtain a more accurate fit of the gas density: the result of this new analysis is evident in the smoother trend of most of the galaxies.

\begin{figure*}
\centering
  \includegraphics[width=80mm]{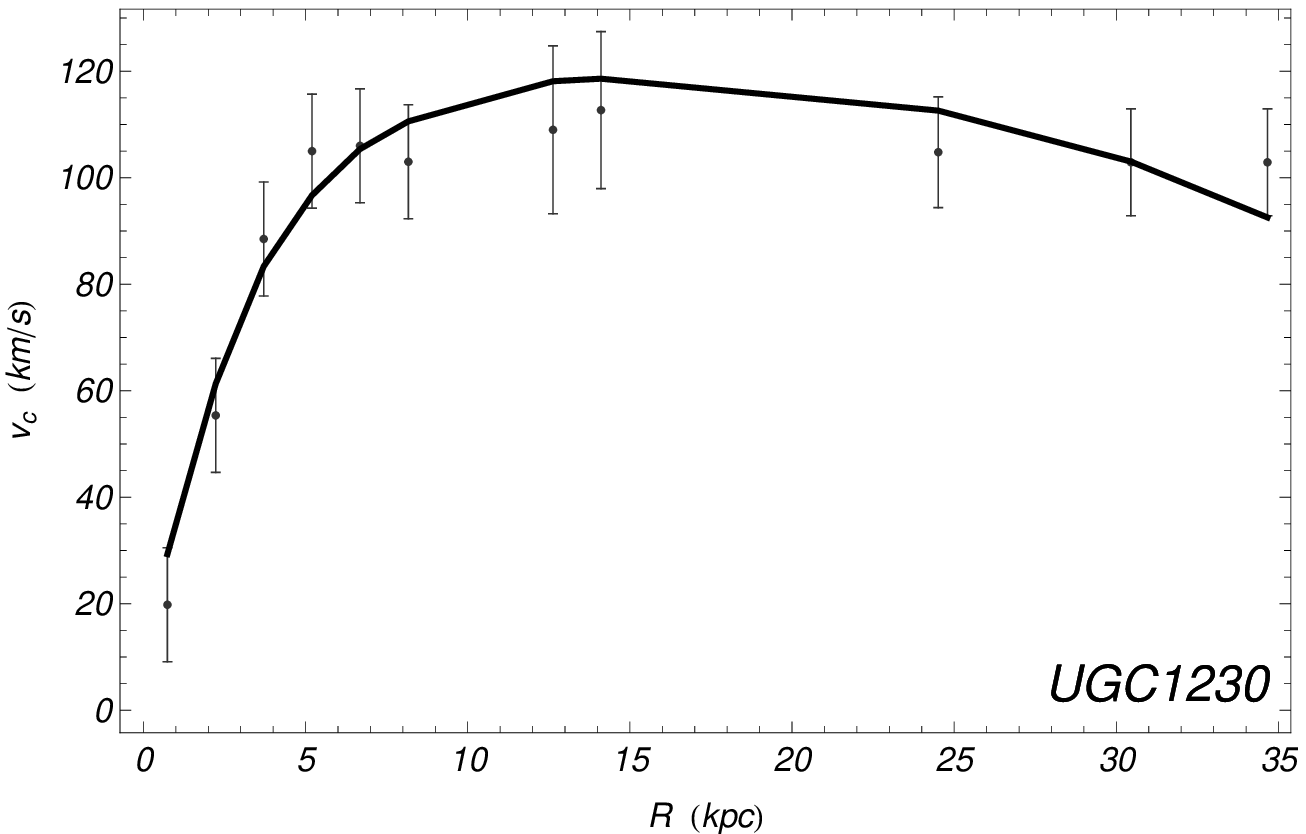}~~~
  \includegraphics[width=80mm]{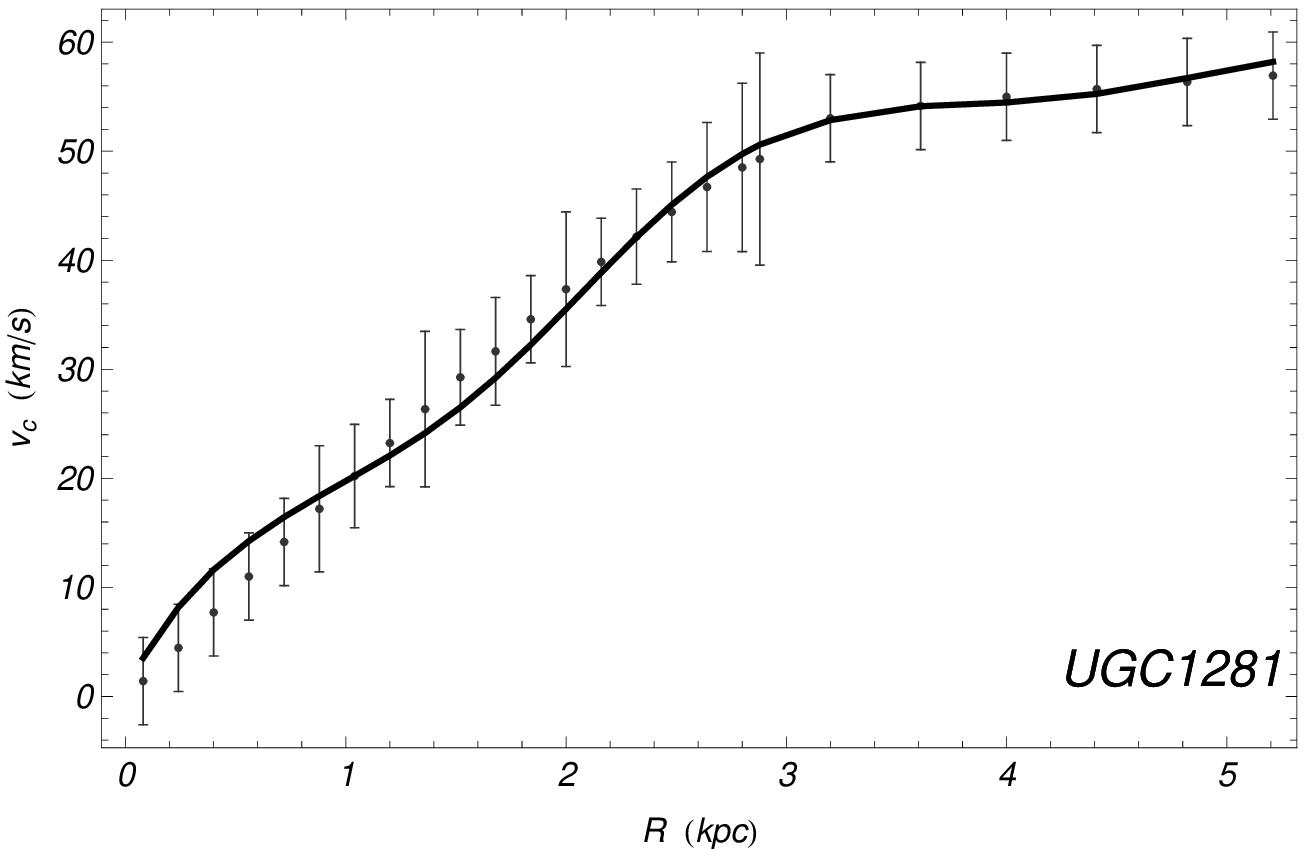}
  \includegraphics[width=80mm]{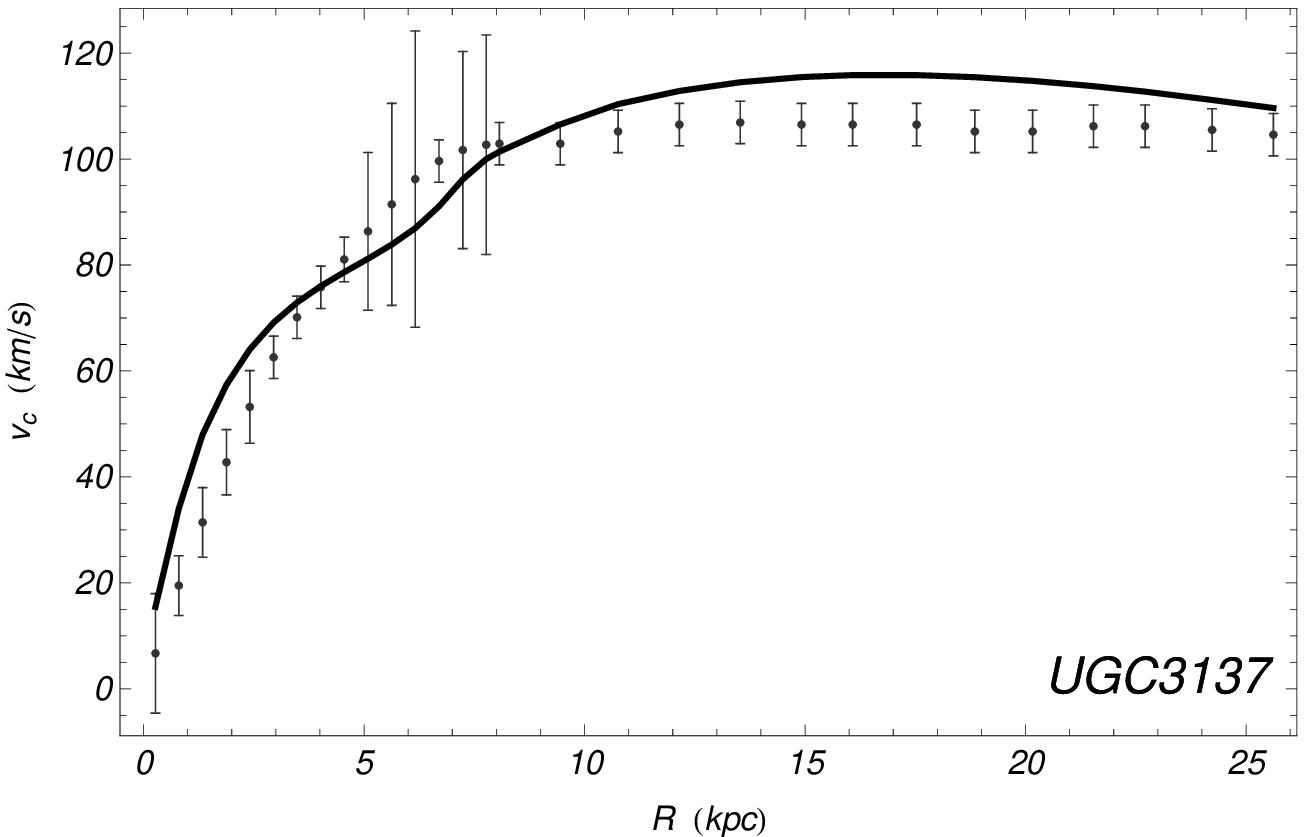}~~~
  \includegraphics[width=80mm]{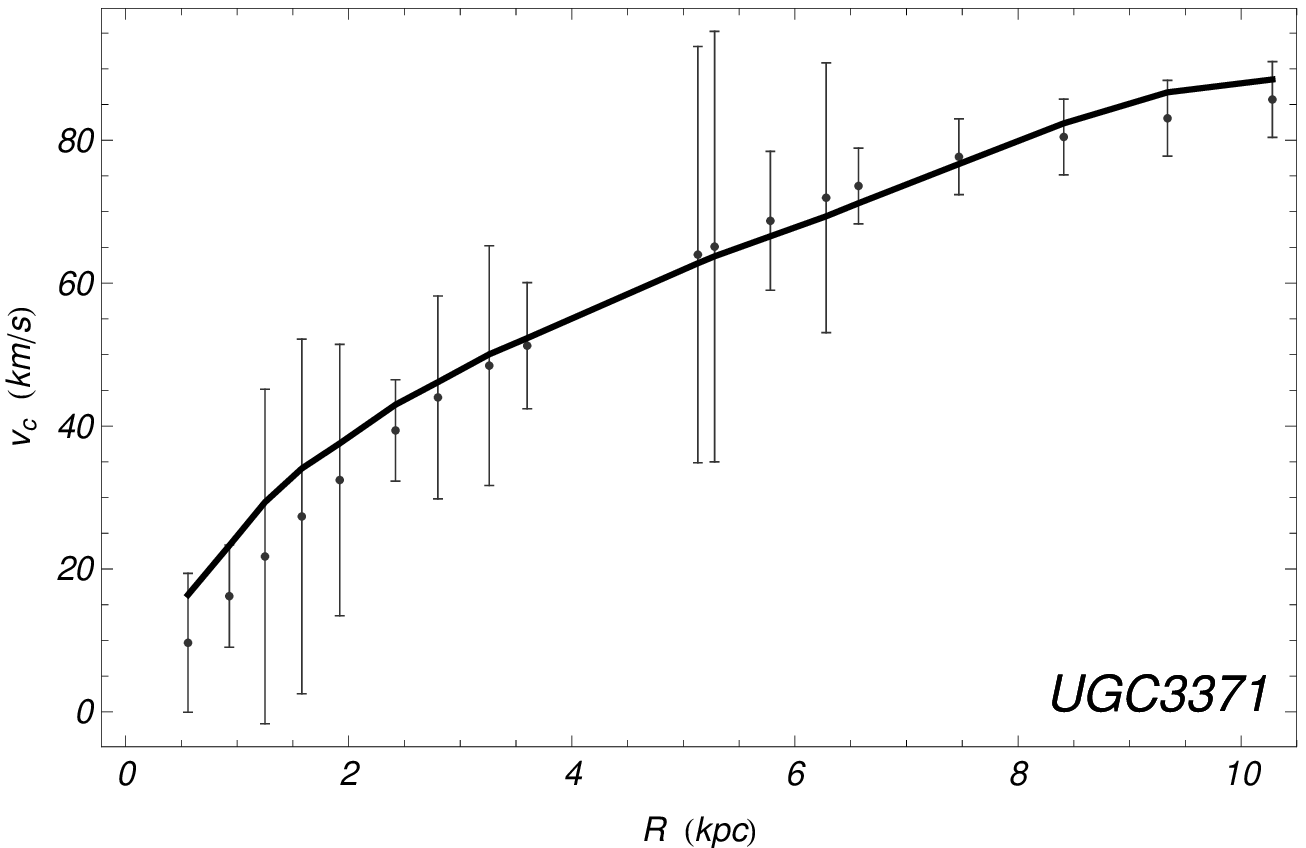}
  \includegraphics[width=80mm]{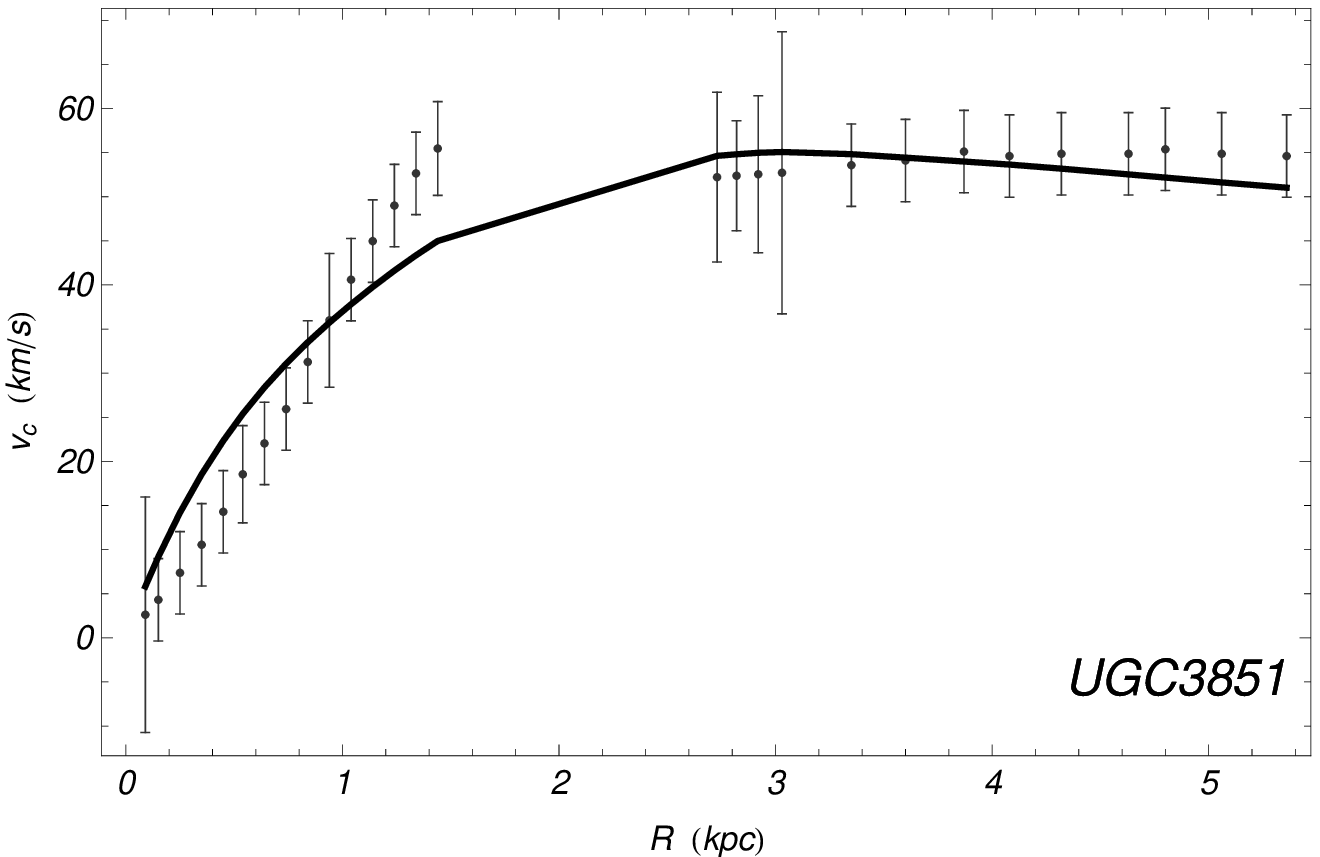}~~~
  \includegraphics[width=80mm]{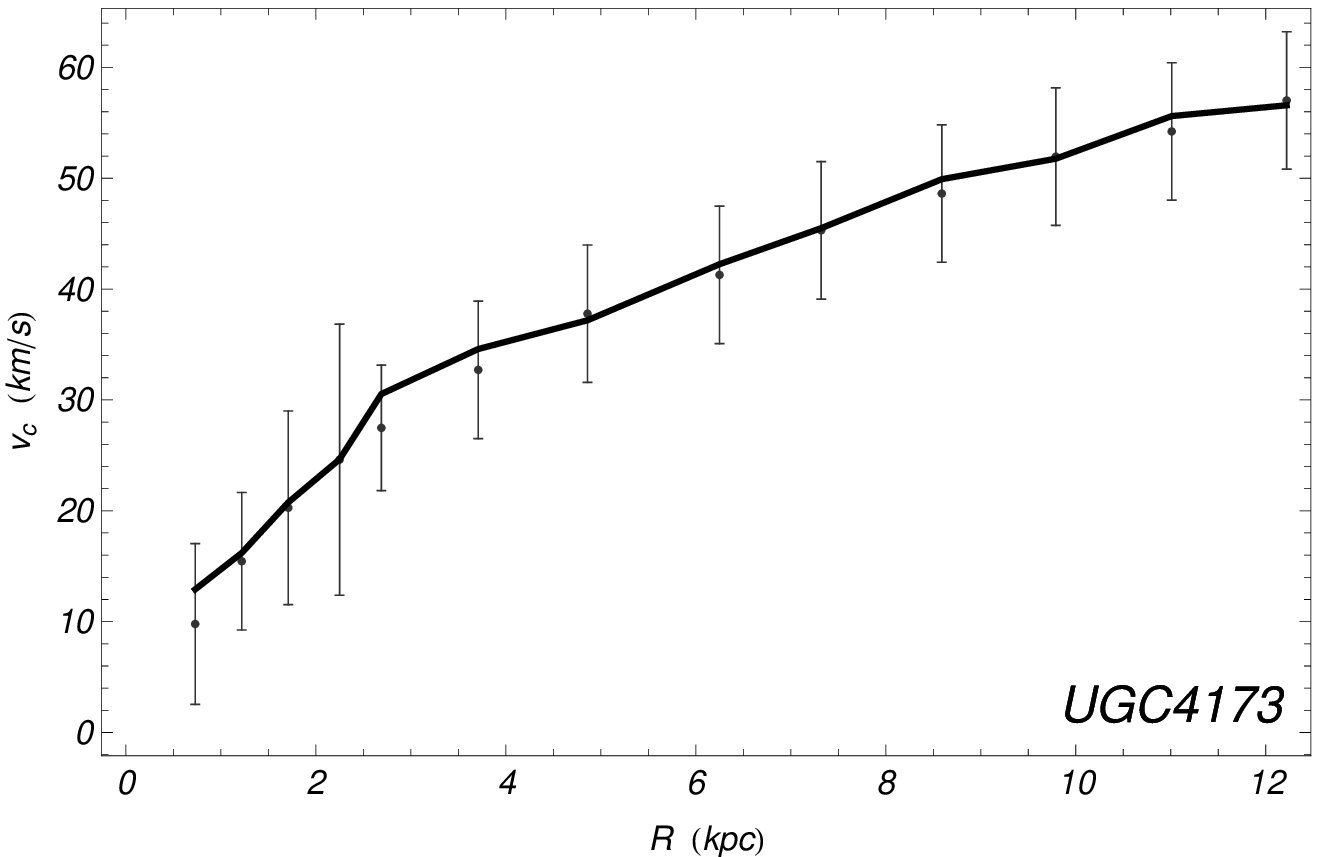}
  \includegraphics[width=80mm]{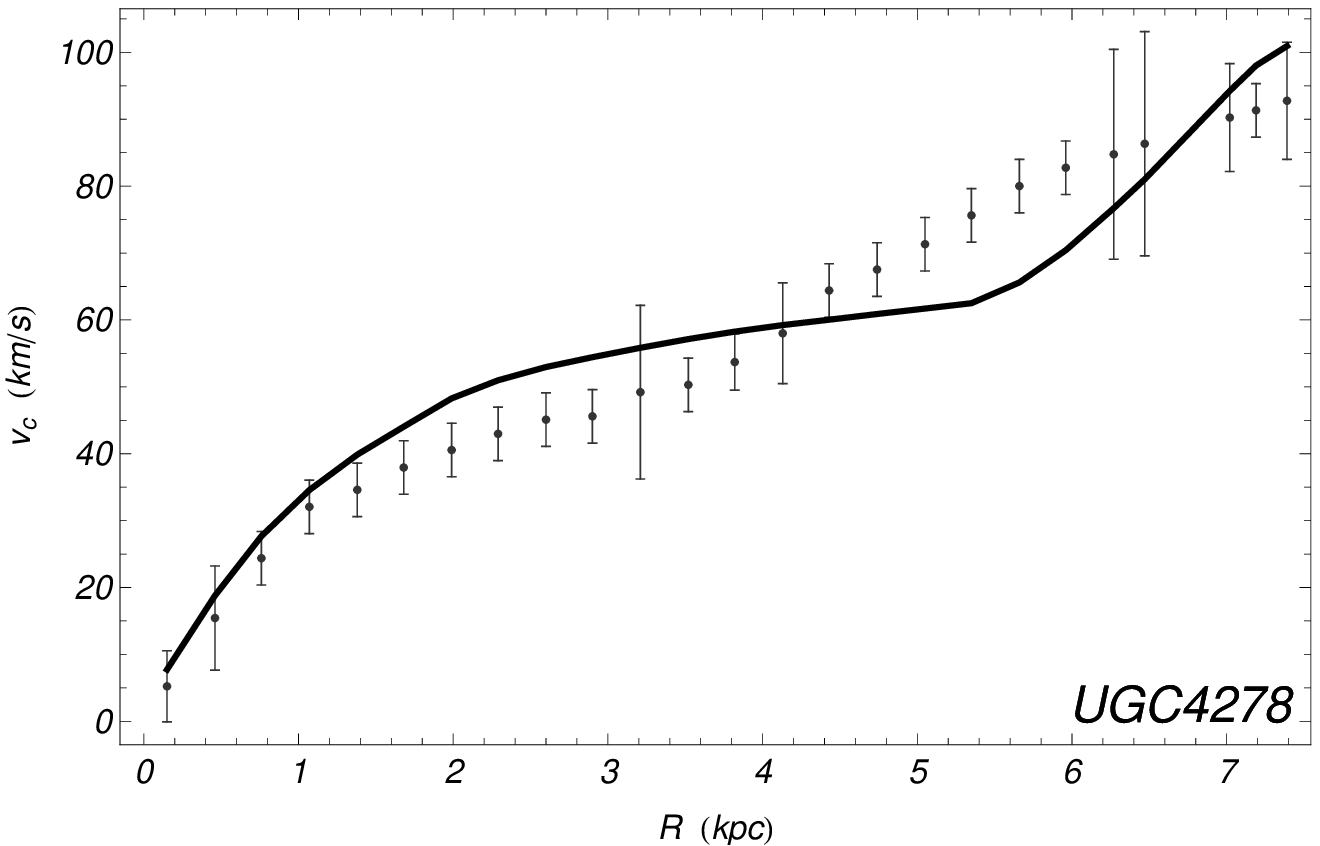}~~~
  \includegraphics[width=80mm]{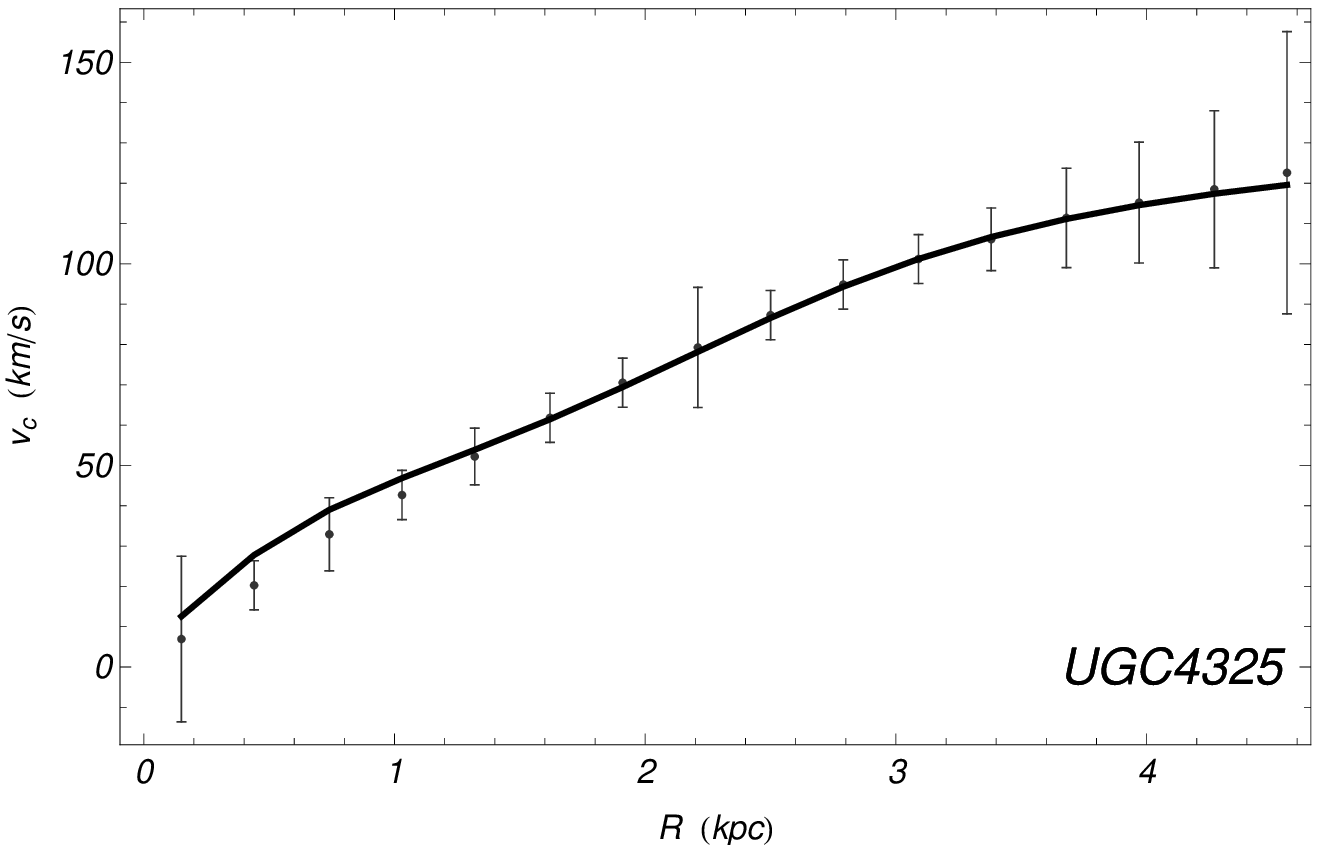}
  \caption{Rotation curves of LSB galaxies. Dots are velocities from data; solid line is the
  theoretical model, $v^{2}_{c} = r \; d \Psi /d r$. \label{fig:cham_gal1}}
\end{figure*}
\begin{figure*}
\ContinuedFloat
\centering
  \includegraphics[width=80mm]{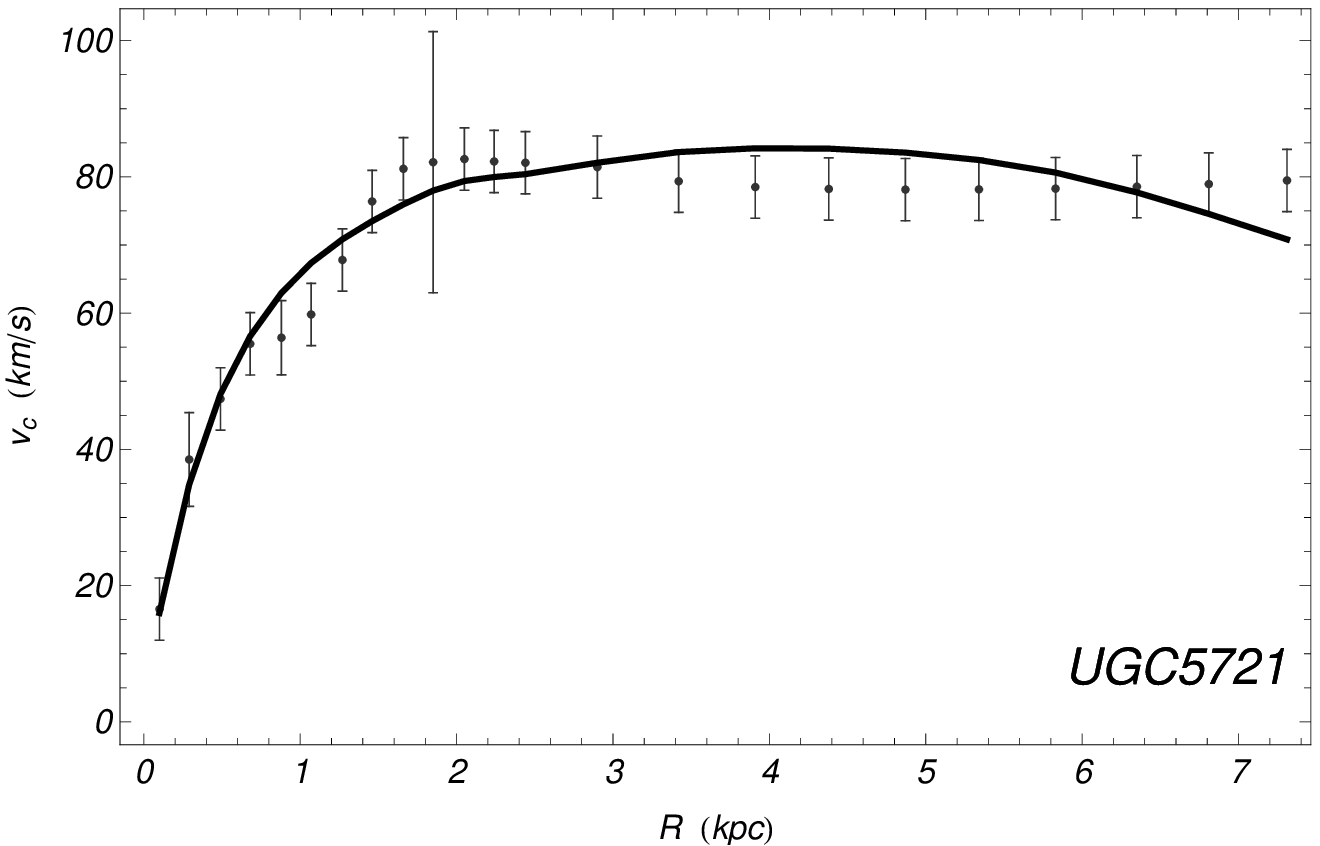}~~~
  \includegraphics[width=80mm]{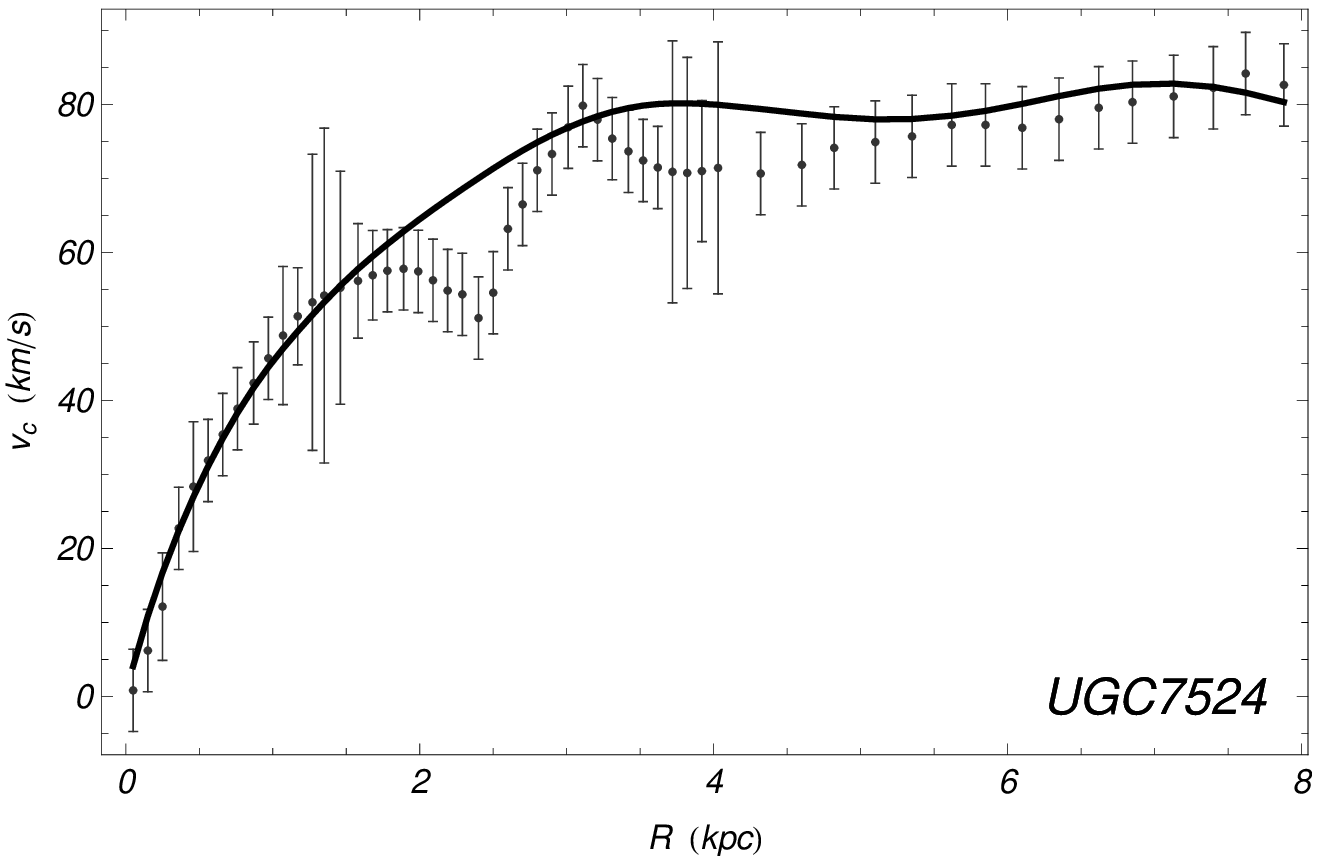}
  \includegraphics[width=80mm]{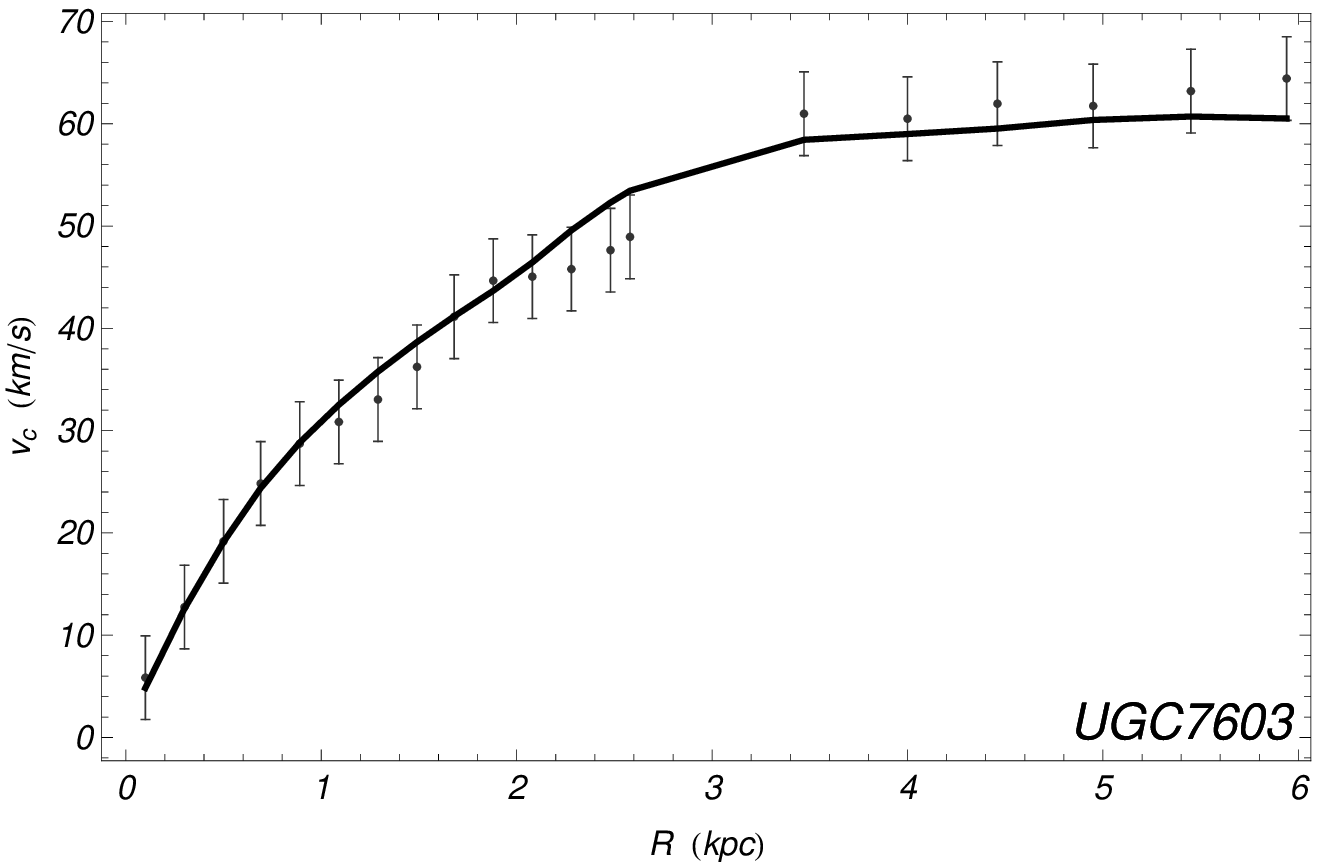}~~~
  \includegraphics[width=80mm]{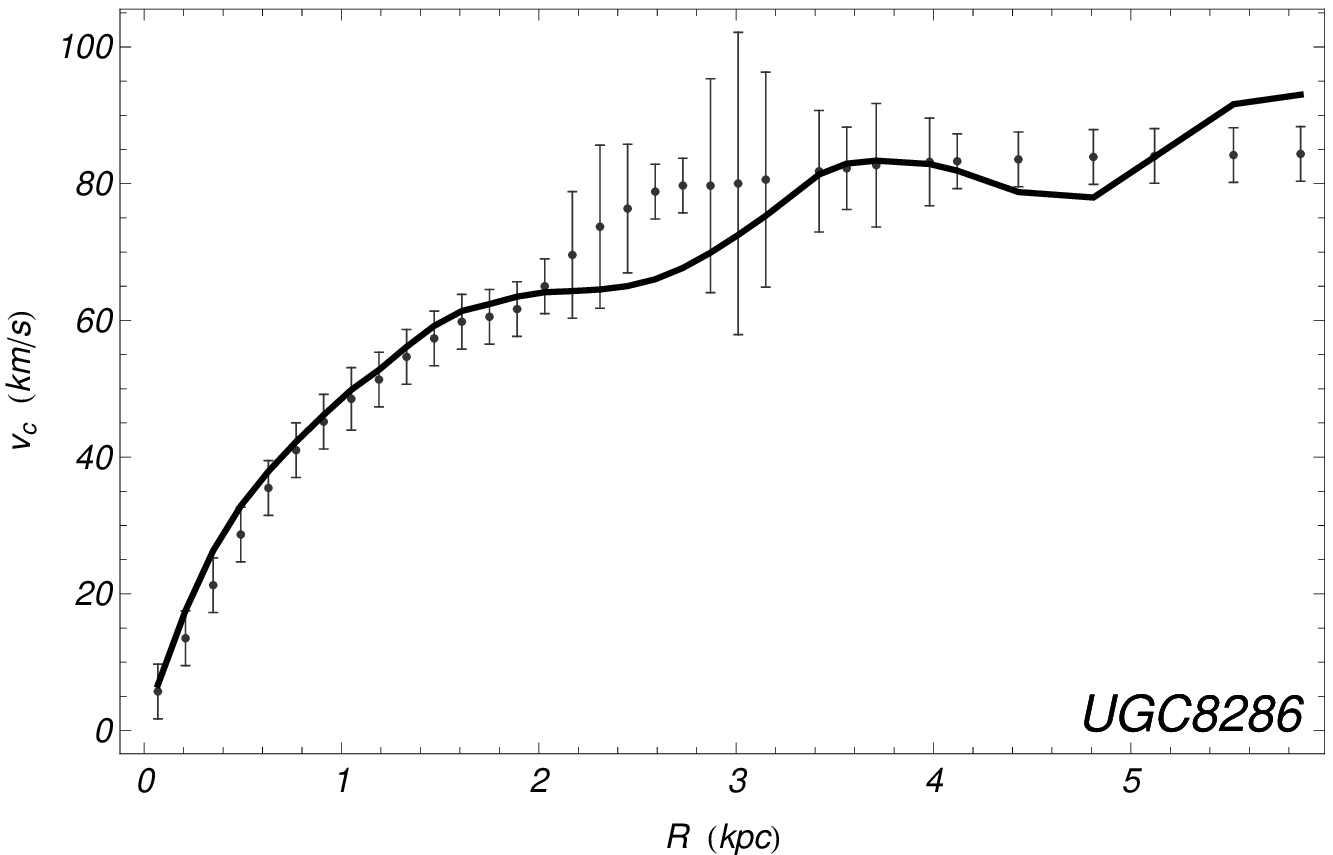}
  \includegraphics[width=80mm]{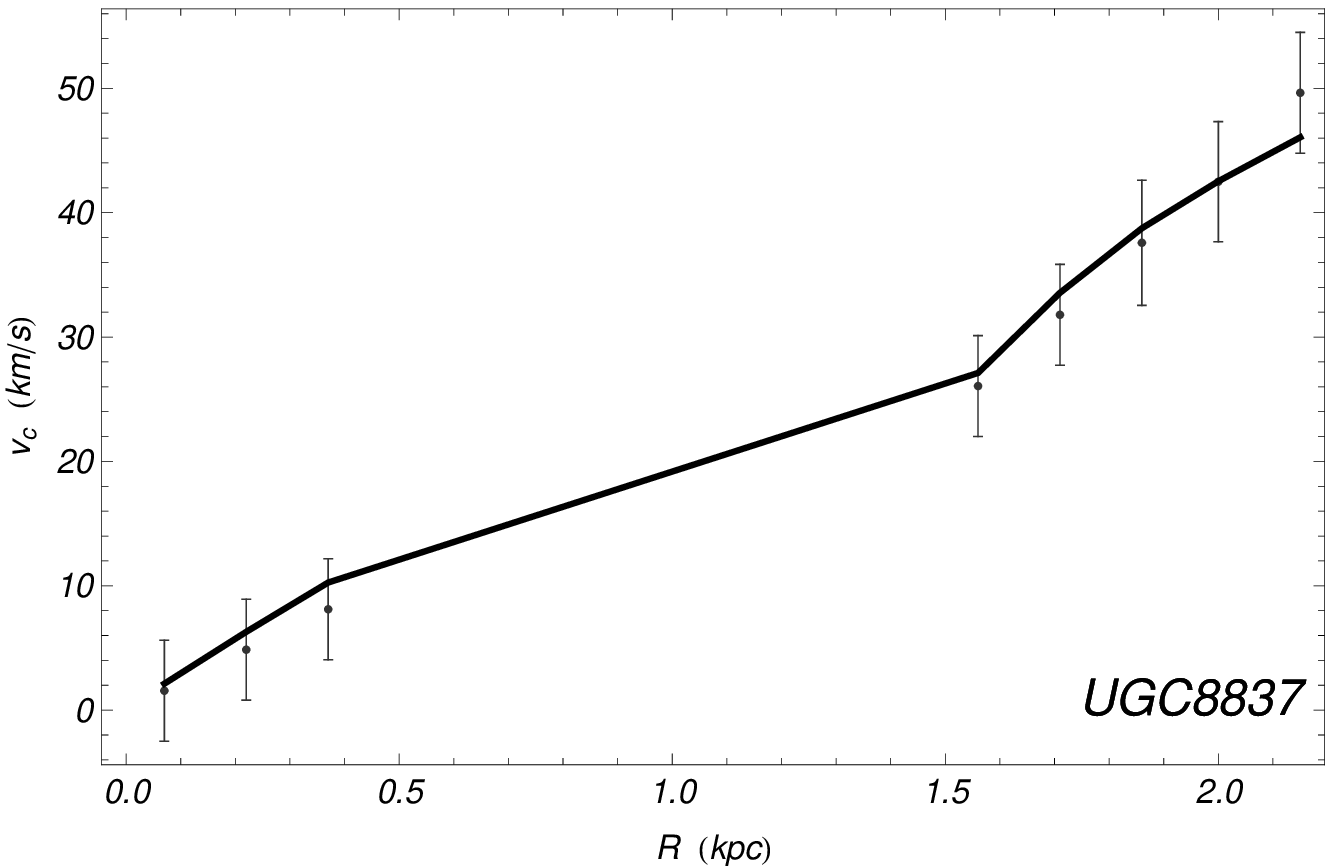}~~~
  \includegraphics[width=80mm]{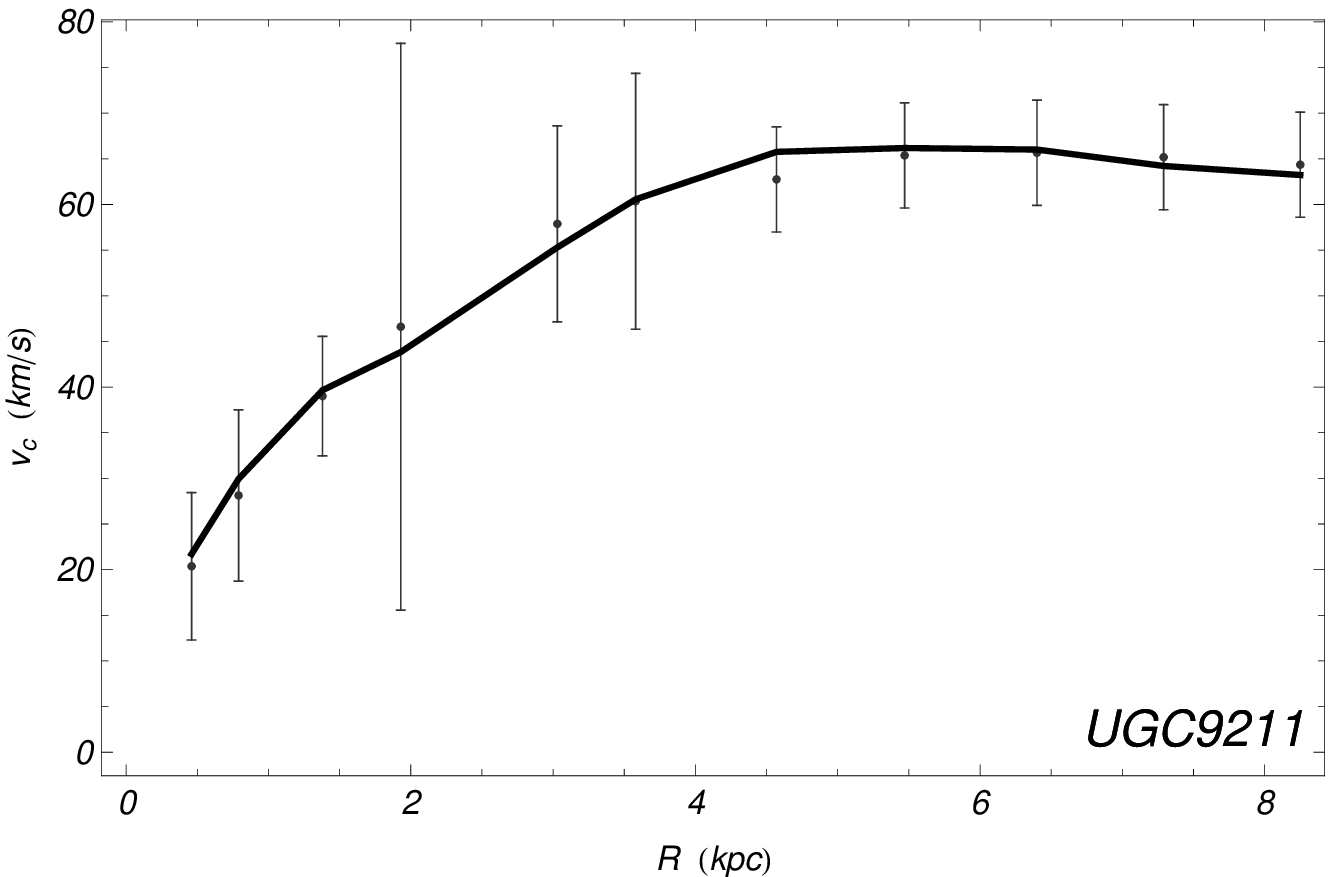}
  \includegraphics[width=80mm]{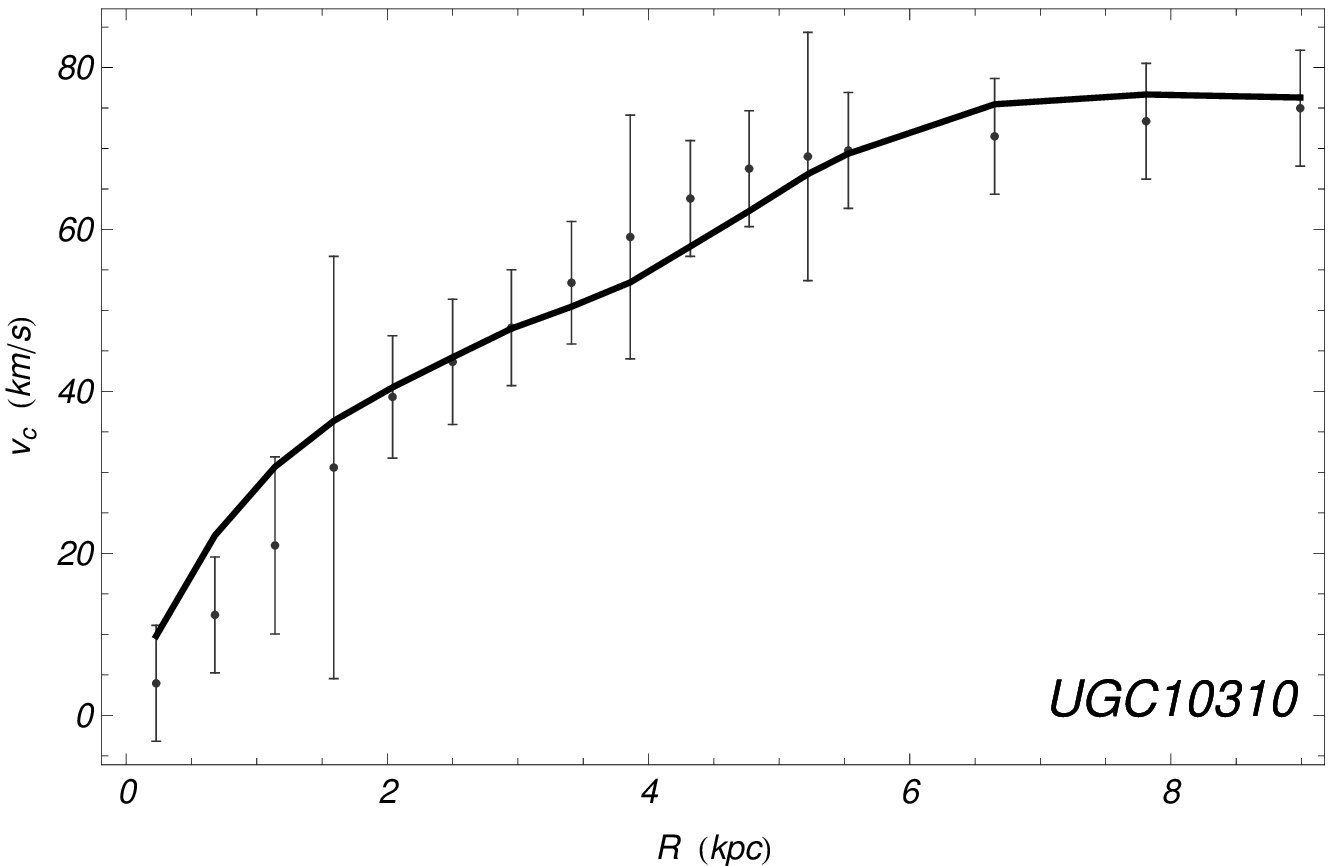}
  \caption{Continued.}
\end{figure*}

The new models of the spiral rotation curves seems to give results that are less clear than the one obtained for clusters in the previous section (see Fig.~\ref{fig:cham_gal1}): by decoupling the gas and the star components, we do not gain much in the quality of the fits (except a slightly better accordance to data of NGC4173 and NGC9211, where the models are closer to the ones obtained in Paper I).
This is the consequence of the evidence showed in the previous section when testing the method on elliptical galaxies. We have seen that the effect of the decoupling of the mass components does not show up in the circular velocity but only the velocity dispersion. In the LSB analysis, as we are modelling the circular velocity we should expect a minimal improvement by the addition of a further coupling constant which implies that we are limited to reliably decouple the different mass components in spiral galaxies because of the unsuitability of the dynamical probe (the rotation curve).

We remind that this \textit{induced} idea has to be considered with caution, since elliptical and spiral galaxies are completely different gravitational structures from a dynamical point of view: elliptical galaxies are dominated by random motions while spiral galaxies are dominated by ordered (rotational) motions. Moreover we can add another element which make our global results more homogeneous and consistent to each other: the decoupling (with an effective improvement in fits) is given for elliptical galaxies and clusters of galaxies, which are both hot systems (i.e. dominated by random motions) and are supposed to be virialized, whereas this does not happen for spiral galaxies. Thus it is possible that the possibility of decoupling the different mass components is related to the reached equilibrium in those two kinds of gravitational structures, while in spiral galaxies stars and gas are again more strictly correlated. In that case, we would have also a correlation of our scalar field analysis with the evolutionary state of the gravitational objects; but to verify this is out of the purpose of this paper.

For all these reasons we have proceeded with the star and gas component decoupling approach also for LSB systems and we have found we have a clear distinct behaviour for the coupling constants. First of all, unsurprisingly the gas coupling constant seems to be very similar to the only one coupling constant of Paper I, and it is very well constrained in the range $[1.1;2.1]$ with only 4 exception: UGC3851, which has the smallest value (as in Paper I); UGC4173, which has a value smaller than one, but still being compatible with the previous range at the highest $1\sigma$ confidence limit; UGC4325 and UGC5721 which have larger values. This means that the dominant coupling constant for LSBs is the one from the gas component.

On the other side, the stellar coupling constant shows a two-fold behaviour: there are galaxies with $\beta_{star}\gg 1$ (UGC1230, UGC3137, UGC3371, UGC3851, UGC4173, UGC4278, UGC8286, UGC9211) and galaxies with $\beta_{star}< 1$ (UGC5721, UGC7524, UGC7603, UGC8837, UGC10310). Two systems (UGC1281 and UGC4325) have $\beta_{star}<1$ but their $1\sigma$ confidence level are compatible with values $\gg 1$.

Noticeably the systems with the larger $\beta_{star}$ are also the ones with the smaller stellar mass-to-light ratio. This is a warning for a possible degeneracy working among these two parameters. Such a degeneracy was already discussed in Paper I where we also noticed an anti--correlation between the $Y_{\ast}$ and the only one $\beta$ adopted. The fact that we have now broken down $\beta$ in two coupling constants, ensures us that $\beta_{gas}$ is unaffected by any degeneracy, while $\beta_{star}$ is not. This is a major benefit we have gained by the adoption of the mass component decoupling approach; we could add that the main coupling constant one should rely on is $\beta_{gas}$. Interestingly the gas status is really what it makes the big difference between cold and hot dynamical systems which correspondingly possess cold and hot gas.

Concerning the best fit stellar mass-to-light ratios, we have 11 of 15 galaxies compatible at $1\sigma$ level with the previously prescribed range; one (UGC7524) has an higher value; and three (UGC7603,UGC8286, UGC8837) have values smaller than 0.5. Among these three, the first two were peculiar in Paper I too, even if they can be considered acceptable, and UGC7603 in particular still has the smallest value, $Y_{\ast} = 0.052$. Only in two cases (UGC4325 and UGC5721) we have more compatible values mainly due to their large errors. Anyway, in general, in Paper I we obtained a better match of $Y_{\ast}$ with the prescribed range in 10 of 15 cases.

Finally, the interaction length shows a wider range than the clusters galaxies case and is not very well constrained as in that case. One particular case being UGC3137: even in Paper I it was one of the galaxies with the largest interaction length, but now it has a too large value, $L \approx 500$ kpc, which is difficult to understand. Anyway, by taking a look to its rotational profile, we can see how it satisfies two main requirements: $1.$ it is the second largest object in the sample (or, better, the second object for which observations of the gas component where done up to a larger distance from the center); $2.$ it is the most clear case in which it seems to have reached the expected plateau in the rotation curve. Lacking more detailed, extended and less disturbed data for the galaxies in our sample we cannot conclude anything; we can only argue that in spiral galaxies, more than clusters and the elliptical galaxies, the degeneracy among parameters due to limited extension of the data could be more important.

\begin{figure*}
\centering
  \includegraphics[width=89mm]{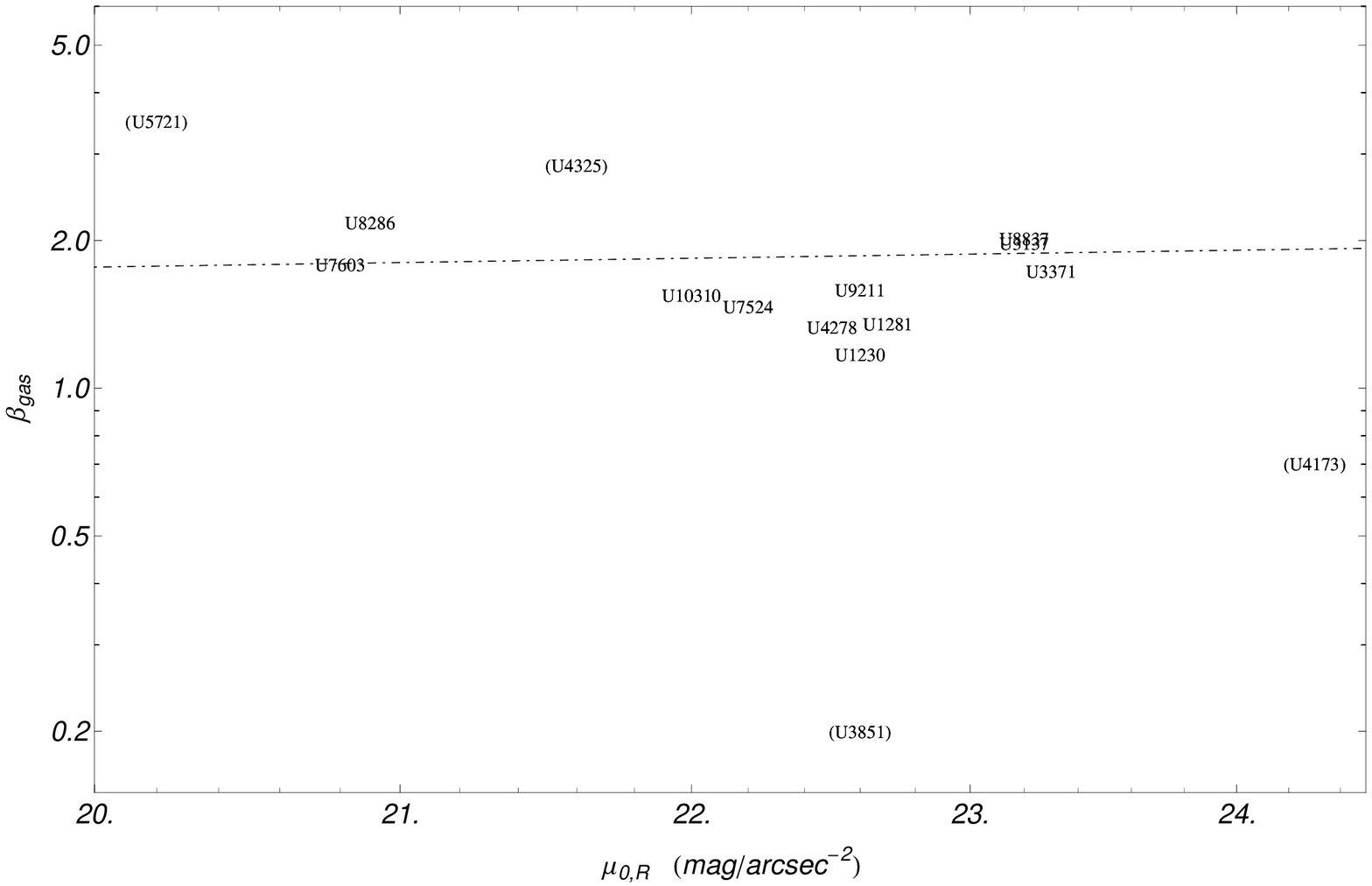}
  \includegraphics[width=89mm]{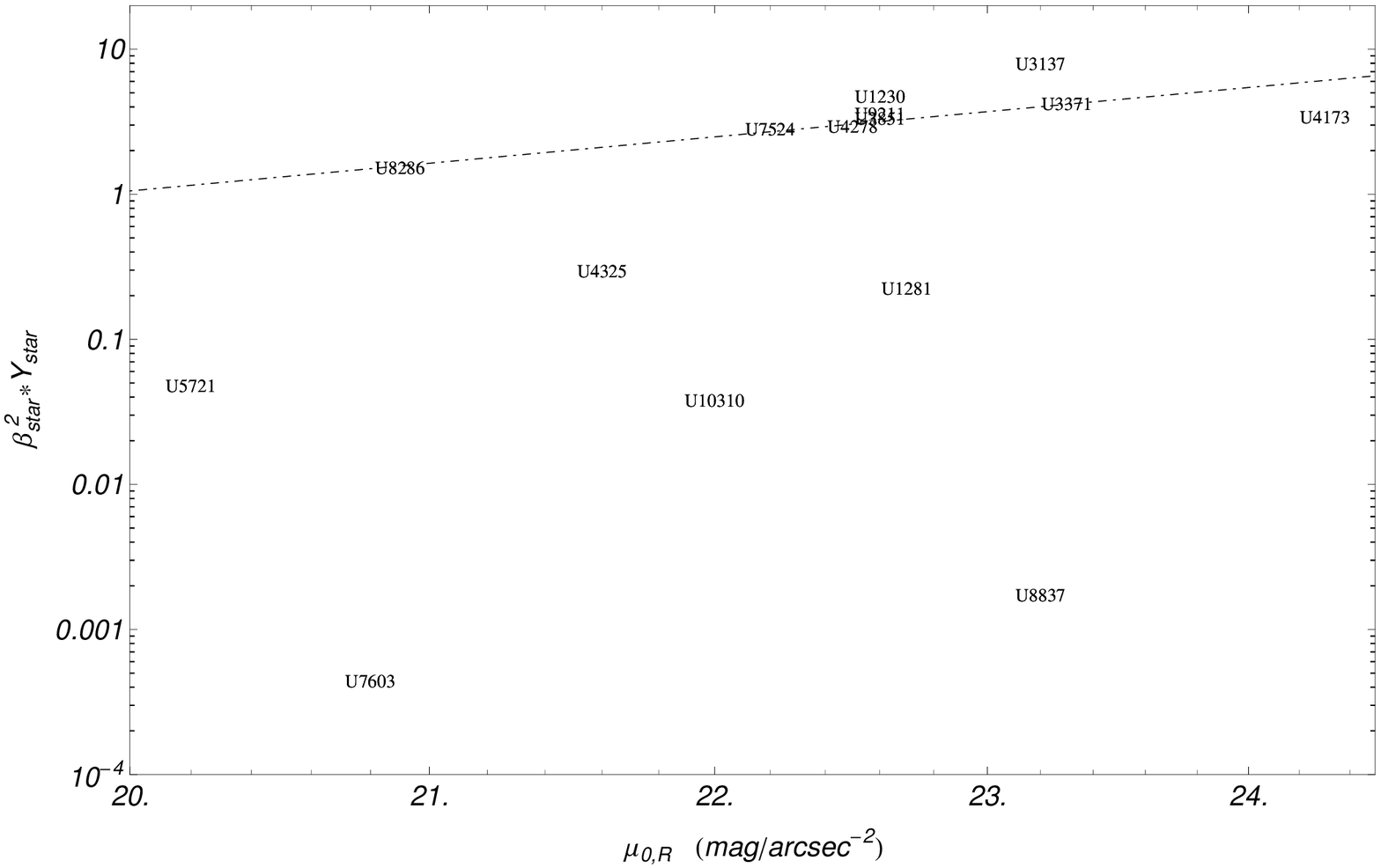}
  \caption{Correlation among the scalar field coupling constants and the central surface brightness of spiral galaxies.
  \label{fig:spiral_bMu0}}
\end{figure*}
\begin{figure*}
\centering
  \includegraphics[width=89mm]{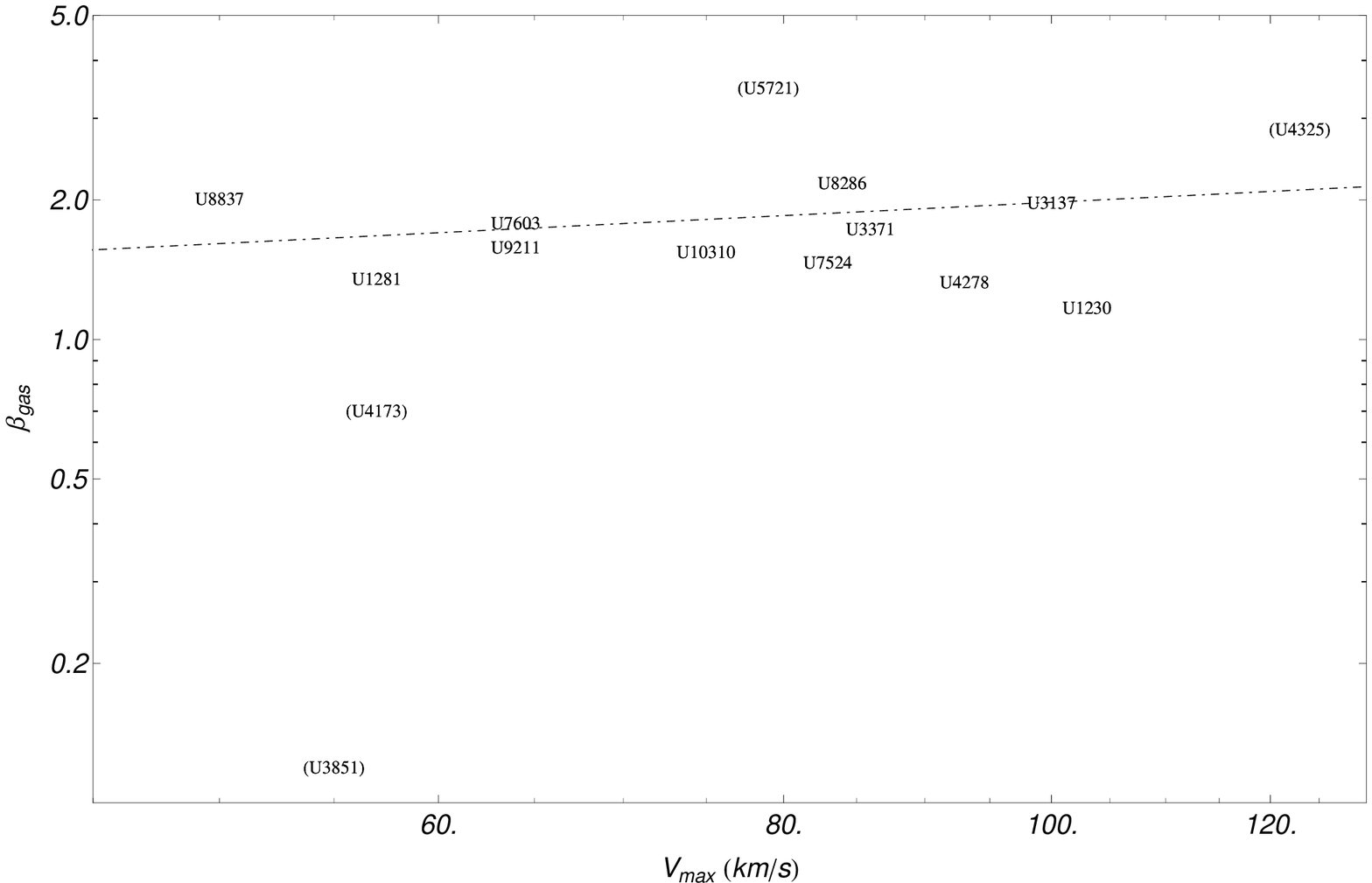}
  \includegraphics[width=89mm]{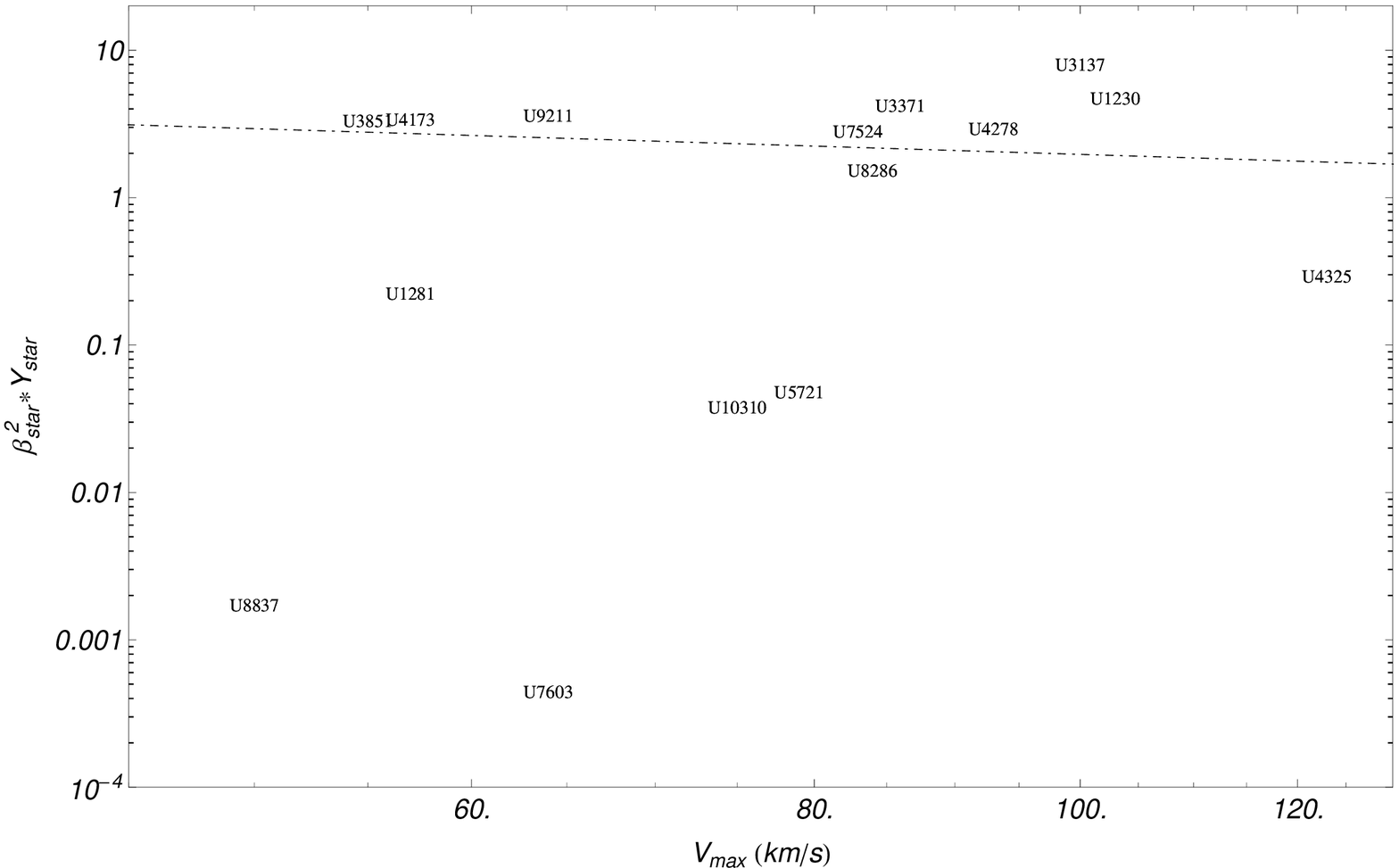}
  \caption{Correlation among scalar field coupling constants and the maximum stellar velocity.
  \label{fig:spiral_bVmax}}
\end{figure*}
\begin{figure*}
\centering
  \includegraphics[width=89mm]{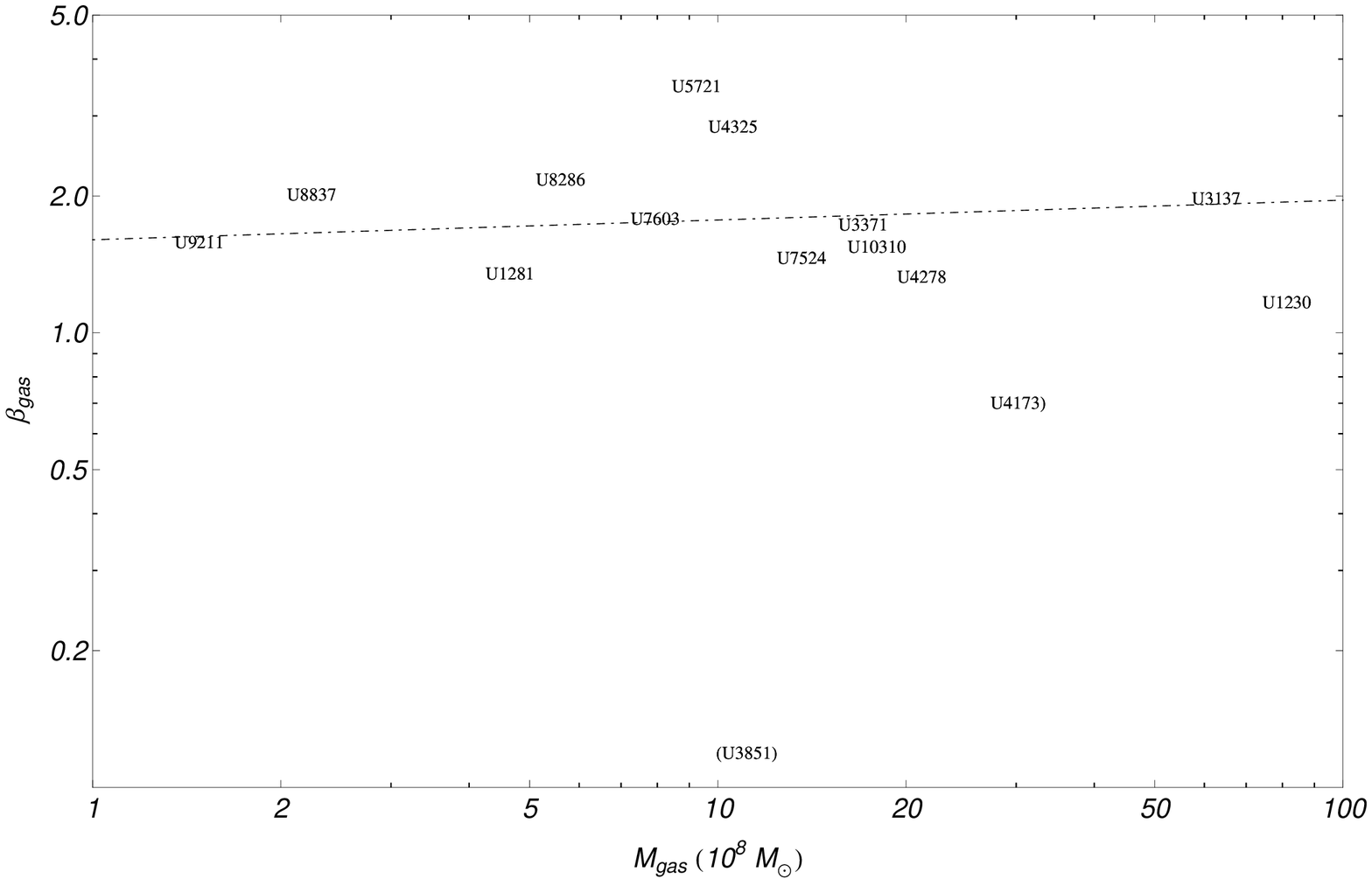}
  \includegraphics[width=89mm]{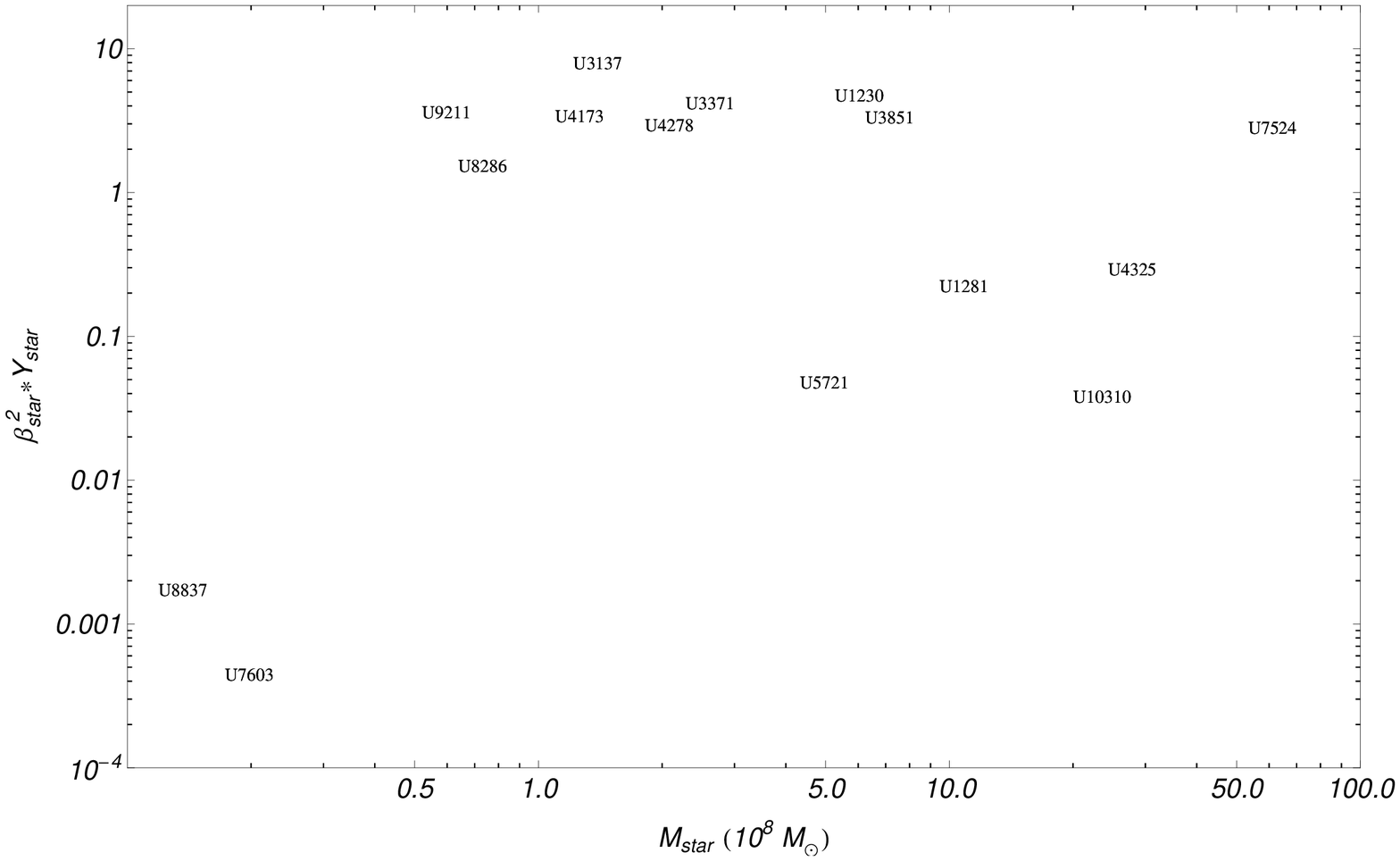}
  \caption{Correlation among scalar field coupling constants and the gas mass.
  \label{fig:spiral_bMgas}}
\end{figure*}
\begin{figure*}
\centering
  \includegraphics[width=89mm]{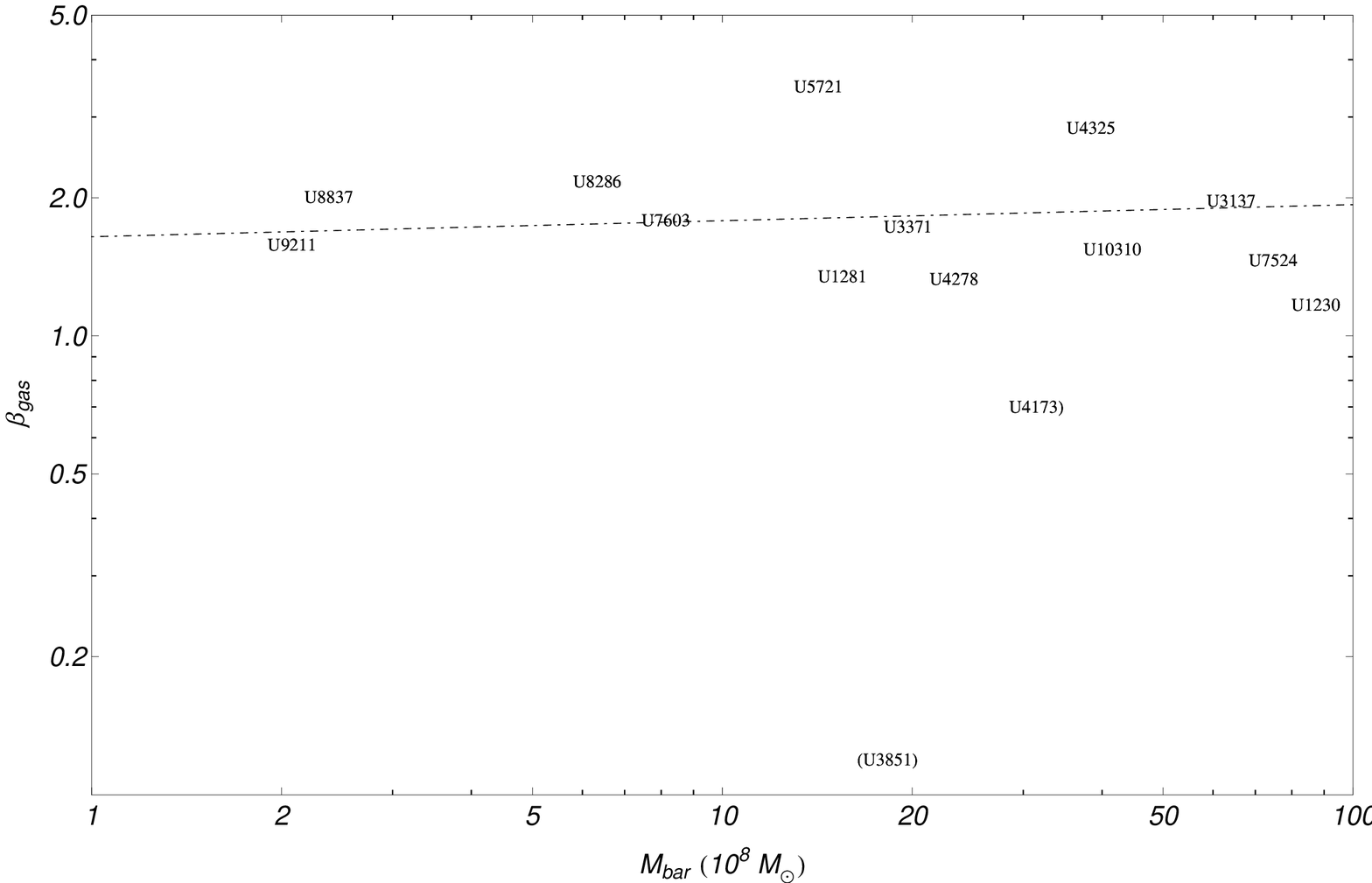}
  \includegraphics[width=89mm]{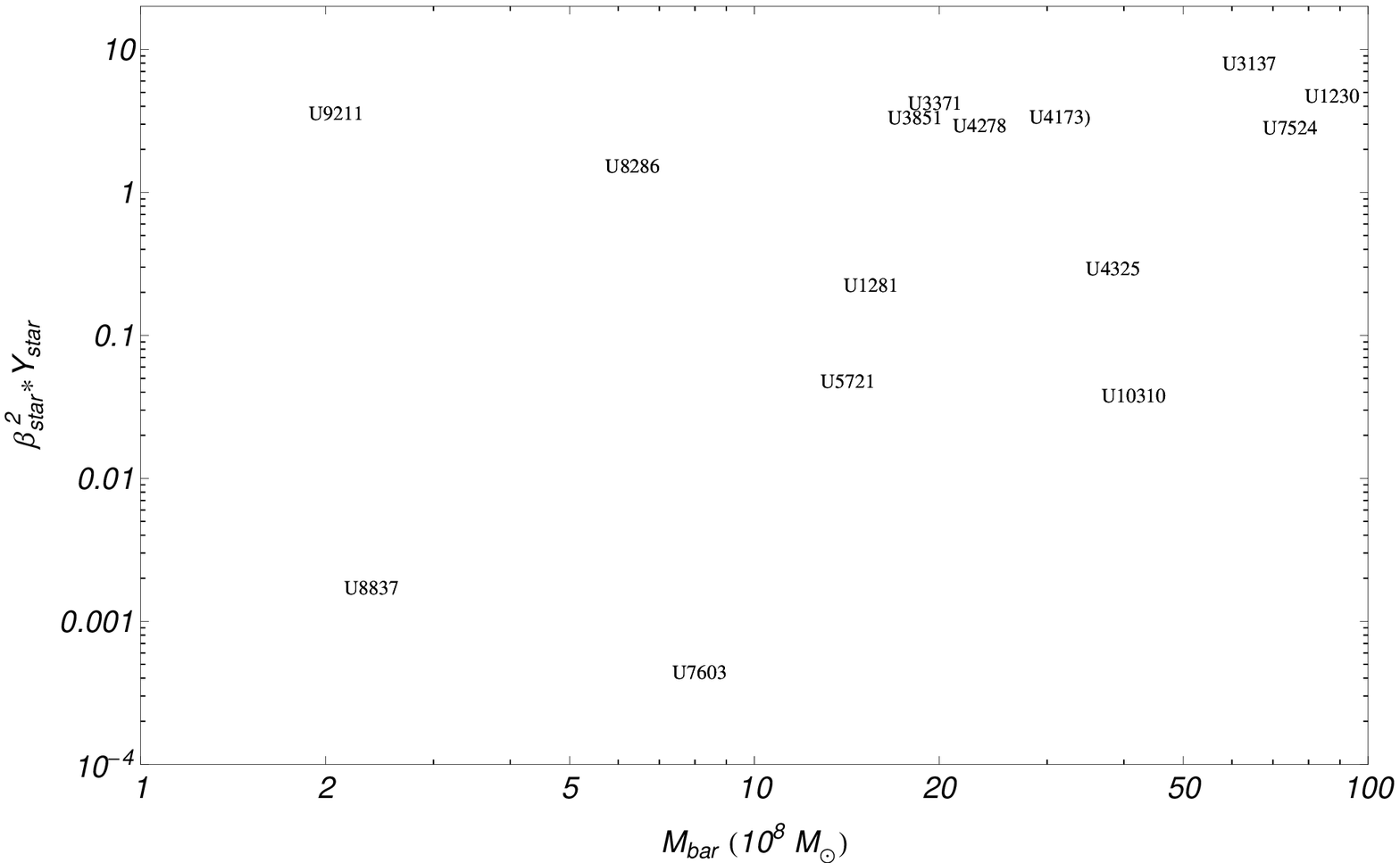}
  \caption{Correlation among scalar field coupling constants and the total baryonic mass.
  \label{fig:spiral_bMbar}}
\end{figure*}
\begin{figure*}
\centering
  \includegraphics[width=89mm]{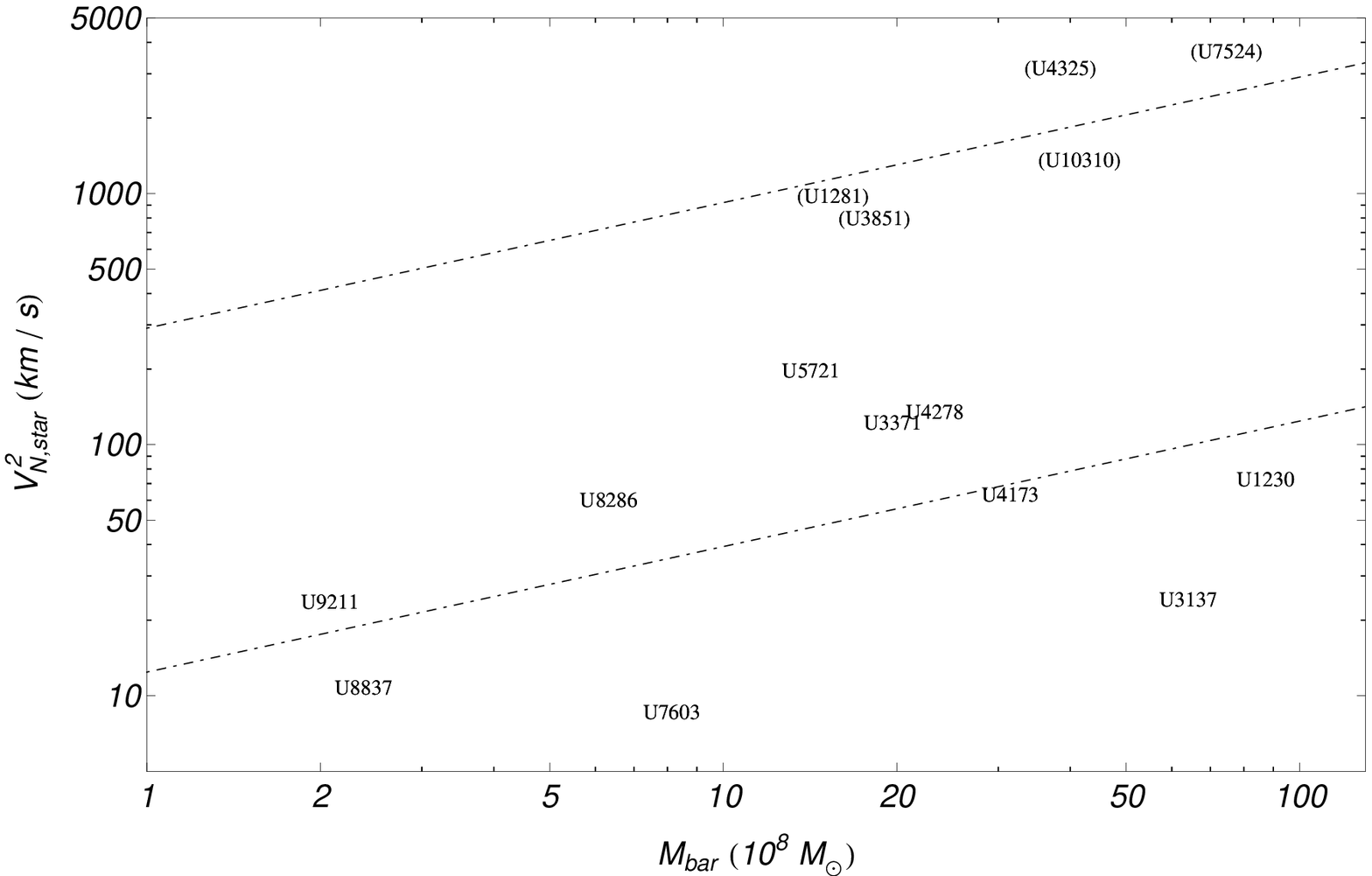}
  \includegraphics[width=89mm]{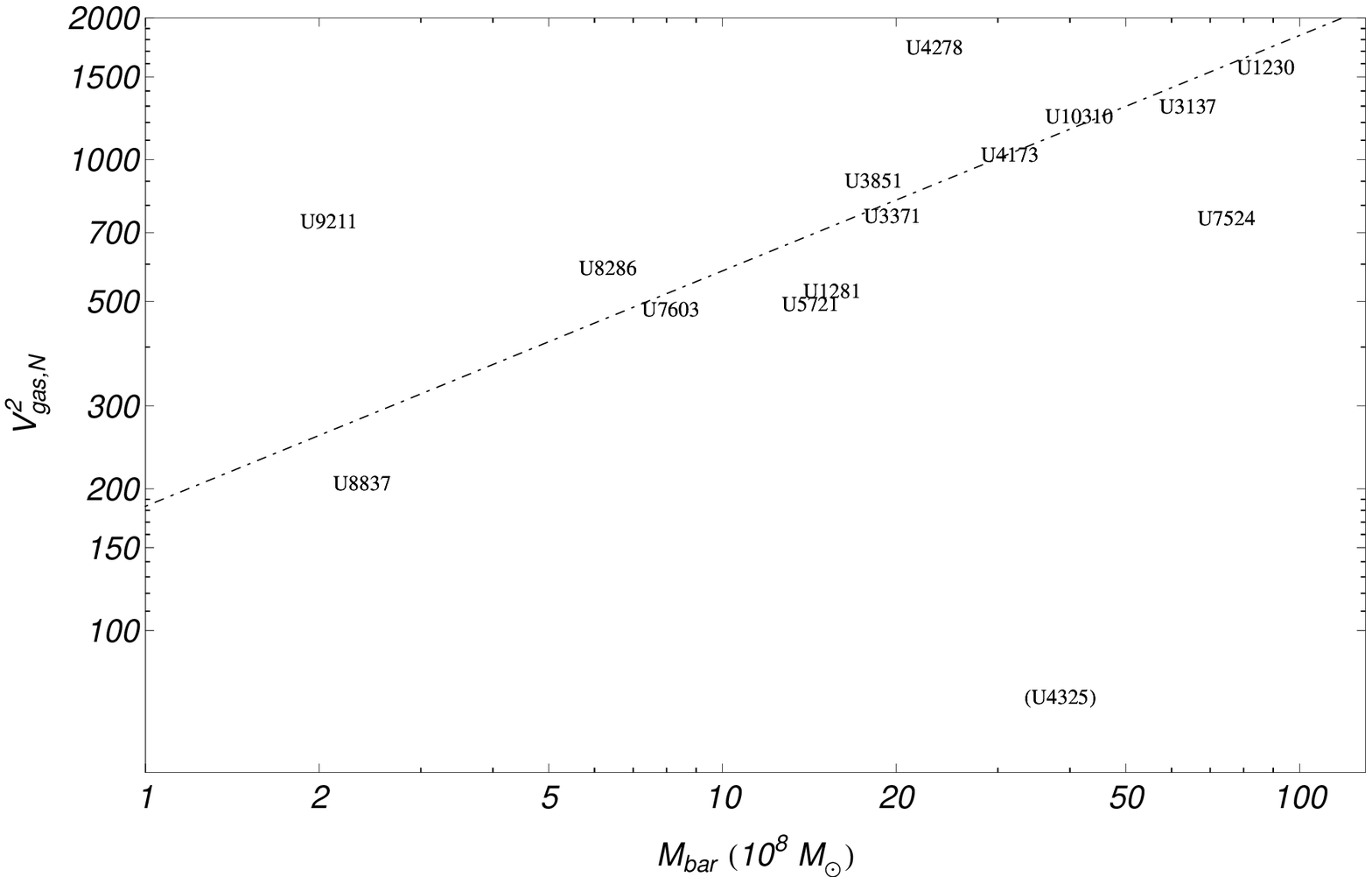}
  \includegraphics[width=89mm]{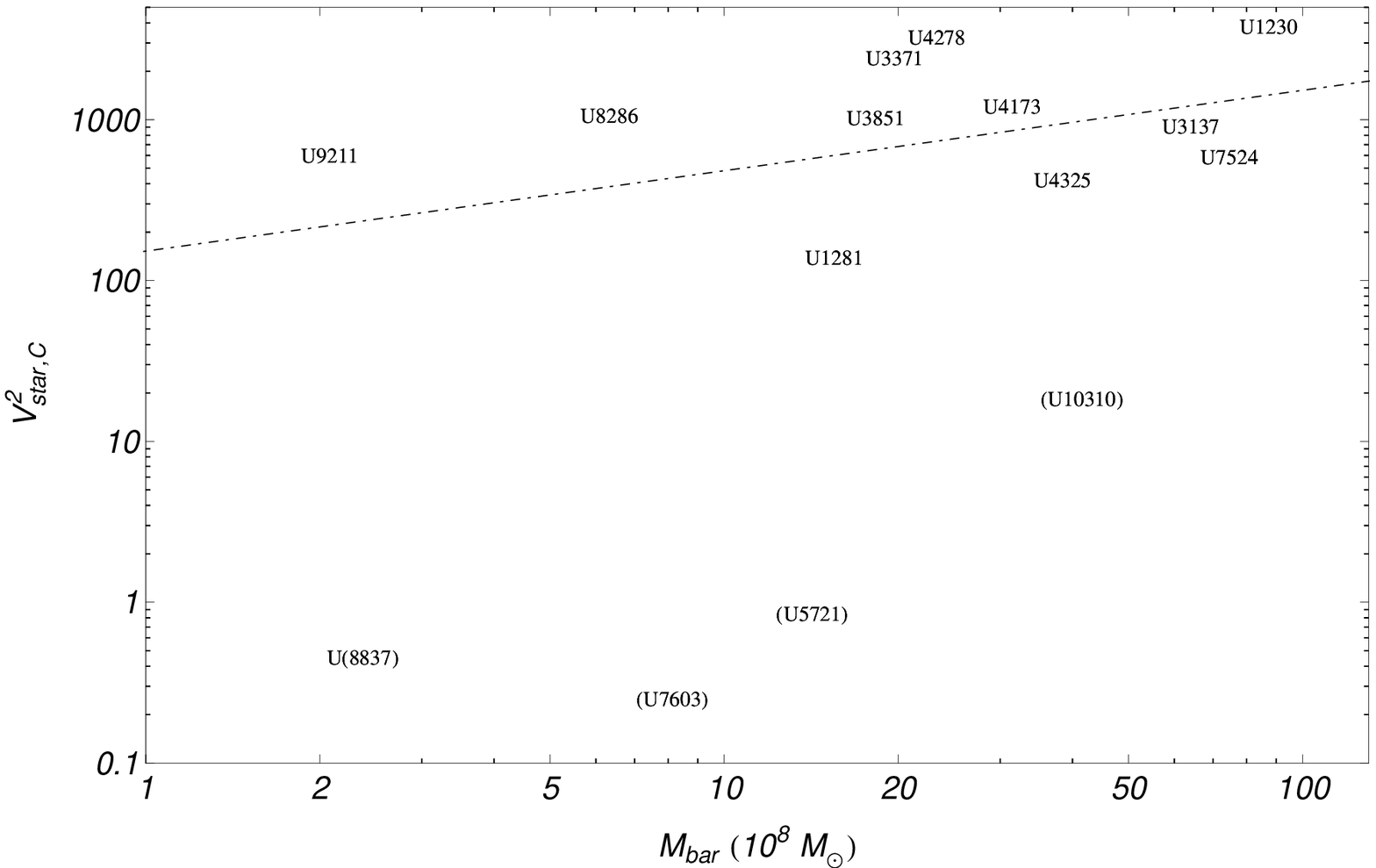}
  \includegraphics[width=89mm]{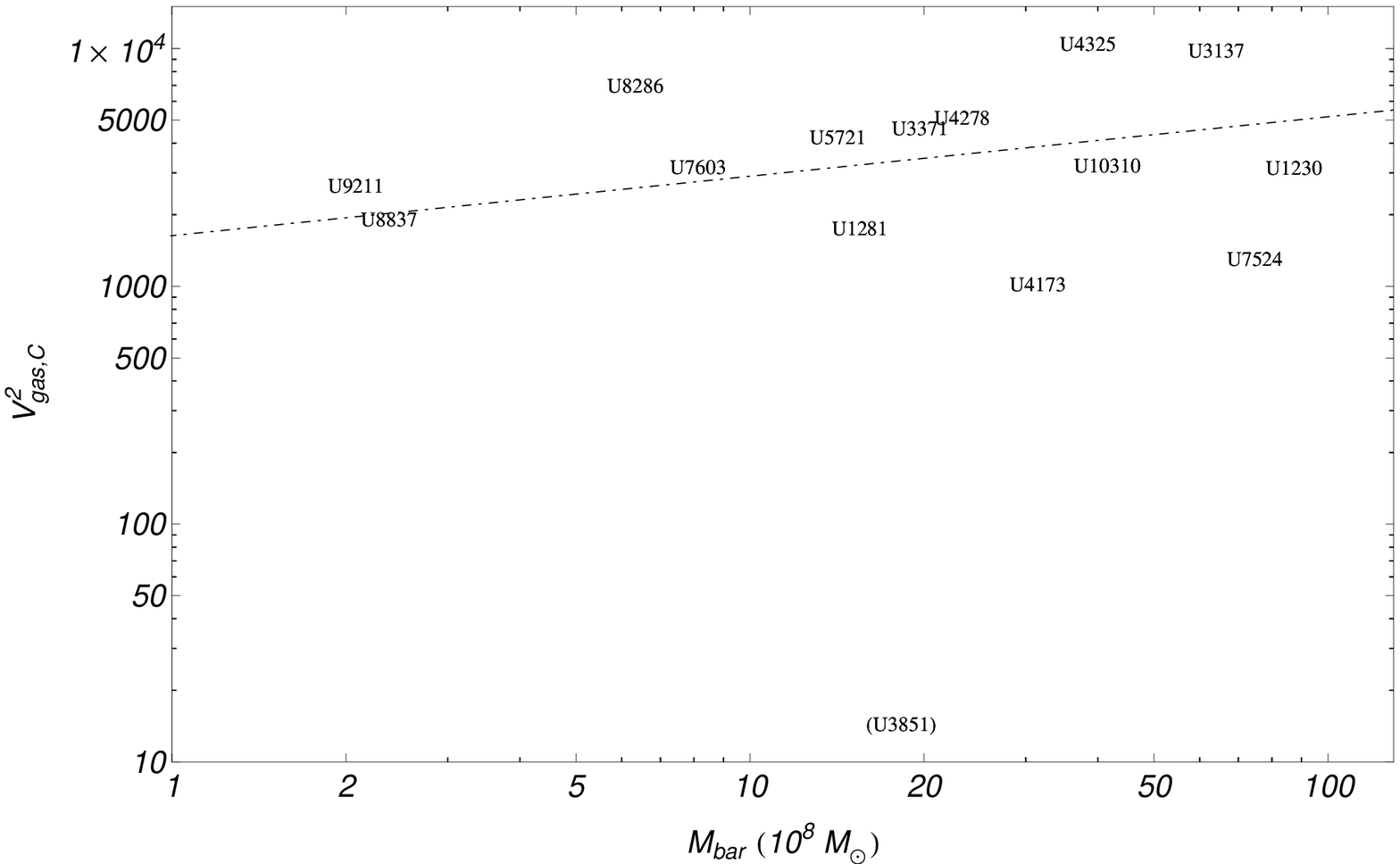}
  \caption{Baryonic Tully-Fisher relation for our galaxy sample: contributions from different velocity components.
  \label{fig:spiral_TF1}}
\end{figure*}
\begin{figure*}
\centering
  \includegraphics[width=90mm]{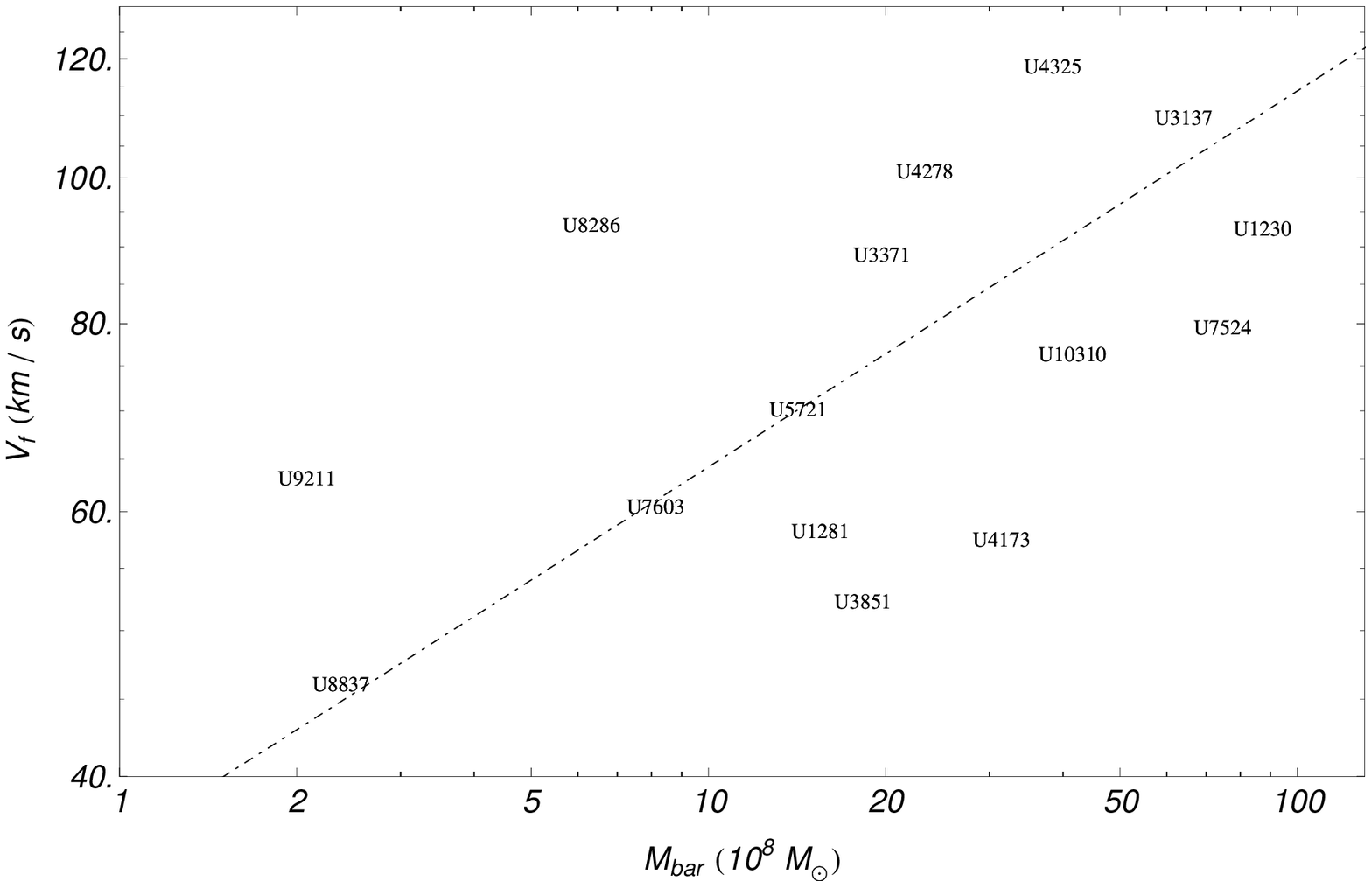}
  \caption{Baryonic Tully-Fisher relation for our galaxy sample: total rotational velocity.
  \label{fig:spiral_TF2}}
\end{figure*}

As we did in Paper I, we search for possible scaling relations, also for comparing our alternative approach with more popular theories, like MOND \citep{Swaters10}.

A correlation among the gas coupling constant (probably more related to the global gravitational structure properties than the star related parameter) and the extrapolated central disk surface brightness $\mu_{0,R}$ (reported in Table~3 of Paper I) is possible, see left panel of Fig.~\ref{fig:spiral_bMu0}:
\begin{equation}
\log \beta_{gas} = (-0.32 \pm 1.29) + (0.43 \pm 0.96) \cdot \log \mu_{0,R} \; ;
\end{equation}
as long as between the same quantity and the maximum rotational velocity, see left panel of Fig.~\ref{fig:spiral_bVmax}:
\begin{equation}
\log \beta_{gas} = (-0.30 \pm 0.28) + (0.30 \pm 0.14) \cdot \log V_{max} \; .
\end{equation}
Excluding only peculiar cases with respect of $\beta_{gas}$ (UGC3851, UGC4173, UGC4325, UGC5721) we can see that the dispersion around such relations is quite low. We also show in the left panel of Fig.~\ref{fig:spiral_bMgas} the relation between $\beta_{gas}$ and the total gas mass of each galaxy:
\begin{equation}
\log \beta_{gas} = (0.20 \pm 0.06) + (0.04 \pm 0.04) \cdot \log M_{gas} \; ;
\end{equation}
clearly the parameter is quite independent of such quantity, giving a further confirmation of our preliminary hypothesis $\mathrm{d}\beta/\mathrm{d}r \approx 0$ (see \S\ \ref{sec:scalar_theory}). In the left panel of Fig.~\ref{fig:spiral_bMbar}, we show the relation with the total baryonic (stars and gas) mass:
\begin{equation}
\log \beta_{gas} = (0.22 \pm 0.06) + (0.03 \pm 0.04) \cdot \log M_{bar} \; ;
\end{equation}
in this case, in the estimation of the stellar counterpart, we have to take into account the stellar mass-to-light ration, $Y_{\ast}$, which is actually one of the fit parameters. No change is detected when moving from the gas to the total baryonic.

This does not happen when considering the coupling constant with stars; as it is possible that a degeneracy between the coupling constant and the stellar mass-to-light ratio is working, we consider more properly the quantity $\beta^2_{star} \cdot Y_{\ast}$ which appears in the stellar rotational contribution, finally having (right panel of Fig.~\ref{fig:spiral_bMu0}):
\begin{equation}
\log \beta_{star}^2 \cdot Y_{\ast} = (-11.69 \pm 1.71) + (9.01 \pm 1.28) \cdot \log \mu_{0,R} \; ,
\end{equation}
and (right panel of Fig.~\ref{fig:spiral_bVmax})
\begin{equation}
\log \beta_{star}^2 \cdot Y_{\ast} = (1.45 \pm 2.44) + (-0.58 \pm 1.27) \cdot \log V_{max} \; ,
\end{equation}
which holds only for the galaxies with $\beta_{star} \gg 1$, while the others show a more scattered distribution.

The relation between $\beta^2_{star} \cdot Y_{\ast}$ and $M_{gas}$ or $M_{bar}$ are less evident than the previous case; taking a look to the right panels of Figs.~\ref{fig:spiral_bMgas}~-~\ref{fig:spiral_bMbar} we can only detect (more clearly in the $M_{bar}$ case) three sub-groups: two galaxies in the bottom-left corner ($UGC7603$ and $UGC8837$), corresponding to very low values for both $\beta_{star}$ and $Y_{\ast}$; four galaxies in the middle ($UGC1281$, $UGC4325$, $UGC5721$ and $UGC10310$), corresponding to low values of $\beta_{star}$; and an almost constant $\beta^2_{star} \cdot Y_{\ast}$ group made of the remaining objects.

On the other side, the length parameter, $L$, does not show a clear ordered pattern when compared with the same quantities as before.

An even more interesting relation to be tested here is the well-known Baryonic Tully-Fisher (BTF) relation \citep{McGaugh12}, which relates the total baryonic mass of spiral galaxies, $M_{bar}$, with the maximum observed velocity, $V_{max}$, that in the outer regions becomes approximately flat, $V_{f}$, and is actually quantified to be:
\begin{equation}
M_{bar} = A \cdot V_{f}^4 \; ,
\end{equation}
with $A = 47 \pm 6$ $M_{\odot}$ km$^{-4}$ $s^{4}$. In order to make the comparison with our results more clear, we can re-write the previous in the equivalent expression:
\begin{equation}
\log V_{f} = (1.582 \pm 0.003) + 0.25 \log M_{bar} \; .
\end{equation}
As it is possible to verify from previously plotted rotational curves, not all the galaxies in our sample seem to have reached the flatness regime in the rotational curve; so that the $V_{f}$ appearing in the previous formulas will be more exactly $V_{max}$ for us, i.e., the maximum velocity evaluated at the maximum distance from the center available from the data.

In our alternative scenario, the total velocity can be written as the sum of different terms:
\begin{equation}
V_{f,theo}^{2} = V_{N,star}^{2} + V_{N,gas}^{2} + V_{C,star}^{2} + V_{C,gas}^{2} \; ,
\end{equation}
where the suffix $N$ and $C$, as explained in previous sections, are related to the (extended) Newtonian and corrective terms in our gravitational potential model.

If we first consider the quantity $V_{f,theo}$ made up with $V_{N,gas}$, entirely derived from direct velocity observations; $V_{N,star}$, which mixes direct velocity observations and a fit parameter, $Y_{\ast}$ (which also enters in the calculation of $M_{bar}$); and $V_{C,star}^{2}$ and $V_{C,gas}^{2}$ which completely depend on our theoretical model, we obtain:
\begin{equation}
\log V_{f} = (1.558 \pm 0.030) + 0.25 \log M_{bar} \; ,
\end{equation}
which agree very well with the expected BTF relation. We can also check that the BTF relation is verified for any of the components of the total velocity (of course, we will consider the maximum velocity for any of these). For the pseudo-newtonian-term-derived stellar velocity we can identify two groups: galaxies with $Y_{\ast}>2$ give
\begin{equation}
\log V_{N,star} = (0.547 \pm 0.067) + 0.25 \log M_{bar} \; ,
\end{equation}
and galaxies with $Y_{\ast}<2$ give
\begin{equation}
\log V_{N,star} = (1.232 \pm 0.043) + 0.25 \log M_{bar} \; .
\end{equation}
The pseudo-newtonian-term-derived gas velocity (excluding $UGC4325$) gives:
\begin{equation}
\log V_{N,gas} = (1.132 \pm 0.026) + 0.25 \log M_{bar} \; .
\end{equation}
The corrective-term-derived gas velocity (excluding $UGC5721$, $UGC7603$, $UGC8837$ and $UGC10310$) gives:
\begin{equation}
\log V_{C,gas} = (1.092 \pm 0.064) + 0.25 \log M_{bar} \; .
\end{equation}
The corrective-term-derived stellar velocity (excluding $UGC3851$) gives:
\begin{equation}
\log V_{C,star} = (1.448 \pm 0.051) + 0.25 \log M_{bar} \; .
\end{equation}
All the previous relations are shown in Figs.~\ref{fig:spiral_TF1}~-~\ref{fig:spiral_TF2}.
It is interesting to note that, for these last two quantities, if we relax the condition for which the coefficient of $M_{bar}$ has to be equal to $0.25$, and we leave it free, we obtain, respectively, the values $0.037$ and $0.033$, which are quite consistent with a scenario where the contribution to the rotation curve coming from the effective mass produced by the modified gravity of the chosen scalar field is practically independent of the baryon mass (at least, for the family of spiral galaxies considered), so arguing in favor of a more universal task for our mechanism.

\subsection{Unified picture}

In this section we will qualitatively draw some conclusions from the results that we have derived from the different gravitational structures discussed above.
In particular we refer to the new evidence that a scalar field can mimic dark matter at different scales by changing its properties depending either on the local physical conditions, namely the matter density, or the matter status.

As done in Paper I, we consider the relation among the scalar field parameters and the gas density/mass of any structure in order to check the presence of a global correlation among parameters at all scales, from galaxies to galaxy clusters.

For galaxy clusters this is natural to check because the gas is the main contribution to the mass (under
the assumption that there is no dark matter, as in the current work). The same argument it is not intuitive
for galaxies, as gas is not the main component at galactic scale: however as we have found that the  coupling constant of the gas with the scalar field can be larger than the one of stars, this component turns out to have a significant impact on the observational quantities too, sometimes even larger than the stellar one.

\begin{figure*}
\centering
  \includegraphics[width=89mm]{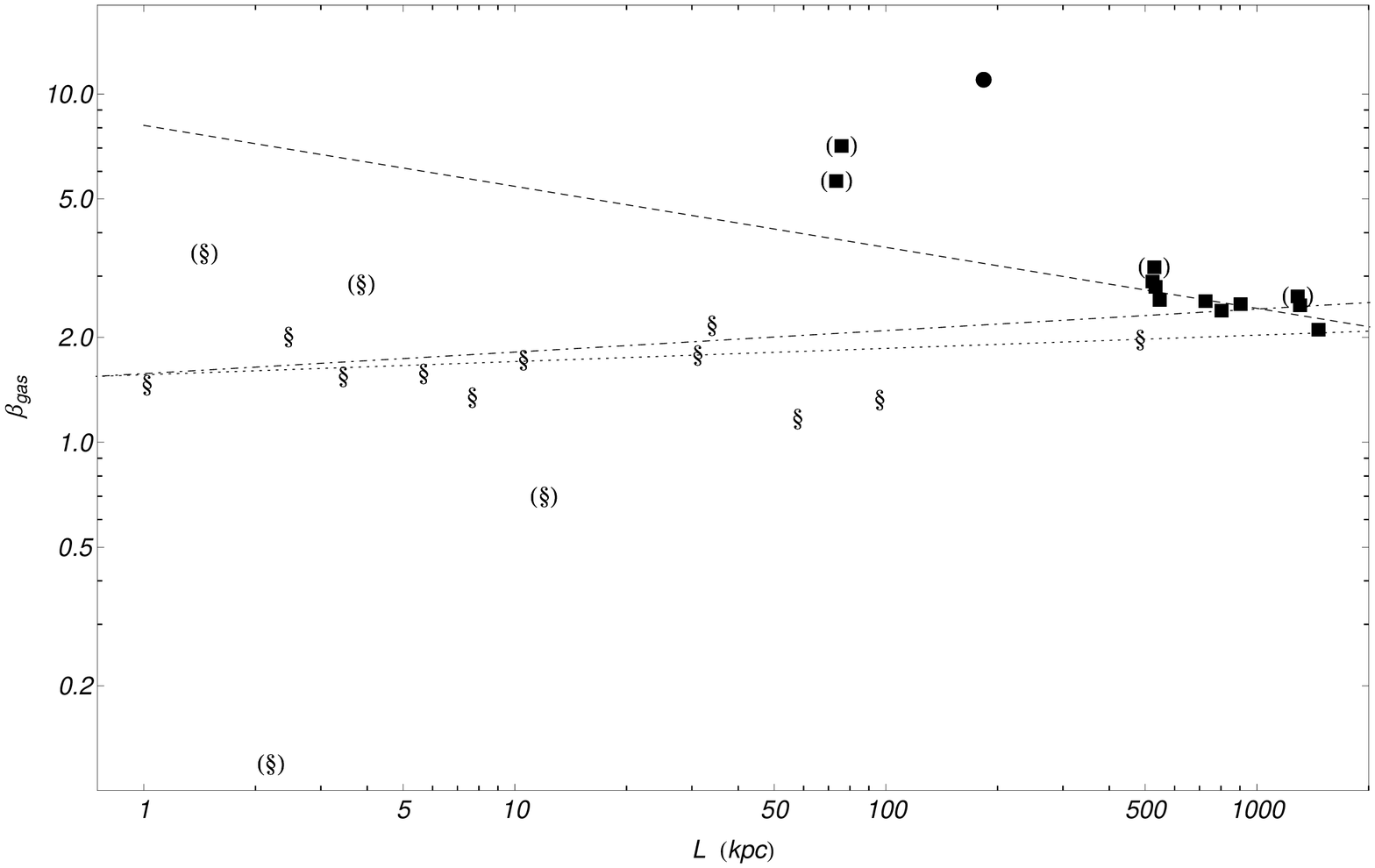}
  \includegraphics[width=89mm]{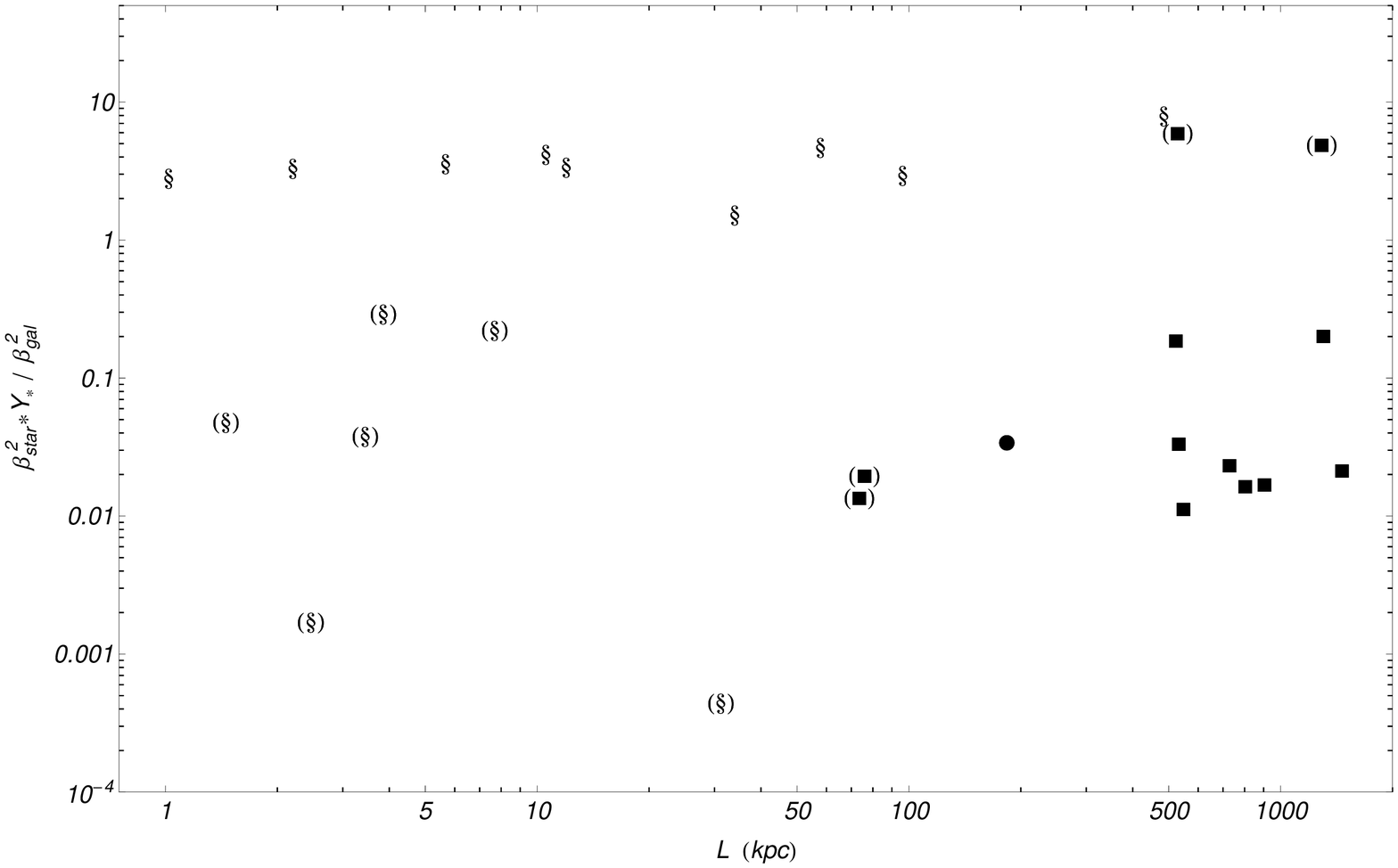}
  \caption{Correlation among the scalar field parameters: spirals are LSB spiral galaxies; filled squares are clusters of galaxies; filled circle is the elliptical galaxy NGC4374. Dashed line is the best fit for clusters only; dotted line is the best fit for spiral galaxies only; dot-dashed line is the best fit for the total sample. Objects in brackets are the peculiar cases described in the text and not considered in the fits.
  \label{fig:cluster_spiral_bL}}
\end{figure*}

Let us start by verifying how the scalar field parameters are correlated with each other. One of our main hypothesis is that the coupling constants $\beta$ are independent of the scale or, more precisely, that $d\beta/dr \sim 0$. We can verify a posteriori that this hypothesis is quite well satisfied by taking a look to the top pane of Fig.~\ref{fig:cluster_spiral_bL}. Here a linear regression of the spiral galaxy parameters gives:
\begin{equation}
\log \beta_{gas} = (0.19 \pm 0.05) + (0.04 \pm 0.02) \cdot \log L \; ;
\end{equation}
while for clusters of galaxies we obtain:
\begin{equation}
\log \beta_{gas} = (0.91 \pm 0.18) - (0.17\pm 0.06) \cdot \log L \; ;
\end{equation}
and using all data together we finally get:
\begin{equation}
\log \beta_{gas} = (0.20 \pm 0.05) + (0.06\pm 0.02) \cdot \log L \; .
\end{equation}
We can verify how coherently each gravitational structure family locate in the parameter space:
the clusters of galaxies lay in a very well constrained region on the right side of the left panel in
Fig.~\ref{fig:cluster_spiral_bL}; spiral galaxies are on the opposite side with a larger dispersion
mainly  due to the uncertainties on the interaction length.
The only elliptical galaxy that we have analysed is interestingly located right in between the spirals and the galaxy cluster regions very close to the area where the smallest clusters and/or group of galaxies are located.
This evidence seems to suggest that the properties of the scalar field are related to the dynamical
properties of the gravitational systems, as cold dynamical systems like spirals are clearly separated
by hot dynamical systems (elliptical galaxies and galaxy clusters). This result is interesting as the
connection between dynamics (especially anisotropy of the orbits) and scalar field parameters
has been found also in other $f(r)$ formulations \citep{Napolitano12}.

For the coupling constant related to stars (in galaxies) and galaxies (in clusters), in the right panel of Fig.~\ref{fig:cluster_spiral_bL} we can see a larger dispersion in the parameter space, but also a clear segregation: the left half of the diagram is dominated by spiral galaxies, the right half by clusters of galaxies, and the central region populated by elliptical galaxies (although a larger sample is required) and smaller clusters and/or group of galaxies.

\begin{figure*}
\centering
  \includegraphics[width=89mm]{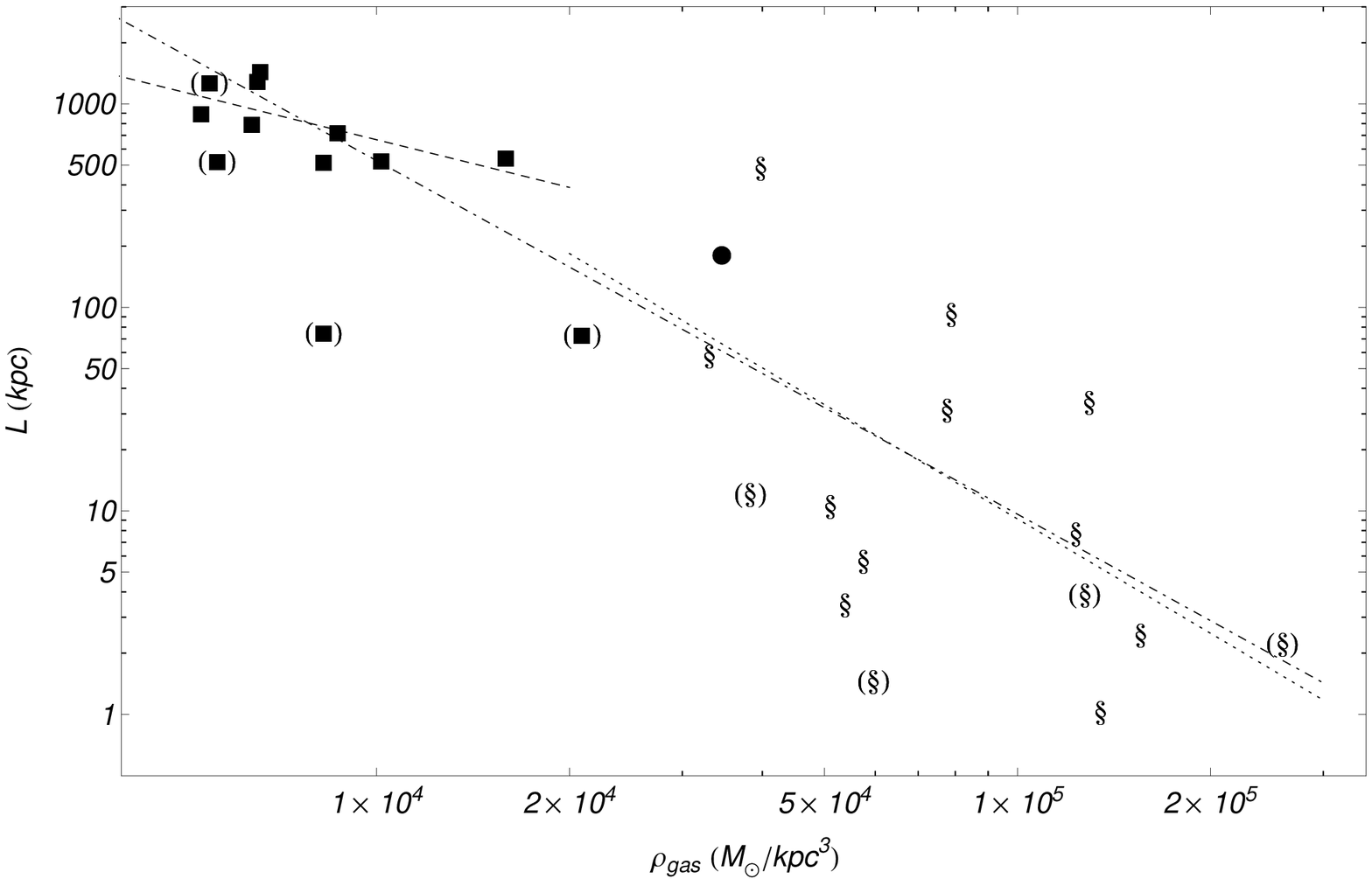}
  \includegraphics[width=89mm]{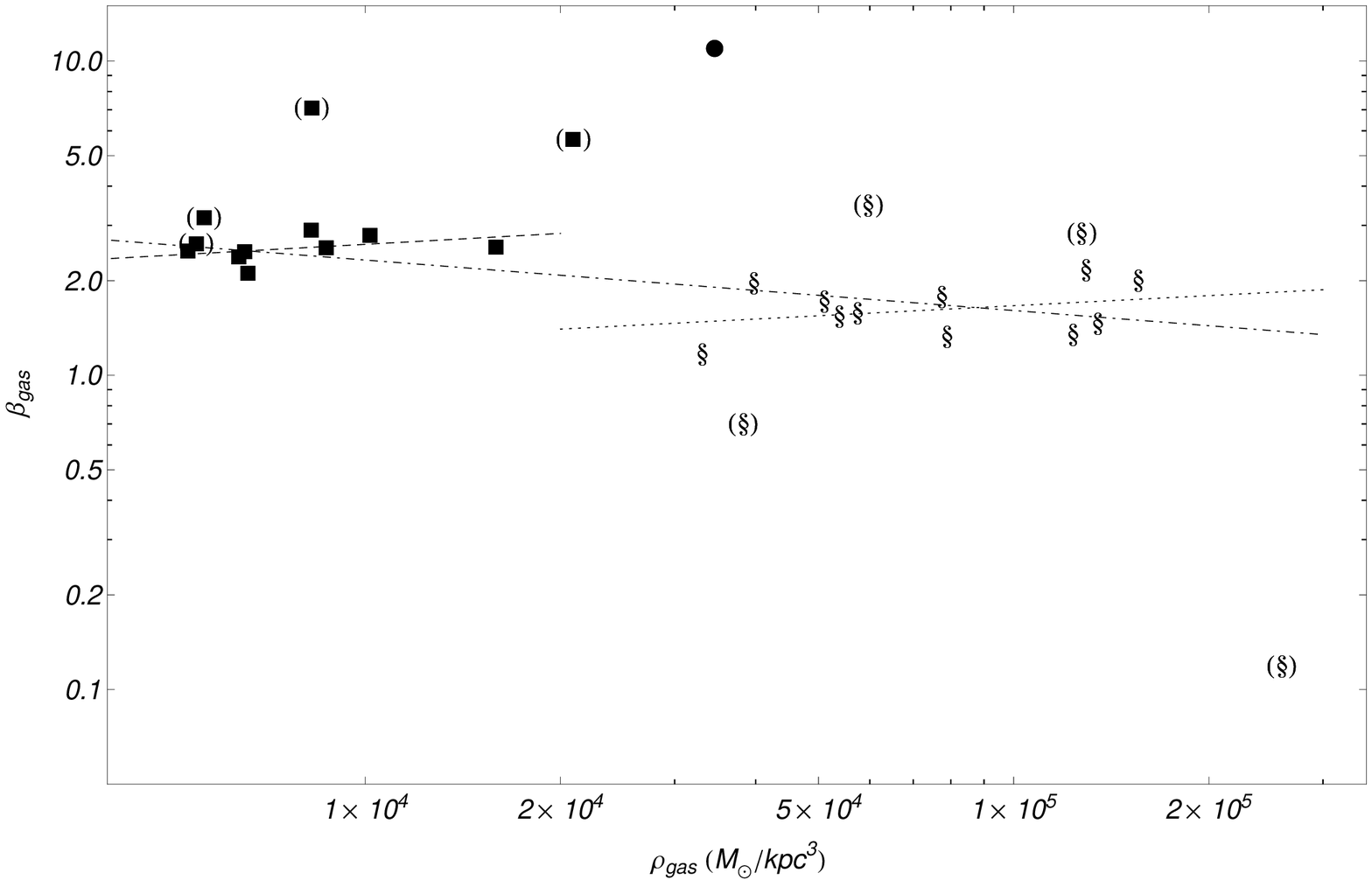}
  \includegraphics[width=90mm]{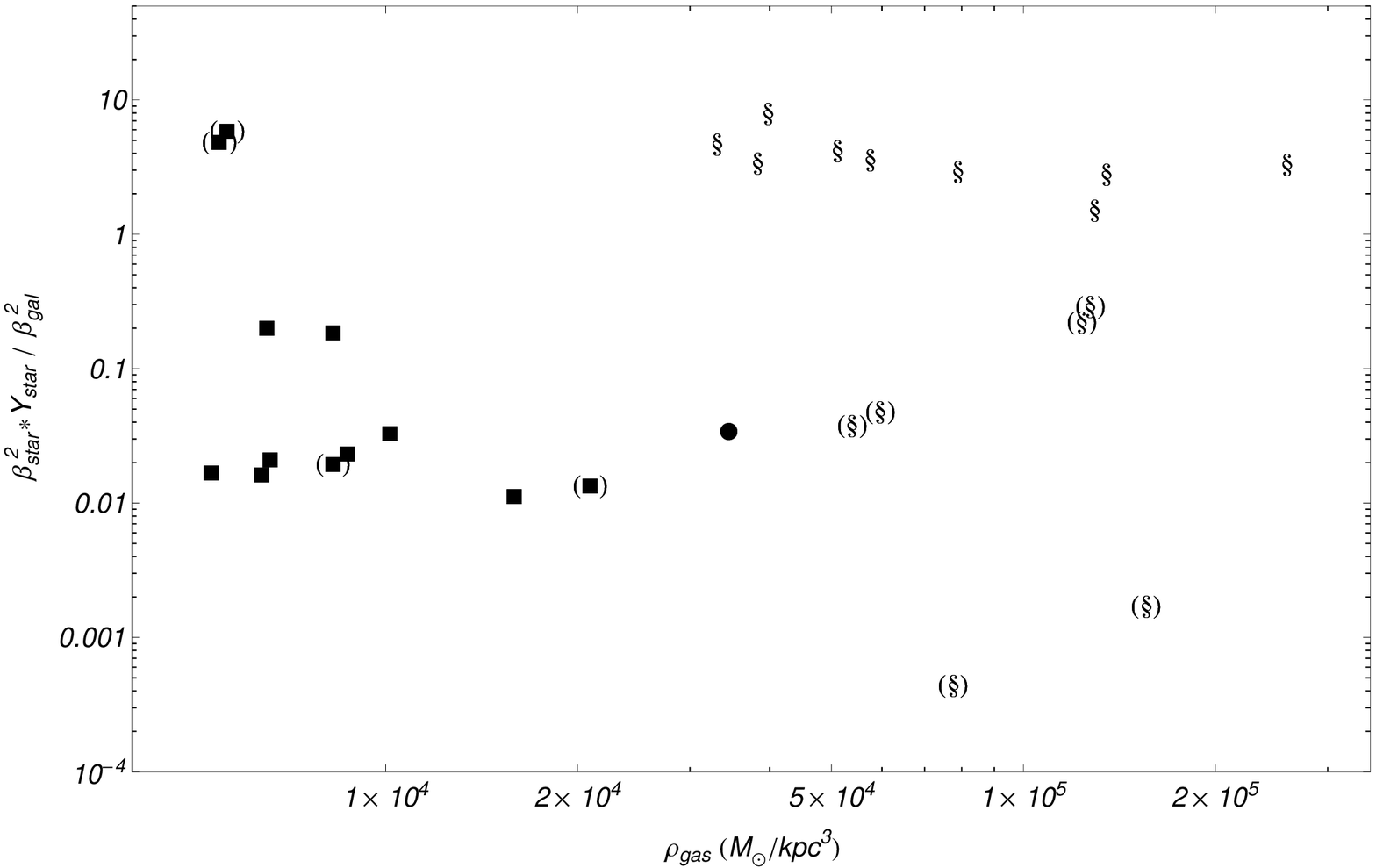}
  \caption{Correlation among the scalar field parameters and the gas density: spirals are LSB spiral galaxies; filled squares are clusters of galaxies; filled circle is the elliptical galaxy NGC4374. Dashed line is the best fit for clusters only; dotted line is the best fit for spiral galaxies only; dot-dashed line is the best fit for the total sample. Objects in brackets are the peculiar cases described in the text and not considered in the fits.
  \label{fig:cluster_spiral_density}}
\end{figure*}

We can eventually reach the same conclusion looking at the correlation between scalar field parameters and the gas density as in Fig.~\ref{fig:cluster_spiral_density}: when considering the top left and right panels of this figure, we can see that the parameters do not distribute randomly, but seem to follow a trend. For the interaction length, the top left corner is populated by clusters of galaxies, the bottom right by spiral galaxies, and the center possibly by elliptical galaxies and group of galaxies, with a clear linear trend and a dispersion which is larger for spiral galaxies than for clusters.
A linear regression in the log-log space where the interaction length $L$ is the dependent variable give, for spiral galaxies:
\begin{equation}
\log L = (10.29 \pm 4.68) - (1.87\pm 0.96) \cdot \log \rho_{gas} \; ;
\end{equation}
for clusters of galaxies:
\begin{equation}
\log L = (5.94 \pm 1.37) - (0.78\pm 0.35) \cdot \log \rho_{gas} \; ;
\end{equation}
and for all systems together:
\begin{equation}
\log L = (9.67 \pm 1.04) - (1.74\pm 0.23) \cdot \log \rho_{gas} \; .
\end{equation}
For the gas coupling constant, on the other side, we have a linear trend almost compatible with a constant $\beta_{gas}$, thus confirming our previous hypothesis ($\mathrm{d}\beta/\mathrm{d}r \approx 0$); in particular for spiral galaxies we have:
\begin{equation}
\log \beta_{gas} = (-0.31 \pm 0.57) + (0.11\pm 0.12) \cdot \log \rho_{gas} \; ;
\end{equation}
for clusters of galaxies we have:
\begin{equation}
\log \beta_{gas} = (-0.04 \pm 0.41) - (0.11\pm 0.10) \cdot \log \rho_{gas} \; ;
\end{equation}
end for the total sample:
\begin{equation}
\log \beta_{gas} = (1.01 \pm 0.18) - (0.16\pm 0.04) \cdot \log \rho_{gas} \; .
\end{equation}

When considering the combination of $\beta_{star}$ and $Y_{\ast}$ or $\beta_{gal}$, the large spread does not allow to define a clear trend and no data regression is produced.

\section{Conclusions and Discussion}
\label{sec:conclusions}

In this work we have investigated the dynamical properties of several astrophysical systems, from galaxies to clusters of galaxies, within the theoretical framework of scalar-tensor theories as alternative to the dark matter paradigm, and looked for observational signatures in galaxy kinematics and X--ray equilibrium.

The crucial ingredients of our work are some novel properties of the scalar field with respect to previous analyses (Paper I): (i) a new parametrization of the theory where we have introduced the coupling of the scalar field with the different baryonic components of the gravitational systems (stars and gas) and (ii) its mass (or interaction length) can vary with scale.

This new parametrization with basically two coupling constants, allowed the scalar field to mimic dark matter, i.e. reproduce remarkably well
observations without introducing a new kind of matter and by a modification of gravity.

Going in more details of the results in the different astrophysical tests we have performed here, results have shown that (i) the velocity dispersion of elliptical galaxies can be fit very well by a scalar field, even better than assuming a classical NFW profile for the dark matter component; (ii) a
scalar field can fairly well reproduce the matter profile in clusters of galaxies, estimated by X-ray observations and without the need for any additional dark matter; (iii) good fits to the rotation curves of low surface brightness galaxies are obtained.

All these results show that the scalar field gravity theories with the peculiar properties we have assumed, can be compatible with a wide range of astrophysical tests and can be considered as a viable alternative to dark matter. In particular, our results show the possibility that the scalar field can couple with ordinary matter with different strengths (different coupling constants) depending on the clustering state of matter components in the considered gravitational structures and that a possible correlation with the evolutionary state of gravitational systems is possible. These results are extremely important when compared to the theoretical scenario they are inspired by, namely the chameleon or the symmetron theory: they have been considered so far only at cosmological scales, as possible explanations for dark energy; we have shown in this work that their properties also make them suitable to explain astrophysical phenomena at smaller scales, thus mimicking dark matter.

We point out that this work is part of a larger project where we want to test our model and parametrization by using other different cosmological and astrophysical probes and whose main goal is to strengthen the idea that such a model can work as a consistent unifying scenario for all the dark sectors.

A challenging but decisive test will be the treatment of the lensing theory in the Newtonian and post-Newtonian formalism. A comprehensive and formally corrected approach, from the theoretical point of view, is needed in order to fully test the theory against lensing observables. In fact, by simply adopting the Newtonian limit and the  standard of  general relativity, it is easily  shown that  the same outcome of  Einstein theory is obtained  \citep{lubini}. This is straightforward  because the Newtonian limit recovered from general relativity does not take into account the further degrees of freedom of alternative gravities.  On the other hand, considering that alternative gravities usually  contain further degrees of freedom with respect to general relativity, a more accurate treatment has to be adopted for lensing problem. As shown in \citep{stabile}, differences emerge as soon as as the Newtonian and the post-Newtonian formalisms are fully considered assuming also the perturbation theory of further gravitational degrees of freedom. In this perspective, differences between alternative gravity and dark matter pictures could emerge giving the signature for the final theory. Due to these reasons we have searched and collected from literature data regarding the mass profiles of clusters of galaxies, reconstructed by both strong and weak lensing and we will use them in the next future.

Moreover, on the astrophysical side, we will also explore the possible correlations between classical models (NFW or generalised NFW profiles) for dark matter and the effective contribution from our scalar field; concerning spiral galaxies, it will be interesting to verify if and how our model is able to reproduce the Universal Rotation Curve scenario \citep{Persic97,Salucci11}.

After this preliminary analysis of the static properties of gravitational structures, we will go deeper in the study of the nature of a scalar field with the properties we have considered in this work, studying its influence on the formation and the evolution of the same structures: we will explore the non-linear effects in clustering processes; and, on the cosmological side, we will study the feasibility of this model with the cosmic microwave background radiation data and matter power spectrum features.

\section*{Acknowledgments}

D.F.M. thanks the Research Council of Norway FRINAT grant 197251/V30. D.F.M. is also partially supported by project CERN/FP/123618/2011 and CERN/FP/123615/2011. V. S. is now working at UPV/EHU under the project ``Gravitaci$\mathrm{\acute{o}}$n, Cosmolog$\mathrm{\acute{i}}$a Relativista y otros aspectos de la Estructura del Universo''. S. C. acknowledges the support of INFN (Sez. di Napoli, iniziative specifiche NA12 and OG51) and the ERASMUS/SOCRATES European program. V.S. thanks L. Coccato for giving important suggestions in managing elliptical galaxies data, Y. Fukazawa for providing data about X-ray emission of NGC4374 and P. Brownstein for giving interesting comments and suggestions.


\begin{thebibliography}{}

\bibitem[Acquaviva et al.(2005)]{Acquaviva05} Acquaviva, V., Baccigalupi, C., Leach, S.M., Liddle, A.R., Perrotta, F., 2005, Phys. Rev. D, 71, 104025

\bibitem[Ade, Aghanim, Armitage-Caplan et al.(2013)]{planck} Ade, P.A.R., Aghanim, N., Armitage-Caplan, C., et al., 2013, arXiv:1303.5076

\bibitem[Bekenstein(2004)]{Bekenstein04} Bekenstein, J.~D., 2004,
Phys. Rev. D, 70, 083509
\bibitem[Bekenstein(2005)]{Bekenstein05} Bekenstein, J.~D., 2005,
Phys. Rev. D, 71, 069901
\bibitem[Bergond et al.(2006)]{Bergond06} Bergond, G., Zepf, S.E.,
Romanowsky, A.J., Sharples, R.M., Rhode, K.L., 2006, A\&A, 448, 155
\bibitem[Bertone et al.(2005)]{Bertone05} Bertone, G., Hooper, D.,
Silk, J., 2005, Phys. Rept., 405, 279
\bibitem[Bertschinger \& Zukin(2008)]{Bertschinger08} Bertschinger,
E., Zukin, P., 2008, Phys. Rev. D, 78, 024015
\bibitem[Binetruy(2006)]{Binetruy} Binetruy P., 2006, Supersymmetry:
Theory, experiment and cosmology, {\it Oxford, UK: Oxford Univ. Pr.
520 p}
\bibitem[Brax et al.(2004)]{Brax04} Brax, P., van de Bruck, C., Davis,
A.-C., Khoury, J., Weltman, A., 2004, Physical Review D, 70, 123518
\bibitem[Brax et al.(2012)]{Brax12} Brax. P., et al., 2012, accepted to JCAP
\bibitem[Bruneton \& Esposito-Farese(2007)]{Bruneton07} Bruneton, J.~P., Esposito-Farese, G., 2007, Phys. Rev. D, 76, 124012
\bibitem[Bryan \& Norman(1998)]{Bryan98} Bryan, G.L., ApJ, 495, 80
\bibitem[Buote \& Humphrey(2011)]{Buote11} Buote, D.A., Humphrey,
P.J., 2011, arXiv:1104.0012

\bibitem[Capaccioli et al.(1992)]{Capaccioli92} Capaccioli, M., Caon,
N., D'Onofrio, M., 1992, \mnras, 259, 323
\bibitem[Capozziello(2002)]{Cap02} Capozziello, S., 2002,
Int. Jou. Mod. Phys., D 11, 483
\bibitem[Capozziello et al.(2003)]{fr3} Capozziello, S., Carloni, S.,
Troisi, A., 2003, Recent Res.\ Dev.\ Astron.\ Astrophys.\ {\bf 1}, 625
\bibitem[Capozziello \& Tsujikawa (2008)]{tsuji} Capozziello, S. and
Tsujikawa, S., 2008, Phys. Rev. D 77, 107501
\bibitem[Capozziello \& Francaviglia(2008)]{CapFra} Capozziello,
S. and Francaviglia, M., 2008, Gen. Rel. Grav., 40, 357
\bibitem[Capozziello \& De Laurentis(2011)]{mfrev} Capozziello, S. and
De Laurentis, M., 2011, Phys. Repts., 509, 167
\bibitem[Cappellari et al.(2006)]{Cappellari06} Cappellari, M., et
al., 2006, MNRAS, 366, 1126
\bibitem[Carroll et al.(1992)]{CarLam} Carroll, S.M., Press, W.H.,
Turner, E.L. 1992, ARA\&A, 30, 499
\bibitem[Cavaliere \& Fusco-Femiano(1978)]{Cavaliere78} Cavaliere, A.,
Fusco-Femiano, R., 1978, A\&A, 70, 677
\bibitem[Clifton et al.(2005)]{Clifton05} Clifton, T., Mota, D.,
Barrow, J., 2005, Mon. Not. R. Astron. Soc., 358, 601
\bibitem[Coccato et al.(2009)]{Coccato09} Coccato, L., Gerhard, O.,
Arnaboldi, M., Das, P., Douglas, N.G., Kuijken, K., Merrifield, M.R.,
Napolitano, N.R., Noordermeer, E., Romanowsky, A.J., Capaccioli, M.,
Cortesi, A., De Lorenzi, F., Freeman, K.C., 2009, MNRAS, 394, 1249
\bibitem[Copeland et al.(2006)]{Copeland06} Copeland E.J., Sami M.,
Tsujikawa S., 2006, Int. J. Mod. Phys. D 15, 1753

\bibitem[Davis et al.(2011)]{sym3} Davis et al., arXiv:1108.3081
[astro-ph.CO]
\bibitem[Dekel et al.(2005)]{Dekel05} Dekel, A., Stoehr, F., Mamon,
G.A., Cox, T.J., Novak, G.S., Primack, J.R., 2005, Nature, 437, 707
\bibitem[de Blok \& Bosma(2002)]{deBlok02} de Blok, W.J.G., Bosma, A.,
2002, A\&A, 385, 816
\bibitem[De Lorenzi et al.(2008)]{DeLorenzi08} De Lorenzi, F.,
Gerhard, O., Saglia, R.P., Sambhus, N., Debattista, V.P., Pannella,
M., M$\mathrm{\acute{e}}$ndez, R.H., 2008, MNRAS, 385, 1729
\bibitem[Diehl \& Statler(2007)]{Diehl07} Diehl, S., Statler, T.S.,
2007, ApJ, 668, 150
\bibitem[Douglas et al.(2002)]{Douglas02} Douglas, N.G., Arnaboldi,
M., Freeman, K.C., Kuijken, K., Merrifield, M.R., Romanowsky, A.J.,
Taylor, K., Capaccioli, M., Axelrod, T., Gilmozzi, R., Hart, J.,
Bloxham, G., Jones, D., 2002, PASP, 114, 1234
\bibitem[Douglas et al.(2007)]{Douglas07} Douglas, N.G., 2007, ApJ,
664, 257
\bibitem[Dunkley et al.(2005)]{Dunkley05} Dunkley, J., Bucher, M.,
Ferreira, P.~G., Moodley, K., Skordis, C., 2005, MNRAS, 356, 925

\bibitem[Esposito-Farese \& Polarski(2001)]{Esposito01}
Esposito-Farese, G., Polarski, D., Physical Review D, 63, 2001, 063504
\bibitem[Evrard et al.(1996)]{Evrard96} Evrard, A.E., Metzler, C.A.,
Navarro, J.F., 1996, ApJ, 469-494

\bibitem[Fukazawa et al.(2006)]{Fukazawa06} Fukazawa, Y.,
Betoya-Nonesa, J.G., Pu, J., Ohto, A., Kawano, N., 2006, ApJ, 636,
698

\bibitem[Gannouji et al.(2009)]{Gannouji09} Gannouji, R., Moraes, B.,
Polarski, D., arXiv:0907.0393
\bibitem[Gultekin et al.(2009)]{Gultekin09} Gultekin, K., Richstone,
D.O., Gebhardt, K., Lauer, T.R., Tremaine, S., Aller, M.C., Bender,
R., Dressler, A., Faber, S.M., Filippenko, A.I., Green., R., Ho, L.C.,
Kormendy, J., Magorrian, J., Pinkney, J., and Siopis, C., 2009, ApJ,
698, 198

\bibitem[Hinterbichler \& Khoury(2010)]{sym1} Hinterbichler, K.,
Khoury, J. 2010, Physical Review Letters, 104, 231301
\bibitem[Hu \& Sawicki(2007)]{fr1} W.~Hu, I.~Sawicki, Phys.\ Rev.\
{\bf D76}, 064004 (2007).
\bibitem[Humphrey et al.(2006)]{Humphrey06} Humphrey, P.J., Buote,
D.A., Gastaldello, F., Zappacosta, L., Bullock, J.S., Brighenti, F.,
Mathews, W.G., 2006, ApJ, 646, 899

\bibitem[Jing \& Suto(2000)]{JingSuto} Jing, Y.P., Suto, Y., 2000,
ApJ, 529, L69

\bibitem[Khoury \& Weltman(2004)]{Khoury04} Khoury, J., Weltman, A.,
2004, Physical Review D, 69, 044026
\bibitem[Klypin \& Prada(2009)]{Klypin09} Klypin, A., Prada, F., 2009,
ApJ, 690, 1488
\bibitem[Koivisto \& Mota (2006)]{dfm3} Koivisto T., Mota D. F.,
Phys.\ Rev.\ 2006, D73, 083502.
\bibitem[Koivisto, Mota \& Pitrou (2009)]{dfm5} Koivisto, T. S., Mota
D. F., Pitrou C., 2009, JHEP 0909, 092.
\bibitem[Kormendy et al.(2009)]{Kormendy09} Kormendy, J., Fisher,
D.B., Cornell, M.E., Bender, R., 2009, ApJs, 182, 216
\bibitem[Krauss(2006)]{Krauss06} Krauss, L.M., 2006, Invited Review
talk, Neutrino 2006, to appear in Proceedings
[arxiv.org/abs/hep-ph/0702051]
\bibitem[Kroupa(2001)]{Kroupa01} Kroupa, P., 2001, MNRAS, 322, 231

\bibitem[Li, Barrow \& Mota(2007)]{dfm4} Li B., Barrow J.D., Mota
D. F. , 2007, Phys.\ Rev.\ D76, 104047
\bibitem[Lima Neto, Gerbal \& Marquez(1999)]{LimaNeto99} Lima Neto,
G.B., Gerbal, D., M$\mathrm{\acute{a}}$rquez, I., 1999, MNRAS, 309,
481
\bibitem[Linde(2008)]{Linde} Linde A.D., 2008 , Lect.\ Notes Phys.\
738, 1
\bibitem[Lubini et al. (2011)] {lubini} 	
 Lubini M.,  Tortora C.,  Naef J.,  Jetzer Ph.,  Capozziello S., 2011.
  Eur.Phys.J. C71, 1834.
\bibitem[McGaugh(2012)]{McGaugh12} S.~S. McGaugh, 2012, Astronomical Journal, 143, 40
\bibitem[Mamon \& Lokas(2005A)]{MamonLokas1} Mamon, G.A., Lokas, E.L.,
2005A, MNRAS, 362, 95
2005A, MNRAS, 362, 95
\bibitem[Mamon \& Lokas(2005B)]{MamonLokas2} Mamon, G.A., Lokas, E.L.,
2005B, MNRAS, 363, 705
\bibitem[Mamon \& Lokas(2006)]{MamonLokas3} Mamon, G.A., Lokas, E.L.,
2006, MNRAS, 370, 1581
\bibitem[Manera \& Mota (2006)]{dfm1} Manera M. \& Mota D., 2006,
MNRAS 371, 1373
\bibitem[Mathews \& Brighenti (2003)]{Mathews03} Mathews, W.G.,
Brighenti, F., 2003, ARA\&A, 41, 191
\bibitem[Mendez et al.(2001)]{Mendez01}
M$\mathrm{\acute{e}}$ndez, R.H., Riffeser, A., Kudritzki, R.-P.,
Matthias, M., Freeman, K.C., Arnaboldi, M., Capaccioli, M., Gerhard,
O.E., 2001, ApJ, 563, 135
\bibitem[Milgrom(1983)]{Milgrom83} Milgrom, M., 1983, ApJ, 270, 365
\bibitem[Mota \& Shaw(2006)]{shaw} Mota, D. F., Shaw, D. J. 2006,
Phys.Rev.Lett., 97, 151102
\bibitem[Mota (2008)]{dfm2} Mota D. F., 2008, JCAP 09006
\bibitem[Mota et al.(2011)]{MotaSalzano} Mota, D.F., Salzano, V.,
Capozziello, S., 2011, Phys. Rev. D, 83, 084038 (Paper I)

\bibitem[Napolitano et al.(2001)]{Napolitano01} Napolitano, N.R.,
Arnaboldi, M., Freeman, K.C., Capaccioli, M., 2001, A\&A, 377, 784
\bibitem[Napolitano et al.(2002)]{Napolitano02} Napolitano, N.R.,
Arnaboldi, M., Capaccioli, M., 2002, A\&A, 383, 791
\bibitem[Napolitano et al.(2009)]{Napolitano09} Napolitano, N.R.,
Romanowsky, A.J., Coccato, L., Capaccioli, M., Douglas, N.G.,
Noordermeer, E., Gerhard, O., Arnaboldi, M., De Lorenzi, F., Kuijken,
K., Merrifield, M.R., O'Sullivan, E., Cortesi, A., Das, P., and
Freeman, K.C., 2009, MNRAS, 393, 329
\bibitem[Napolitano et al.(2011)]{Napolitano11} Napolitano, N.R.,
Romanowsky, A.J., Capaccioli, M., Douglas, N.G., Arnaboldi, M.,
Coccato, L., Gerhard, O., Kuijken, K., Merrifield, M.R., Bamford,
S.P., Cortesi, A., Das, P., Freeman, K.C., 2011, MNRAS, 411, 2035
(N+11)
\bibitem[Napolitano et al.(2012)]{Napolitano12} Napolitano, N.R., Capozziello, S., Romanowsky, A.J., Capaccioli, M., Tortora, C.,  2012, ApJ, 748, 87
\bibitem[Navarro, Frenk \& White(1996)]{NFW96} Navarro, J.F., Frenk, C.S., White, S.D.M., 1996, ApJ, 462, 563
\bibitem[Navarro et al.(2004)]{Navarro04} Navarro, J.F., Hayashi, E., Power, C., Jenkins, A.R., Frenk, C.S., White, S.D.M., Springel, V., Stadel, J., Quinn, T.R., 2004, MNRAS, 349, 1039

\bibitem[Olive \& Pospelov(2008)]{sym2} Olive, K. A.,  Pospelov, M., 2008, Phys. Rev. D, 77, 043524
\bibitem[Oyaizu et al.(2008)]{Oyaizu08} Oyaizu, H., et al., 2008, Phys. Rev. D, 78, 123524

\bibitem[Padmanabhan(2003)]{Pad03} Padmanabhan, T. 2003, Physics Report, 380, 235
\bibitem[Peebles \& Rathra(2003)]{PB03} Peebles, P.J.E., Rathra, B. 2003,  Rev. Mod. Phys., 75, 559
\bibitem[Peng et al.(2004)]{Peng04} Peng, E.W., Ford, H.C., Freeman, K.C., 2004, ApJ, 602, 685
\bibitem[Persic et al.(1997)]{Persic97} Persic, M., Salucci, P., Stel, F., 1997, MNRAS, 281, 27
\bibitem[Prugniel \& Simien(1994)]{Prugniel94} Prugniel, P., Simien, F., 1994, A\&A, 321, 111
\bibitem[Puzia et al.(2004)]{Puzia04} Puzia, T.H., et al., 2004, A\&A, 415, 123

\bibitem[Romanowsky et al.(2003)]{Romanoswky03} Romanowsky, A.J., Douglas, N.G., Arnaboldi, M., Kuijken, K., Merrifield, M.R., Napolitano, N.R., Capaccioli, M., Freeman, K.C., 2003, Science, 301, 1696
\bibitem[Romanowsky et al.(2009)]{Romanowsky09} Romanowsky, A.J., Strader, J., Spitler, L.R., Johnson, R., Brodie, J.P., Forbes, D.A., Ponman, T., 2009, Astronomical Journal, 137, 4956

\bibitem[Sahni \& Starobinski(2000)]{Sahni} Sahni, V., Starobinski, A. 2000, Int. J. Mod. Phys. D, 9, 373
\bibitem[Salucci(2011)]{Salucci11} Salucci, P., Frigerio Martins, C., Lapi, A., arXiv:1102.1184
\bibitem[Sanchez et al.(2006)]{sanch05} Sanchez, A.G. et al. 2006, MNRAS, 366, 189
\bibitem[Sanders(2005)]{Sanders05} Sanders, R.~H., 2005, MNRAS, 363, 459
\bibitem[Schuberth(2010)]{Schuberth10} Schuberth, Y., Richtler, T., Hilker, M., Dirsch, B., Bassino, L.P., Romanowsky, A.J., Infante, L., 2010, A\&A, 513, A52
\bibitem[Seljak et al.(2005)]{Sel04} Seljak, U. et al. 2005, Physical Review D, 71, 103515
\bibitem[Shen \& Gebhardt(2010)]{Shen10} Shen, J., Gebhardt, K., 2010, ApJ, 711, 484
\bibitem[Stabile \& Stabile (2012)]{stabile} Stabile, A.,  Stabile, An.,  2012, PRD, 85, 044014
\bibitem[Starobinsky(2007) ]{fr2} A.~A.~Starobinsky, 2007, JETP Lett.\  86, 157-163
\bibitem[Swaters(2010) ]{Swaters10} R.~A.~Swaters, R.~H.~Sanders, S.~S.~McGaugh, 2010, ApJ, 718, 380
\bibitem[Tegmark et al.(2004)]{Teg03} Tegmark, M. et al. 2004, Physical Review D, 69, 103501
\bibitem[Teodorescu et al.(2010)]{Teo10} Teodorescu, A.M., M$\mathrm{\acute{e}}$ndez, R.H., Bernardi, F., Riffeser, A., Kudritzki, R.P., 2010, ApJ, 721, 369
\bibitem[Tonry et al.(2001)]{Tonry01} Tonry, J.L., Dressler, A., Blakeslee, J.P., Ajhar, E.A., Fletcher, A.B., Luppino, G.A., Metzger, M.R., and Moore, C.B., 2001, ApJ, 546, 681

\bibitem[Vikhlinin et al.(2005)]{Vikhlinin05} Vikhlinin, A., Markevitch, M., Murray, S.S., Jones, C., Forman, W., Van Speybroeck, 2005, ApJ, 628, 655

\bibitem[Woodley et al.(2010)]{Woodley10} Woodley, K.A, G$\mathrm{\acute{o}}$mez, M., Harris, W.E., Geisler, D., Harris, G.L.H., 2010, Astronomical Journal, 139, 1871

\end{thebibliography}

{\renewcommand{\tabcolsep}{2.mm}
{\renewcommand{\arraystretch}{1.5}
\begin{table*}
\begin{center}
\caption{\textit{Clusters of galaxies: Scalar field.} Column 1: Name of the cluster. Column 2: Coupling constant of scalar field and galaxy component.
  Column 3: Coupling constant of scalar field and gas component. Columns 4: Scalar field interaction length. \label{tab:clusterdata}}
\begin{tabular}{cccc}
  \hline
  &  $\beta_{gal}$                      & $\beta_{gas}$                     & $L$  \\
  &                                     &                                   &(kpc) \\
  \hline
  \hline
  A133 & $2.230^{+0.778}_{-1.080}$ & $2.645^{+0.202}_{-0.212}$ & $1289.18^{+145.62}_{-188.13}$ \\
  A262 & $0.117^{+0.308}_{-0.086}$ & $5.687^{+1.719}_{-1.099}$ & $73.628^{+33.572}_{-22.954}$ \\
  A383 & $0.107^{+0.337}_{-0.079}$ & $2.584^{+0.106}_{-0.125}$ & $547.703^{+84.917}_{-50.069}$ \\
  A478 & $0.129^{+0.415}_{-0.097}$ & $2.410^{+0.214}_{-0.165}$ & $802.668^{+292.035}_{-205.815}$ \\
  A907 & $0.435^{+3.472}_{-0.377}$ & $2.922^{+0.254}_{-2.500}$ & $523.430^{+693.474}_{-78.797}$ \\
  A1413 & $0.452^{+0.830}_{-0.390}$ & $2.501^{+0.064}_{-0.096}$ & $1305.87^{+76.33}_{-83.37}$ \\
  A1795 & $0.154^{+0.532}_{-0.118}$ & $2.568^{+0.207}_{-0.198}$ & $727.922^{+389.727}_{-174.47}$ \\
  A1991 & $0.184^{+0.635}_{-0.145}$ & $2.823^{+0.144}_{-0.138}$ & $532.531^{+76.564}_{-33.074}$ \\
  A2029 & $0.131^{+0.473}_{-0.096}$ & $2.516^{+0.210}_{-0.180}$ & $905.584^{+275.600}_{-271.873}$ \\
  A2390 & $0.147^{+0.390}_{-0.105}$ & $2.123^{+0.076}_{-0.083}$ & $1465.02^{+94.02}_{-104.32}$ \\
  MKW4 & $0.141^{+0.449}_{-0.106}$ & $7.169^{+2.554}_{-1.566}$ & $75.932^{+36.494}_{-25.118}$ \\
  RXJ1159 & $2.454^{+0.355}_{-0.273}$ & $3.206^{+0.124}_{-0.162}$ & $528.456^{+45.442}_{-27.393}$ \\
  \hline
  \hline
\end{tabular}
\end{center}
\end{table*}}}

{\renewcommand{\tabcolsep}{2.mm}
{\renewcommand{\arraystretch}{1.5}
\begin{table*}
\begin{center}
\caption{\textit{Spiral galaxies: Scalar field.} Column 1: Name of the LSB galaxy. Column 2: Coupling constant of scalar field and star component.
  Column 3: Coupling constant of scalar field and gas component. Columns 4: Scalar field interaction length. Column 5: Stellar mass-to-light ratio in the V-band. \label{tab:spiraldata}}
\begin{tabular}{ccccc}
  \hline
  &  $\beta_{star}$                      & $\beta_{gas}$                     & $L$                      & $Y_{\ast}$ \\
  &                                      &                                   & (kpc)                    & $(Y_{\odot})$   \\
  \hline
  \hline
  UGC1230 & $4.831^{+3.910}_{-2.243}$ & $1.167^{+0.159}_{-0.167}$ & $57.778^{+94.884}_{-23.604}$ & $0.201^{+0.474}_{-0.139}$ \\
  UGC1281 & $0.283^{+2.620}_{-0.234}$ & $1.347^{+0.090}_{-0.077}$ & $7.687^{+13.748}_{-3.482}$   & $2.775^{+0.493}_{-2.574}$  \\
  UGC3137 & $4.400^{+4.032}_{-1.884}$ & $1.970^{+0.021}_{-0.019}$ & $484.99^{+2106.55}_{-382.01}$ & $0.409^{+0.781}_{-0.294}$ \\
  UGC3371 & $3.452^{+3.644}_{-1.985}$ & $1.726^{+0.316}_{-0.178}$ & $10.543^{+28.614}_{-6.046}$ & $0.350^{+1.283}_{-0.268}$ \\
  UGC3851 & $1.243^{+2.258}_{-0.476}$ & $0.119^{+0.315}_{-0.086}$ & $2.203^{+3.199}_{-0.933}$ & $2.144^{+1.866}_{-1.806}$ \\
  UGC4173 & $4.096^{+3.490}_{-2.389}$ & $0.700^{+0.368}_{-0.528}$ & $11.995^{+51.799}_{-7.377}$ & $0.201^{+0.657}_{-0.138}$ \\
  UGC4278 & $3.479^{+3.347}_{-1.573}$ & $1.328^{+0.063}_{-0.062}$ & $96.241^{+102.113}_{-33.083}$ & $0.242^{+0.498}_{-0.176}$ \\
  UGC4325 & $0.316^{+2.485}_{-0.268}$ & $2.838^{+0.407}_{-0.291}$ & $3.853^{+6.053}_{-1.901}$ & $2.924^{+1.420}_{-2.716}$ \\
  UGC5721 & $0.180^{+0.512}_{-0.132}$ & $3.486^{+1.049}_{-0.459}$ & $1.454^{+0.567}_{-0.586}$ & $1.468^{+0.866}_{-1.116}$ \\
  UGC7524 & $0.721^{+0.278}_{-0.222}$ & $1.462^{+0.292}_{-0.547}$ & $1.022^{+0.394}_{-0.156}$ & $5.395^{+1.356}_{-1.537}$ \\
  UGC7603 & $0.092^{+0.204}_{-0.066}$ & $1.782^{+0.038}_{-0.037}$ & $31.176^{+105.046}_{-19.245}$ & $0.052^{+0.071}_{-0.035}$ \\
  UGC8286 & $3.069^{+2.092}_{-0.923}$ & $2.169^{+0.073}_{-0.054}$ & $34.052^{+167.546}_{-25.431}$ & $0.161^{+0.151}_{-0.101}$ \\
  UGC8837 & $0.124^{+0.327}_{-0.092}$ & $2.009^{+0.263}_{-0.189}$ & $2.460^{+6.386}_{-0.121}$ & $0.111^{+0.289}_{-0.082}$ \\
  UGC9211 & $3.575^{+3.367}_{-1.883}$ & $1.579^{+0.439}_{-0.192}$ & $5.670^{+13.011}_{-2.933}$ &  $0.280^{+0.858}_{-0.204}$ \\
  UGC10310 & $0.122^{+0.582}_{-0.091}$ & $1.543^{+0.701}_{-0.278}$ & $3.450^{+9.622}_{-2.397}$ &  $2.535^{+0.465}_{-1.234}$ \\
  \hline
  \hline
\end{tabular}
\end{center}
\end{table*}}}

\end{document}